\documentclass[a4paper,aps,pre,twocolumn,showpacs,nofootinbib,superscriptaddress,fleqn]{revtex4-1}

\usepackage{mathtools}
\usepackage{subfigure}
\usepackage{lipsum}
\usepackage{amsmath}
\usepackage{amssymb}
\usepackage{graphicx}
\usepackage{etoolbox} 
\usepackage{amsmath}
\usepackage{dsfont}
\usepackage{amsthm}
\usepackage{bold-extra}
\usepackage[title]{appendix}
\usepackage[usenames,dvipsnames,svgnames,table]{xcolor}
\usepackage[margin=30pt,font=small,labelfont=bf,labelsep=endash]{caption}
\usepackage{amsthm}
\usepackage{rotating}
\usepackage{hyperref}
\usepackage{slashed} 
\usepackage{array,multirow}
\usepackage{shuffle}
\usepackage{empheq}
\usepackage{marvosym}
\usepackage{yfonts}
\usepackage[title]{appendix}
\usepackage[bottom]{footmisc}
\usepackage{tikz}
\usetikzlibrary{calc,matrix}
\usepackage{stackrel}
\usepackage{textcomp}
\usepackage{fancyvrb}
\usepackage{tcolorbox}
\usepackage{scalefnt}
\usepackage{setspace}
\usepackage{enumitem}
\usepackage{csquotes}
\usepackage{url}
\usepackage{hhline}
\usepackage[final]{pdfpages}

\newcommand{\cI}{{\cal I}}

\newcommand{\cN}{{\cal N}}

\newcommand{\dR}{\mathds{R}}
\newcommand{\dQ}{\mathds{Q}}
\newcommand{\dN}{\mathds{N}}

\newcommand{\rd}{\text{d}}

\newcommand{\ri}{\text{i}}

\newcommand{\ie}{{i.e.}}
\newcommand{\viz}{{viz.}}
\newcommand{\eg}{{e.g.}}

\newcommand{\Erdos}{{Erd\H{o}s}}

\newcommand{\tdel}{t_{\rm \scriptscriptstyle del}}
\newcommand{\kA}{k_{\rm \scriptscriptstyle A}}
\newcommand{\kB}{k_{\rm \scriptscriptstyle B}}
\newcommand{\nA}{n_{\rm \scriptscriptstyle A}}
\newcommand{\nB}{n_{\rm \scriptscriptstyle B}}
\newcommand{\nAB}{n_{\rm \scriptscriptstyle AB}}
\newcommand{\dotnA}{\dot n_{\rm \scriptscriptstyle A}}
\newcommand{\dotnB}{\dot n_{\rm \scriptscriptstyle B}}

\newcommand{\MA}{\text{M}_{\rm \scriptscriptstyle A}}
\newcommand{\MB}{\text{M}_{\rm \scriptscriptstyle B}}
\newcommand{\TA}{T_{\rm \scriptscriptstyle A}}
\newcommand{\TB}{T_{\rm \scriptscriptstyle B}}
\newcommand{\nAbar}{\bar n_{\rm \scriptscriptstyle A}}
\newcommand{\nBbar}{\bar n_{\rm \scriptscriptstyle B}}
\newcommand{\nXbar}{\bar n_{\rm \scriptscriptstyle X}}

\newcommand{\rmod}{\,\text{mod}\,}

\newcommand{\alphaB}{\alpha_{\tiny\text{\tiny B}}}
\newcommand{\alphaAB}{\alpha_{\tiny\text{\tiny AB}}}
\newcommand{\betaB}{\beta_{\tiny\text{\tiny B}}}
\newcommand{\betaAB}{\beta_{\tiny\text{\tiny AB}}}
\newcommand{\gammaA}{\gamma_{\tiny\text{\tiny A}}}
\newcommand{\gammaAB}{\gamma_{\tiny\text{\tiny AB}}}
\newcommand{\tauA}{\tau_{\tiny\text{\tiny A}}}
\newcommand{\tauB}{\tau_{\tiny\text{\tiny B}}}
\newcommand{\nuA}{\nu_{\tiny\text{\tiny A}}}
\newcommand{\nuB}{\nu_{\tiny\text{\tiny B}}}

\numberwithin{equation}{section}
\definecolor{shadecolor}{rgb}{0.97,0.97,0.97}
\definecolor{myblue}{rgb}{0.97,0.97,0.97}

\makeatletter
\AtBeginDocument{\let\LS@rot\@undefined}
\makeatother

\begin{document}

\title{\bf\textcolor{black}{Influence of periodic external fields in multiagent models \\ with language dynamics}}

\author{Filippo Palombi}
\email{filippo.palombi@enea.it}
\affiliation{ENEA---Italian National Agency for New Technologies, Energy and Sustainable Economic Development\\ Via Enrico Fermi 45, 00044 Frascati -- Italy\\[1.0ex]}

\author{Stefano Ferriani}
\affiliation{ENEA---Italian National Agency for New Technologies, Energy and Sustainable Economic Development\\ Via Martiri di Monte Sole 4, 40129 Bologna -- Italy\\[1.0ex]}

\author{Simona Toti}
\affiliation{ISTAT---Italian National Institute of Statistics\\ Via Cesare Balbo 16, 00184 Rome -- Italy}

\date{\today}

\begin{abstract}
  We investigate large-scale effects induced by external fields, phenomenologically interpreted as mass media, in multiagent models evolving with the microscopic dynamics of the binary naming game. In particular, we show that a single external field, broadcasting information at regular time intervals, can reverse the majority opinion of the population, provided the frequency and the effectiveness of the sent messages lie above well-defined thresholds. We study the phase structure of the model in the mean field approximation and in numerical simulations with several network topologies. We also investigate the influence on the agent dynamics of two competing external fields, periodically broadcasting different messages. In finite regions of the parameter space we observe periodic equilibrium states in which the average opinion densities are reversed with respect to naive expectations. Such equilibria occur in two cases: (i) when the frequencies of the competing messages are different but close to each other; (ii) when the frequencies are equal and the relative time shift of the messages does not exceed half a period. We interpret the observed phenomena as a result of the interplay between the external fields and the internal dynamics of the agents and conclude that, depending on the model parameters, the naming game is consistent with scenarios of first- or second-mover advantage (to borrow an expression from the jargon of business strategy).
\end{abstract}

\maketitle

\section{Introduction}\label{sect:1}

The cultural debate on the role of media in modern societies has received contributions from social and behavioral sciences as well as from the humanities in the past century. Following the advent of network science~\cite{Barabasi}, the influence of mass communication on opinions, attitudes and actions of people has likewise attracted the interest of scholars working in statistical mechanics of complex systems. As known, at the heart of this research field lies a great variety of agent-based models, specifically designed to quantitatively explore diversified aspects of diffusive phenomena, such as the spread of opinions across social networks (see Ref.~\cite{Castellano:1} for a review). It is remarkable that many such models display convergence to a macroscopic \emph{consensus} state, in which all agents share the same opinion as an ultimate consequence of their local interactions. Investigating whether and how the path to consensus is favored or hampered by the presence of media has been the subject of a significant body of research in various theoretical setups. Yet, the up-to-date analysis is still limited to a partial subset of the models populating the scientific literature. 

The matter has been recently reviewed in Ref.~\cite{Sirbu1}. Here, agent-based models are classified into groups with one-dimensional or multidimensional opinions and models within each group are then further divided depending on whether opinions are discrete or continuous. Relevant studies of the influence of media on opinion dynamics include Refs.~\cite{Crokidakis,Sznajd,Hu,Gonzalez:1,Laguna,Lu,Colaiori}~(one-dimensional discrete models), \cite{Carletti,Gargiulo:1,Martins,Hegselmann:1,Kurz,Quattrociocchi:1}~(one-dimensional continuous models), \cite{Gonzalez:2,Gonzalez:3,Peres:1,Peres:2,Gonzalez:4,Gonzalez:5,Rodriguez:1,Mazzitello,Rodriguez:2,Candia,Gandica}~(multidimensional discrete models) and \cite{Quattrociocchi:2,Sirbu:2,Sirbu:3}~(multidimensional continuous models). In all of these studies, media are represented as an \emph{external} source of information, variously affecting the \emph{internal} dynamics of the agents.  

\begin{figure*}[t!]
  \begin{minipage}{0.54\textwidth}
    \centering
    \includegraphics[width=1.0\textwidth]{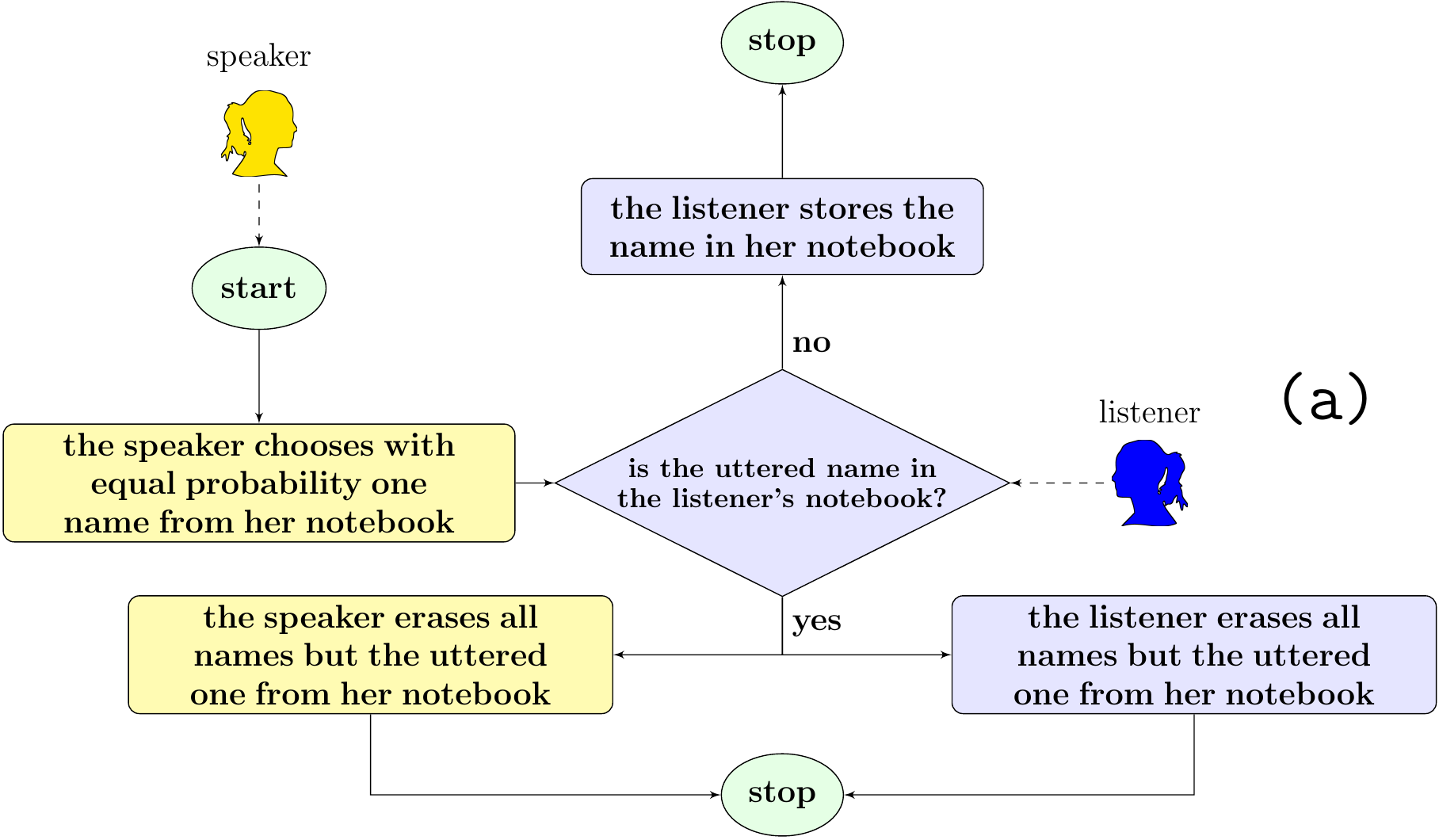}
  \end{minipage}
  \begin{minipage}{0.34\textwidth}
    \centering
    \scriptsize
    \begin{tabular}{|c|c|}
      \hline
      before interaction  \raisebox{8pt} & after interaction \\
      \hline
      \\[-2.8ex]
      ${\rm S} \to {\rm L}$ \raisebox{10pt} & ${\rm S} - {\rm L}$ \\
      \hline\\[-2.8ex]
      $\phantom{B}{\rm A}\stackbin{\rm A}{\to}{\rm A}\phantom{A}$ \raisebox{12pt} & $\phantom{B}{\rm A} - {\rm A}\phantom{B}$ \\
      $\phantom{B}{\rm A}\stackbin{\rm A}{\to}{\rm B}\phantom{A}$ \raisebox{0pt} & $\phantom{B}{\rm A} - {\rm AB}$ \\
      $\phantom{B}{\rm A}\stackbin{\rm A}{\to}{\rm AB}$ \raisebox{0pt} & $\phantom{B}{\rm A} - {\rm A}\phantom{B}$ \\
      $\phantom{A}{\rm B}\stackbin{\rm B}{\to}{\rm A}\phantom{B}$ \raisebox{0pt} & $\phantom{A}{\rm B} - {\rm AB}$ \\
      $\phantom{A}{\rm B}\stackbin{\rm B}{\to}{\rm B}\phantom{A}$ \raisebox{0pt} & $\phantom{A}{\rm B} - {\rm B}\phantom{A}$ \\
      $\phantom{A}{\rm B}\stackbin{\rm B}{\to}{\rm AB}$ \raisebox{0pt} & $\phantom{A}{\rm B} - {\rm B}\phantom{A}$ \\
      ${\rm AB}\stackbin{\rm A}{\to}{\rm A}\phantom{B}$ \raisebox{0pt} & $\phantom{B}{\rm A} - {\rm A}\phantom{B}$ \\
      ${\rm AB}\stackbin{\rm A}{\to}{\rm B}\phantom{A}$ \raisebox{0pt} & ${\rm AB} - {\rm AB}$ \\
      ${\rm AB}\stackbin{\rm A}{\to}{\rm AB}$ \raisebox{0pt} & $\phantom{B}{\rm A} - {\rm A}\phantom{B}$ \\
      ${\rm AB}\stackbin{\rm B}{\to}{\rm A}\phantom{B}$ \raisebox{0pt} & ${\rm AB} - {\rm AB}$ \\
      ${\rm AB}\stackbin{\rm B}{\to}{\rm B}\phantom{A}$ \raisebox{0pt} & $\phantom{A}{\rm B} - {\rm B}\phantom{A}$ \\
      ${\rm AB}\stackbin{\rm B}{\to}{\rm AB}$ \raisebox{0pt} & $\phantom{A}{\rm B} - {\rm B}\phantom{A}$ \\
      \hline
    \end{tabular}{\large\tt \ \ (b)}
  \end{minipage}
  \vskip -0.1cm
  \caption{\footnotesize {\color{red} [color online]} (a) NG interaction flowchart; (b) table of interactions (S\,=\,speaker, L\,=\,listener).\label{fig:gamerule}}
  \vskip -0.3cm
\end{figure*}

A quick glance to the above literature is sufficient to realize that external fields have never been considered in the context of multiagent models expressly conceived to study the emergence of spoken languages, except for Ref.~\cite{Lu} (see below). The naming game ({\bf NG}) is the simplest and most popular of such models. Inspired by the pioneering work of Refs.~\cite{Luc:1,Luc:2}, it was first proposed in Ref.~\cite{Baronchelli:1} after several attempts had been previously made to ascribe the origin of language conventions to evolutionary mechanisms~\cite{Niyogi,Nowak:1,Nowak:2,Nowak:3,Smith,Komarova}. The NG is a language game in the sense of Ref.~\cite{Wittgenstein}. Its goal is to name an initially unnamed object. Each agent partaking in the game is endowed with a notebook, in which he/she writes names. Elementary interactions involve two agents, playing respectively as \emph{speaker} and \emph{listener}. In each interaction, the speaker is chosen randomly among the agents, while the listener is chosen randomly among the speaker's neighbors. The binary version of the model, introduced in Ref.~\cite{Baronchelli:4}, assumes only two competing names, conventionally denoted by letters A and~B. In this case, the basic interaction rule is equivalently described by either the flowchart reported in Fig.~\ref{fig:gamerule} (a) or the table reported in Fig.~\ref{fig:gamerule} (b). An in-depth bibliographic analysis of articles concerning the NG would take up this whole paper, since a considerable literature was produced on the subject over the past decade. To avoid this, we refer the reader to Ref.~\cite{Baronchelli:10}, where a recent review of the model is provided.

The aim of this work is to investigate the influence of media on language dynamics.\footnote{Throughout the paper we use interchangeably the terms \emph{media} and \emph{external field(s)}.} More precisely, we study their effects on a population of agents playing the binary NG. There are at least three good reasons for doing this.

\indent (i) Languages never remain static. They keep on evolving in various ways, \eg via the introduction of neologisms that describe tools, concepts and so forth, first entering common use. In modern societies media may favor or discourage the adoption of a given neologism, either directly or indirectly. Take, for example, the verb \emph{tweet}. An anecdote concerning it circulates on the Web. In 2010, P.~B.~Corbett, standards editor at the NYT, advised his colleagues to abstain from using that word in news articles in the social-networking sense. As he wrote in his blog, \emph{``I~had suggested that outside of ornithological contexts, {\sl tweet} should still be treated as colloquial rather than as standard English. It can be used for special effect, or in places where a colloquial tone is appropriate, but should not be used routinely in straight news articles.''}~\cite{Tweet}. Corbett's position, once it became of public domain, triggered a swift and widespread negative reaction outside the NYT, as he himself observed. The epilogue of the story is well known: Corbett's opposers had the best. Despite the cascade effects that the opinion of an influential NYT editor might have had on journalists throughout the world, the verb \emph{tweet}, which was already in the Oxford English Dictionary in relation to the chirp of small birds, got an additional definition in the June 2013 release~\cite{OED}. This marked the end of the debate.     
   
\indent (ii) The application domain of the NG goes beyond the mere understanding of how language conventions arise. If we interpret letters A and B as opposite arguments rather than as names, one argument being \emph{in favor of} and one \emph{against} a given choice, we can regard an agent in state A or B as one who has made the choice (A and B are certainty states) and an agent in state AB as one who has not made it yet and holds both arguments for further consideration (AB is an uncertainty state). In other words, language dynamics is a valid example of discrete opinion dynamics.\footnote{\raggedbottom Notice that the notion of opinion uncertainty conveyed by the agent state AB is different from the one characterizing bounded confidence models, such as the Deffuant-Weisbuch~\cite{Deffuant} or the Hegselmann-Krause~\cite{Hegselmann:2} ones. While in the NG an agent in state AB can interact with a neighbor independently of the latter's state, in bounded confidence models the uncertainty, measured by a bounded confidence parameter, determines which neighbors of an agent are allowed to interact with him/her.} It is natural to represent media in this framework as an external force favoring one opinion, say opinion A, and acting on the agents so as to (a)~instill uncertainty into those adopting opinion~B and (b)~convince the uncertain ones to abandon opinion~B. There are plenty of real-life situations where such dichotomous dynamics occur in first approximation. Just as a reference example the reader may consider the longstanding debate concerning the legitimacy of the European institutions. Since several years EU political elites are addressing the problem of how improving the level of acceptance of the European governance among citizens. The issue of the EU legitimacy has been the subject of scholarly research (see, for instance, Refs.~\cite{Beetham,Chryssochoou,Banchoff,Arnull}). Specifically, the potential impact of the media coverage in the EU legitimacy process has been discussed in Refs.~\cite{Baetens,Liebert,Medrano,Hurrelmann}. If, in the framework of the NG, we represent citizens supporting the EU institutions as agents in state A and citizens opposing to them as agents in state B, then we are naturally led to investigate the extent to which a media campaign promoting opinion A is able to globally influence the population.

\indent (iii) Theoretical predictions for how the speed of convergence to consensus depends upon the connectivity of agents playing the NG have been recently put to the test in Web-based live games with controlled design and have been shown to correctly reproduce experimental observations~\cite{Centola}. It is tempting to conjecture that the correspondence between model and real world is preserved also when one or more external fields influence the process of consensus formation. If this is correct, what does the NG have to tell us about exogenously directed consensus?

It must be observed that in most of the literature reviewed in Ref.~\cite{Sirbu1} the strength of media is described by a constant parameter $\alpha\in(0,1)$, measuring the probability that an agent selected at a given microscopic time interacts with the external source of information rather than with a neighbor. A significant exception is found in Refs.~\cite{Carletti,Martins}, where media act on the agent dynamics as a periodic perturbation affecting instantaneously the whole population. There is an interpretative difference between these two approaches. In the former, for any macroscopic time interval $\Delta T$ a fraction $\alpha\Delta T>0$ is spent by agents in asynchronous interactions with media; in the latter, the internal dynamics controls the system with no external influence between any two subsequent impulsive perturbations, \ie, in practice it rules all the macroscopic time up to zero-measure intervals. Broadly speaking, the first approach suggests a picture where agents leave TV (or radio) switched on and watch (or listen to) it from time to time, whereas the second one describes a situation where short messages of an advertising campaign are sent on TV (or radio) at peak hours of the day, when a large fraction of the population is watching (or listening to) it. In the example of the EU legitimacy, it would seem more natural to investigate the effects on European societies of a media campaign promoting the EU institutions or, in consideration of the recent rise of Euro-sceptic political movements and parties in several European countries, to study the hypothetical clash of two different media campaigns respectively promoting and denigrating the EU. With this in mind, we choose to follow the approach of Refs.~\cite{Carletti,Martins}. The other one has been discussed in Ref.~\cite{Lu} in the specific case of agents playing the NG on empirical community-based~networks.

To summarize the key points of our study, we first consider a setup in which all agents are initially in state~B and media broadcast messages advertising opinion A at regular macroscopic time intervals. We find that the media campaign is able to convert the whole population to opinion A within a finite macroscopic time only provided frequency and effectiveness of the advertising messages lie above specific thresholds. In other words, under proper conditions, the action of media induces a sharp transition in the NG dynamics from a phase where opposite opinions stay forever in a state of periodic equilibrium to one where consensus on opinion A is reached within a finite time. We study the critical line of this phase transition in the mean field approximation and compare analytic estimates with numerical simulations. Then, we consider a different setup, in which equal fractions of agents are initially in states A and B and two competing media broadcast messages advertising, respectively, opinions A and B at regular macroscopic time intervals. We study how the equilibrium of the system depends on the relative frequency, the relative effectiveness, and the relative time shift of the media campaigns. We find that periodic equilibria exist, in which the average opinion densities are reversed with respect to what one would naively expect. Such states occur either when the frequencies of the competing messages are different but sufficiently close to each other, or when the frequencies are equal and the relative time shift of the messages does not exceed half a period. We argue that these states provide an explicit realization of a general scenario known to marketing strategists, in which the first entrant in a market is not able to capitalize on its advantage, thus leaving a new entrant the chance to compete more effectively and take dominance (second-mover advantage). This feature of the NG dynamics has never been noticed in the literature so far, to the best of our knowledge. We give an interpretation of it in terms of the interplay between the external fields and the internal dynamics of the agents.

\section{Phase transition induced by a single external field}\label{sect:2}

\begin{table}[!t]
  \begin{center}
    \small
    \begin{tabular}{|c|l|}
      \hline
      before interaction   \raisebox{10pt} & after interaction \\
      \hline
      \\[-2.8ex]
      ${\rm M} \to {\rm L}\phantom{A}$ \raisebox{10pt} & ${\rm M} - {\rm L}$ \\
      \hline\\[-2.8ex]
      ${\rm A}\stackbin{\rm A}{\to}{\rm A}\phantom{A}$ \raisebox{8pt} & ${\rm A} - {\rm A}\phantom{B}$ with prob. \ 1 \\
      ${\rm A}\stackbin{\rm A}{\to}{\rm B}\phantom{A}$ \raisebox{0pt} & ${\rm A} - {\rm AB}$ with prob. $\alphaB$ \\
      ${\rm A}\stackbin{\rm A}{\to}{\rm AB}$ \raisebox{0pt} & ${\rm A} - {\rm A}\phantom{B}$ with prob. $\alphaAB$ \\
      \hline
    \end{tabular}
    \vskip 0.0cm
    \caption{\footnotesize Table of media-listener interactions (M\,=\,media, L\,=\,listener).\label{tab:modelrules}}
  \end{center}
  \vskip -0.6cm
\end{table}

We consider $N$ agents, lying on the nodes of a graph (with edges representing acquaintances) and playing the binary NG. Each agent is assumed to be initially in state~B. We represent media as an additional fictitious agent M connected to the whole population, taking part in the game and invariably carrying opinion~A. When interacting, media always play the role of speaker. A media-agent interaction results in a change of the agent's state: agents in state B switch to state AB with probability $\alphaB$; agents in state AB switch to state A with probability $\alphaAB$; agents carrying opinion A retain it after the interaction. These rules are summarized in Table~\ref{tab:modelrules} for the reader's convenience. Often in the sequel we refer to $\alphaB$ and $\alphaAB$ as the effectiveness parameters of the media campaign.\footnote{In the context of the NG the idea of introducing probabilities ruling the outcome of an interaction was first considered in Ref.~\cite{Baronchelli:4}.} As mentioned in Sec.~\ref{sect:1}, we assume that media act like an impulsive periodic perturbation of the internal dynamics of the system, reaching instantaneously all agents. We let $T=\tau N$ be the period of the campaign. If we define a sweep as a number~$N$ of microscopic interactions (each agent plays as speaker once in a sweep on average) and take it as a macroscopic time unit, then $\tau\ge 1$ represents the macroscopic period. In this model, the media action is parametrized by the 3-tuple $(\alphaB,\alphaAB,\nu)$, where $\nu=1/\tau$ is the macroscopic frequency. We also assume that media send the first message at time $t=0$. Accordingly, all subsequent messages are sent at macroscopic times $t_\ell = \ell\tau$ for $\ell=1,2,3,$\,\ldots. Although the model is well defined for $\tau<1$ too, we consider only macroscopic periods in the sequel. 

\subsection{Mean field equations}\label{sect:2.1}

We begin our study from the mean field approximation. In principle, this is expected to correctly reproduce results of numerical simulations on a complete graph ({\bf CG}) in the thermodynamic limit, \ie, for $N\to\infty$. In this limit, stochastic fluctuations become fully negligible. To describe the state of the system at a given time, we introduce opinion densities
\begin{equation}
  n_\text{\tiny X}(t) = \frac{\text{no. of agents in state X at time }t}{N}\,,
\end{equation}
with $ \text{X} = \text{A},\,\text{B},\,\text{AB}\,$. In the mean field approximation, $n_\text{\tiny X}(t)$ is interpreted as an average over different realizations of the dynamics at time $t$. The opinion densities fulfill a simplex constraint, viz.,
\begin{equation}
  \nA(t) + \nB(t)+\nAB(t) = 1\,,\qquad \text{for all } t\ge 0\,.
  \label{eq:simplex}
\end{equation}
It follows that one of them is redundant. In the sequel we choose $\nA(t)$ and $\nB(t)$ as independent quantities, hence, we obtain $\nAB(t)$ from Eq.~(\ref{eq:simplex}). The rate of change of the opinion densities results from a balance of positive and negative contributions arising from microscopic interactions. Mean field equations ({\bf MFEs}) quantify this balance. The case with no media was first examined in Ref.~\cite{Baronchelli:4} and is well known. Media-agent interactions generate additional contributions of impulsive nature, altering the balance only for $t=t_\ell$. As such, they are mathematically represented by Dirac delta distributions. MFEs describing the complete system read  as
\begin{equation}
  \left\{\begin{array}{ll}
  \dotnA & = \left(1 - \nA - 2\nB + \nB^2\right)\\[1.0ex]
         & + \alphaAB\left(1 - \nA - \nB\right)\,\delta\left(t\rmod \tau\right)\,,\\[3.0ex]
  \dotnB & = \left(1 - \nB - 2\nA + \nA^2\right)\\[1.0ex]
         & - \alphaB\, \nB\,\delta\left(t\rmod\tau\right)\,.\end{array}\right.
\label{eq:mfesone}
\end{equation}
We aim at solving them with initial conditions
\begin{equation}
  \nA(0) =0\,,\quad \nB(0)=1\,.
\end{equation}
The reader will notice that we could have equivalently described media contributions by Dirac delta functions $\smash{\sum_{\ell=0}^\infty\delta(t-t_\ell)}$. The latter expression is more general in that it applies also when the impulsive perturbation occurs at arbitrary (\ie, not necessarily periodic) times. We remark that the introduction of media breaks explicitly the exchange symmetry $\nA\leftrightarrow\nB$ characterizing the original model. 

Ordinary differential equations with state-dependent impulsive perturbations are rather common in applied research, \eg, they are found in control theory~\cite{Grizzle,Morris,Jiang,Osipov,Cooke,Han,Bressan}, neuroscience~\cite{Markram,Catlla}, etc. Such equations must be carefully managed because the product of a Dirac delta and a discontinuous function is ill defined in a distributional sense. To make this point clear, we observe that according to Eqs.~(\ref{eq:mfesone})  $n_\text{\tiny X}(t)$ evolves continuously from $n_\text{\tiny X}(t_{\ell-1}^+)$ to $n_\text{\tiny X}(t_\ell^-)$ for $\ell\in\dN$, then it jumps to $n_\text{\tiny X}(t_\ell^+)$. The jump amounts to $\alphaAB(1-\nA-\nB)(t_\ell)$ for X\,=\,A and to $-\alphaB \nB(t_\ell)$ for X\,=\,B. Unfortunately, neither $\nA(t_\ell)$ nor $\nB(t_\ell)$ are well defined. This gives rise to an interpretative ambiguity that is expected to become significant for $\alphaB\simeq 1$ and/or $\alphaAB\simeq 1$, \ie, when the components of the perturbation are at their highest level. Several mathematical aspects of impulsive differential equations were originally discussed in Refs.~\cite{bainov,bainov:2}, but the effects of the ambiguity were first highlighted and quantified only in Ref.~\cite{Catlla}, by means of specific examples. In that paper, it was proposed to replace each $\delta(t-t_\ell)$ by a continuous function $\delta_\eta(t-t_\ell)$ approaching $\delta(t-t_\ell)$ in the limit $\eta\to 0$, and to take this limit only after analytic integration. It was also shown in a case study that the limiting solution thus obtained is independent of the specific shape of $\delta_\eta(t-t_\ell)$. Owing to this kind of robustness, the prescription of Ref.~\cite{Catlla} appears to be the best approach, known in literature, for solving differential equations with impulsive terms. 

\begin{figure*}[t!]
  \centering
  \includegraphics[width=0.78\textwidth]{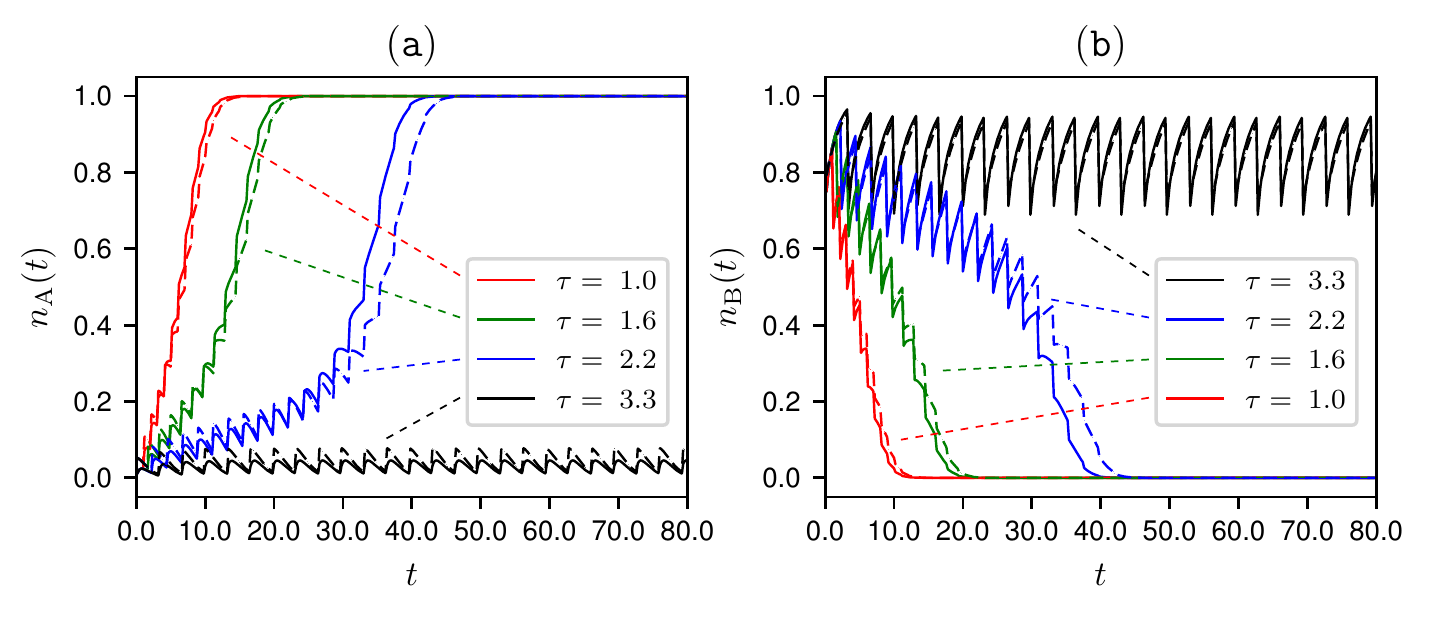}
  \vskip -0.5cm
  \caption{\footnotesize {\color{red} [color online]} (a) Time evolution of $\nA(t)$ for $\alphaB=\alphaAB=0.3$ and $\tau=1.0,1.6,2.2,3.3$; (b) time evolution of $\nB(t)$ for the same choice of parameters. In both plots, continuous lines correspond to the explicit integration scheme [Eqs.~(\ref{eq:explicit}], while dashed lines correspond to the implicit integration scheme [Eqs.~(\ref{eq:implicit})].\label{fig:figtwo}}
  \vskip -0.3cm
\end{figure*}

Unfortunately, we cannot mimic Ref.~\cite{Catlla} because no analytic solution to Eqs.~(\ref{eq:mfesone}) is known for $\alphaB=\alphaAB=0$, owing to the non-linear character of the agent-agent interactions. We can only pursue a numerical solution of the complete equations. In view of the above discussion, we should replace the Dirac delta function $\delta(t\rmod \tau)$ by a smooth counterpart $\delta_\eta(t\rmod \tau)$, then we should integrate the equations according to some numerical scheme with finite integration step $\rd t = \epsilon$ and we should finally perform the double limit $\eta,\,\epsilon\to 0$ numerically in the specified order. This turns out to be computationally expensive, hence, we choose to proceed differently. In particular, we find that discrepancies produced by different numerical integration schemes, corresponding to different interpretations of the impulsive terms, are acceptably small. To integrate Eqs.~(\ref{eq:mfesone}) numerically, we let $\tau = \tau_\epsilon\,\epsilon$ and accordingly $t_\ell = \ell\,\tau_\epsilon\,\epsilon$ for some $\tau_\epsilon\in\dN$.  Then, we replace
\begin{align}
  & \bullet \quad \dot n_\text{\tiny X}(t) \to \frac{n_\text{\tiny X}(t+\epsilon) - n_\text{\tiny X}(t)}{\epsilon}\,,\\[0.0ex]
  & \bullet \quad \delta(t-t_\ell) \to \frac{1}{\epsilon}\delta_{t/\epsilon,t_\ell/\epsilon}\,,
\end{align}
with $\delta_{ab}$ denoting the Kronecker symbol. An explicit integration scheme is represented by the recurrence
\begin{align}
  \left\{\begin{array}{ll}
  \nA\left(t+\epsilon\right) & = \nA\left(t\right) + \epsilon\cdot\left(1-\nA-2\nB+\nB^2\right)(t)\\[1.0ex]
                             & +\,\alphaAB\left(1-\nA-\nB\right)(t)\,\delta_{t/\epsilon,t_\ell/\epsilon}  \,,\\[2.0ex]
  \nB\left(t+\epsilon\right) & = \nB\left(t\right) + \epsilon\cdot\left(1-\nB-2\nA+\nA^2\right)(t)\\[1.0ex]
                             & -\,\alphaB\nB(t)\,\delta_{t/\epsilon,t_\ell/\epsilon}\,,\end{array}\right. \label{eq:explicit}
\end{align}
while a (partially) implicit scheme is given by 
\begin{align}
  \left\{\begin{array}{ll}
  \nA\left(t+\epsilon\right) & = \nA\left(t\right) + \epsilon\cdot\left(1-\nA-2\nB+\nB^2\right)(t)\\[1.0ex]
                             & +\,\alphaAB\left(1-\nA-\nB\right)(t+\epsilon)\,\delta_{t/\epsilon,t_\ell/\epsilon}  \,,\\[2.0ex]
  \nB\left(t+\epsilon\right) & = \nB\left(t\right) + \epsilon\cdot\left(1-\nB-2\nA+\nA^2\right)(t)\\[1.0ex]
                             & -\,\alphaB\nB(t+\epsilon)\,\delta_{t/\epsilon,t_\ell/\epsilon}\,,\end{array}\right. \label{eq:implicit}
\end{align}
both of them being defined for $t = 0,\,\epsilon,\,2\epsilon,$\,\ldots. Formally, the difference between Eqs.~(\ref{eq:explicit}) and (\ref{eq:implicit}) is just that in the former the solution at time $t+\epsilon$ appears only on the left-hand side, while in the latter it appears on both sides. In practice, opinion densities in impulsive terms are evaluated just before the spike in the explicit scheme and just after it in the implicit one, hence, the two methods are substantially different in relation to how they resolve the ambiguity of the impulsive terms. For many ordinary differential equations, explicit and implicit schemes converge as $\epsilon\to 0$. This is not the case with Eqs.~(\ref{eq:explicit}) and (\ref{eq:implicit}). As an example, in Fig.~\ref{fig:figtwo} we plot $\nA(t)$ and $\nB(t)$ vs. $t$ for $\alphaB = \alphaAB = 0.3$ and for a handful of values of~$\tau$ (in the plot we have $\epsilon = 5.0\times 10^{-4}$ and essentially no difference is observed for smaller $\epsilon$). Continuous lines correspond to the explicit scheme, while dashed lines correspond to the implicit one. We see that discrepancies exist, although they are of limited size. For larger values of $\alphaB$ and $\alphaAB$, the behavior of the opinion densities is still qualitatively the same in both schemes, yet we find that divergences gradually increase. We stress that our main motivation for studying the system in the mean field approximation is to gain analytic insight into its dynamics. Since statistical fluctuations in numerical simulations with small $N$ produce effects that can be shown to be comparable in size with those induced by the ambiguity of the impulsive terms in Eqs.~(\ref{eq:mfesone}), we conclude that the predictions of mean field theory are acceptably accurate as far as we are concerned. For this reason, in the sequel we adopt one integration scheme, specifically, the explicit one (it has a better match with numerical simulations, as seen \emph{a posteriori}), unless otherwise stated.

\begin{figure*}[t!]
  \centering
  \mbox{
    \subfigure{\includegraphics[width=0.46\textwidth]{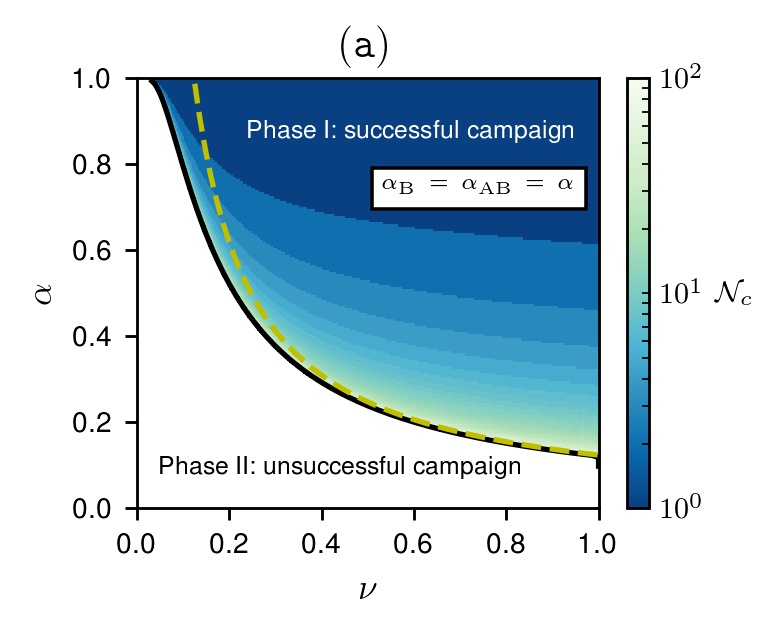}}\,
    \subfigure{\includegraphics[width=0.45\textwidth]{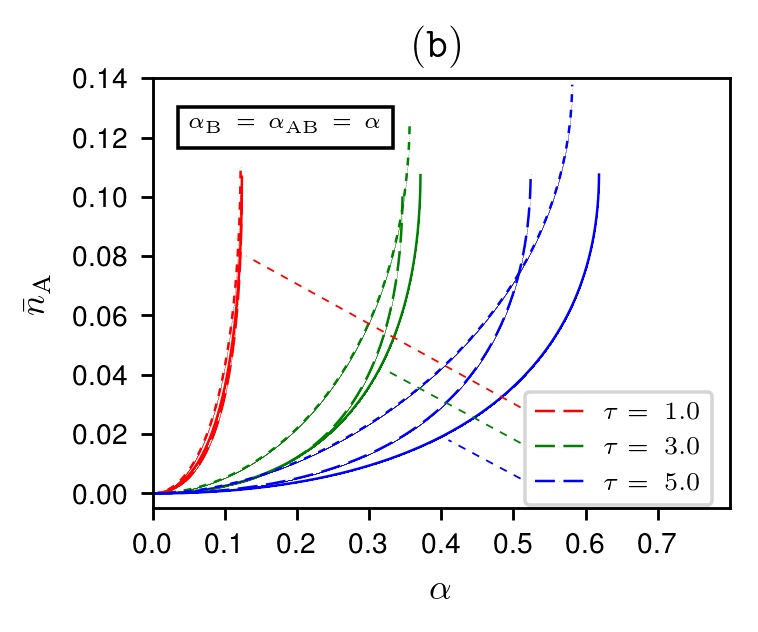}}
  }
  \vskip -0.2cm
  \caption{\footnotesize {\color{red} [color online]} (a) Density plot of $\cN_c$ for $\alphaB=\alphaAB=\alpha$ in the mean field approximation from numerical integration of MFEs; the yellow dashed line represents the approximate analytic solution, Eq.~(\ref{eq:hyperbole}); (b) average asymptotic density $\nAbar$ from numerical integration in the explicit scheme (dashed lines), in the implicit one (dotted lines), and from the approximate analytic solution (continuous lines).\label{fig:figthree}}
  \vskip -0.2cm
\end{figure*}

\vskip -0.05cm
Independently of the adopted scheme, we observe in Fig.~\ref{fig:figtwo} two possible asymptotic behaviors for the opinion densities: either $\nA(t)\to 1$ and accordingly $\nB(t)\to 0$ as $t\to\infty$ (this occurs, \eg, for $\tau = 1.0,1.6,2.2$) or both densities keep on oscillating around finite values without ever converging (this occurs, \eg, for $\tau = 3.3$). For any $(\alphaB,\alphaAB)$ there exists a critical value $\tau_c$ of $\tau$ for which the system switches from converging to consensus to  endlessly oscillating. This critical value is the largest one below which agents in state A overtake those in state B at some point in time. For instance, for $\alphaB = \alphaAB = 0.3$ we have $\tau_c \simeq 2.55$. In full generality, it makes sense to define
\begin{align}
  & t_*(\alphaB,\alphaAB,\tau) = \inf\left\{t\in\dR_+:\right. \nonumber\\[1.0ex]
  & \hskip 2.1cm \left. \nA(t)=\nB(t)\ \bigl|\ \alphaB,\alphaAB,\tau\,\right\}\,.
  \label{eq:t0def}
\end{align}
If $t_*<+\infty$, the system goes to consensus on opinion A with certainty. In principle, we could safely stop the media campaign at $t=t_*$: the internal dynamics of the binary NG would then guarantee that all agents carrying opinion~B are eventually converted to opinion~A. In the jargon of stochastic processes, $t_*$ is commonly referred to  as a \emph{first passage time}. The overall number of messages sent by media since the beginning of the game until $t_*$ amounts to 
\begin{equation}
  \cN_c(\alphaB,\alphaAB,\tau) = \left\lceil\frac{t_*(\alphaB,\alphaAB,\tau)}{\tau}\right\rceil\,,
\end{equation}
where the ceiling function $\lceil\,\cdot\,\rceil$ takes into account that the first message is sent by media at time $t=0$. Intuitively, $\cN_c$ can be taken as a measure of the lowest economic cost of the campaign.\footnote{The reader might wonder why we do not use the macroscopic time to consensus $t_\text{conv}$ in place of $t_*$ to define $\cN_c$. Indeed, for $t_*<+\infty$ and $t>t_*$ the action of media makes the system converge to consensus on opinion A at an exponential rate. Nevertheless, the exact convergence is only asymptotic at the level of MFEs, hence to define $\cN_c$ in terms $t_\text{conv}$ we should formally introduce a cutoff $\epsilon\ll 1$ and define operationally $t_\text{conv}$ as the first time at which $\nA=1-\epsilon$. This would make $\cN_c$ depend explicitly on $\epsilon$. By contrast, $t_*$ is well defined with no need for a cutoff.} Fig.~\ref{fig:figthree}(a) shows a density plot of $\cN_c$ for $\alphaB=\alphaAB=\alpha$. The plot highlights the existence of two sharply separated phases: phase~I corresponds to \emph{successful} campaigns, \ie\ pairs $(\alpha,\nu)$ for which $\cN_c<+\infty$, whereas phase~II corresponds to \emph{unsuccessful} campaigns, \ie\ pairs $(\alpha,\nu)$ for which $\cN_c=+\infty$. We observe that $\cN_c$ keeps low in a wide region of phase~I, while it surges in proximity of the critical line separating the two phases. 

\subsection{Approximate analytic solution for $\alphaB=\alphaAB=\alpha$}\label{sect:2.2}

Although Eqs.~(\ref{eq:mfesone}) cannot be solved analytically, it is possible to derive an approximate formula for the critical line. We said that if $(\alpha,\nu)$ belongs to phase~II, then the opinion densities oscillate endlessly around asymptotic average values. These are formally defined as
\begin{equation}
  \bar n_\text{\tiny X} = \lim_{t\to\infty}\frac{1}{\tau}\int_{t}^{t+\tau}n_\text{\tiny X}(\sigma)\,\rd\sigma\,,\quad \text{X} = \text{A},\,\text{B}\,.
  \label{eq:periodicdens}
\end{equation}
The oscillations can be regarded as periodic equilibrium states of the system, because their width is limited in size. When crossing the critical line, the asymptotic average densities undergo a jump discontinuity. Hence, the formula we are looking for can be read off as a singularity in the analytic behavior of $\nAbar$ and $\nBbar$. Since the system never relaxes to a real dynamical equilibrium, characterized by  steady densities, we are not allowed to impose asymptotic conditions $\rd n_\text{\tiny X}/\rd t = 0$ simultaneously for X$\,=\,$A,\,B in Eqs.~(\ref{eq:mfesone}). However,  if $(\alpha,\nu)$ belongs to phase II, then integrating both sides over a period $\tau$ and letting $t\to\infty$ yields, respectively,
\begin{align}
  0 & = \lim_{t\to\infty}\int_t^{t+\tau}\frac{\rd \nA}{\rd t}(\sigma)\,\rd \sigma = \tau\left(1- \nAbar-2\nBbar + \nBbar^2\right) \nonumber\\[2.0ex]
    & + \tau\lim_{t\to\infty}\frac{1}{\tau}\int_t^{t+\tau}\!\!\!\left[\nBbar-\nB(\sigma)\right]^2\,\rd\sigma\nonumber\\[2.0ex]
    & + \alpha\lim_{t\to\infty}\int_{t}^{t+\tau}\!\!\!\!\left(1 - \nA - \nB\right)(\sigma)\,\delta\left(\sigma\rmod \tau\right)\,\rd\sigma\,,
  \label{eq:periodicone}
\end{align}
and
\begin{align}
  0 & = \lim_{t\to\infty}\int_t^{t+\tau}\frac{\rd \nB}{\rd t}(\sigma)\,\rd \sigma = \tau\left(1- \nBbar-2\nAbar + \nAbar^2\right) \nonumber\\[2.0ex]
    & + \tau\lim_{t\to\infty}\frac{1}{\tau}\int_t^{t+\tau}\!\!\!\left[\nAbar-\nA(\sigma)\right]^2\,\rd\sigma\nonumber\nonumber\\[2.0ex]
    & - \alpha\lim_{t\to\infty}\int_{t}^{t+\tau}\!\!\!\!\nB(\sigma)\,\delta\left(\sigma\rmod \tau\right)\,\rd\sigma\,.
  \label{eq:periodictwo}
\end{align}
The first integral on the right-hand side of Eqs.~(\ref{eq:periodicone}) and (\ref{eq:periodictwo}) measures the width of the oscillations at periodic equilibrium. For small values of $\tau$, these are expected to be negligible in first approximation, hence, dropping the integral from both equations should not affect severely the average densities.  The second integral on the right-hand side is ill defined, as we previously explained. Whatever definition we adopt for it, it will depend somehow on the opinion densities at times just before and after the spike occurring between $t$ and $t+\tau$. In the assumption that the width of the oscillations is small, it looks natural to replace this integral by the corresponding asymptotic average density. The two approximations yield a simplified system of algebraic equations, namely,
\begin{equation}
  \left\{\begin{array}{l}
  0 = \tau\left(1-\nAbar-2\nBbar+\nBbar^2\right) + \alpha\left(1-\nAbar-\nBbar\right)\,,\\[2.0ex]
  0 = \tau\left(1-\nBbar-2\nAbar+\nAbar^2\right) - \alpha\,\nBbar\,.
  \label{eq:approxsys}
  \end{array}\right.
\end{equation}
This system can be solved exactly. To this aim, we observe that the first equation depends linearly upon $\nAbar$, while the second one depends linearly on $\nBbar$. Solving, for instance, the first equation with respect to $\nAbar$ yields by substitution
\begin{align}
  & \nBbar\left[\tau^3\nBbar^3 - 2\tau^2(\alpha+2\tau)\nBbar^2\right. \nonumber\\[0.0ex]
    & \hskip 1.5cm \left. + \tau(\alpha+2\tau)^2\nBbar -(\tau+\alpha)^3\right] = 0\,.
  \label{eq:quartic}
\end{align}
The simple root $\nBbar=0$ yields in turn $\nAbar=1$. This solution does not represent a state of periodic equilibrium, in fact, it describes densities belonging to phase I. Moreover, two of the roots of the cubic polynomial in square brackets are unphysical, hence they must be discarded. The third root, representing the physical solution, yields
\begin{align}
  \label{eq:nAbar}
  \nAbar & = 1 + \frac{\zeta(\alpha,\tau)-\alpha-2\tau}{\tau(\tau+\alpha)}\,\zeta(\alpha,\tau)\,,\\[1.0ex]
  \label{eq:nBbar}
  \nBbar & = \frac{\zeta(\alpha,\tau)}{\tau}\,,
\end{align}
with the auxiliary function $\zeta(\alpha,\tau)$ being given by
\begin{align}
  & \zeta(\alpha,\tau) = \, \frac{4}{3}\tau\, +\, \frac{2}{3}\alpha \nonumber\\[1.0ex]
  & \hskip 1.05cm -\,\frac{1}{12}\left(1+\ri\sqrt{3}\right)\sqrt[3]{\Gamma+12\,\sqrt{\Delta}}\nonumber\\[1.0ex]
  & \hskip 1.05cm - \frac{1}{3}\,\left(1 - \ri\sqrt{3}\right)\frac{(\alpha+2\tau)^2}{\sqrt[3]{\Gamma+12\,\sqrt{\Delta}}}\,,\label{eq:zetaaux} \\[3.0ex]
  & \Gamma = 44\,\tau^3+228\,\tau^2\alpha+276\,\tau\alpha^2+100\,\alpha^3\,, \\[3.0ex]
  & \Delta = 3\,(23\,\alpha^3+57\,\tau\alpha^2+33\,\tau^2\alpha-5\,\tau^3)(\tau+\alpha)^3\,.
\end{align}
It will be noticed that $\zeta(\alpha,\tau)$ is in general a complex function. In Appendix~\ref{sect:appA}, we prove that its imaginary part vanishes exactly provided $\Delta<0$. The latter condition defines phase II within the limits of the approximations yielding Eqs.~(\ref{eq:approxsys}). For $\Delta>0$, $\zeta(\alpha,\tau)$ is truly complex. In this case, Eqs.~(\ref{eq:nAbar}) and (\ref{eq:nBbar}) do not describe anymore physical states. In Fig.~\ref{fig:figthree}(b), we compare the asymptotic average density $\nAbar$ in our analytic approximation with those obtained numerically via the explicit and implicit integration schemes. Each line on the plot stops for $\alpha$ such that $(\alpha,\tau)$ belongs to the (scheme dependent!) critical line on the phase diagram. In the analytic approximation, this line corresponds to $\Delta = 0$, which is a polynomial equation. The root of $\Delta$ describing to the phase transition reads as
\begin{align}
  \alpha_c(\nu) & = \frac{1}{23}(9\cdot 2^{2/3}+6\cdot2^{1/3}-19)\,\frac{1}{\nu}\nonumber\\[2.0ex]
  & =\, \frac{0.123745\ldots}{\nu}\,.
  \label{eq:hyperbole}
\end{align}
In Fig.~\ref{fig:figthree}(a), this formula is represented by a yellow dashed curve. As expected, the approximation is very good for small $\tau$ and deteriorates progressively as $\tau$ increases. In particular, for $\tau\to\infty$, Eq.~(\ref{eq:hyperbole}) is affected by an unphysical divergence. 

\begin{figure*}[t!]
  \centering
  \mbox{
    \subfigure{\includegraphics[width=0.4\textwidth]{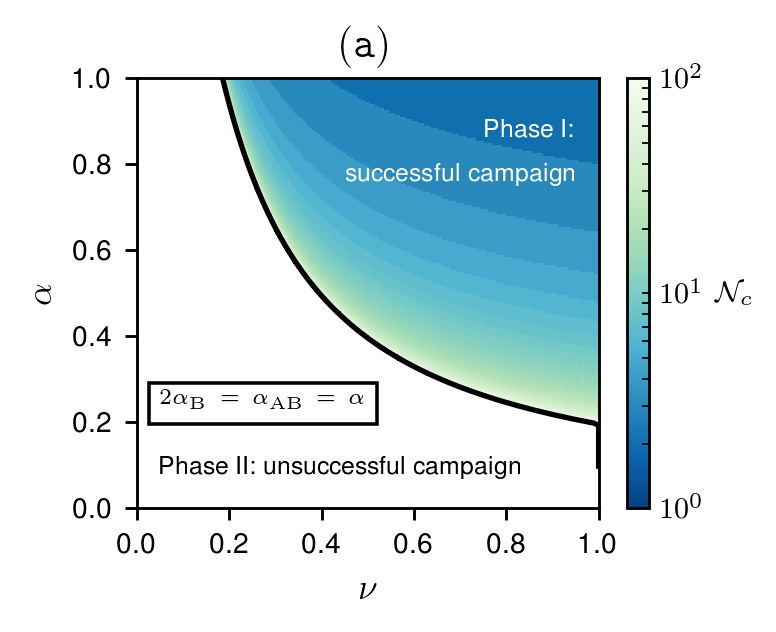}}
    \subfigure{\includegraphics[width=0.4\textwidth]{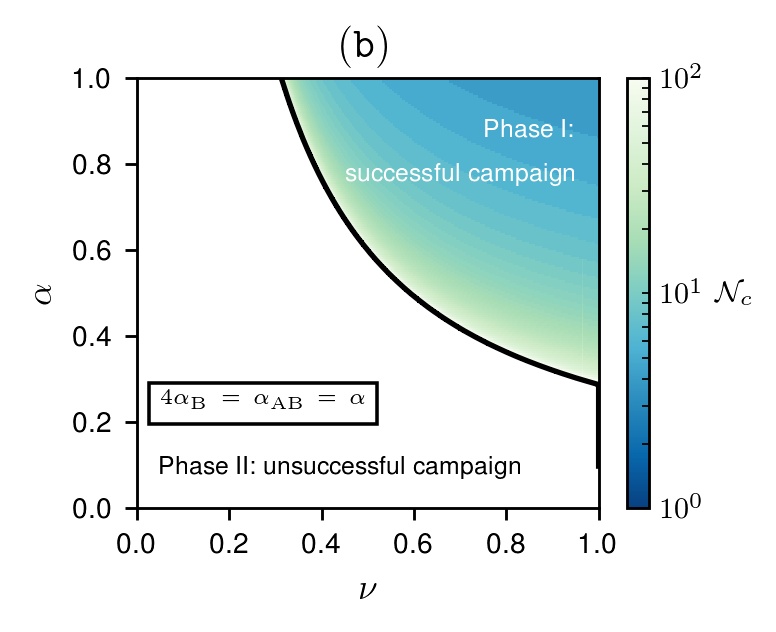}}
  }
  \\[-0.3cm]
  \mbox{
    \subfigure{\includegraphics[width=0.4\textwidth]{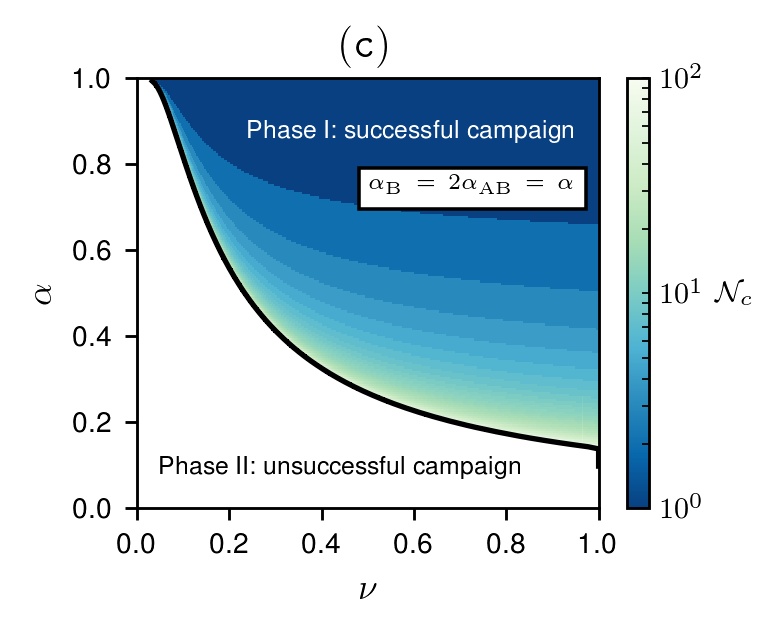}}
    \subfigure{\includegraphics[width=0.4\textwidth]{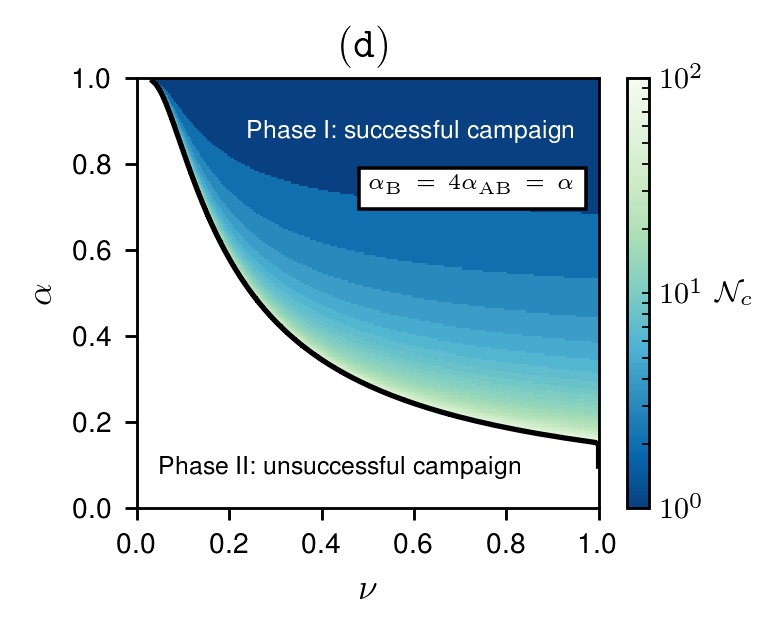}}
  }
  \\[-0.4cm]
  \mbox{
    \subfigure{\includegraphics[width=0.34\textwidth]{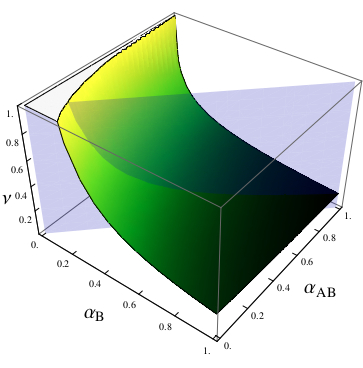}}
  }
  \vskip -3.5cm
  \hskip 8.0cm {\large {\tt (e)}}
  \vskip 2.9cm         
  \caption{\footnotesize {\color{red} [color online]} (a--d) Phase diagram in the mean field approximation from numerical integration of MFEs; (e) critical surface in the analytic approximation of Eq.~(\ref{eq:critsurf}); the intersection of the surface with the plane $\alphaB=\alphaAB$ is described by Eq.~(\ref{eq:hyperbole}).\label{fig:figfour}}
  \vskip -0.2cm
\end{figure*}

\subsection{Approximate analytic solution for $\alphaB\ne\alphaAB$}\label{sect:2.3}

For $\alphaB\ne\alphaAB$ it is equally possible to approximate and solve MFEs as discussed in the previous section. The physics is the same and the generalization is straightforward, yet the algebra is a bit more complicated. The quartic equation analogous to Eq.~(\ref{eq:quartic}) reads as
\begin{align}
  & \nBbar\left[\tau^3\nBbar^3 -2\tau^2(\alphaAB + 2\tau)\nBbar^2 + \tau(\alphaAB+2\tau)^2\nBbar\right.\nonumber\\[1.0ex]
    & \hskip 1.5cm\left. -(\tau+\alphaAB)^2(\tau+\alphaB)\right]=0\,.
\end{align}
The solution has the same analytic structure as in Eqs.~(\ref{eq:nAbar}) and (\ref{eq:nBbar}), \ie, the average densities at periodic equilibrium are given by
\begin{align}
  \nAbar & = 1 + \frac{\zeta(\alphaB,\alphaAB,\tau)-\alphaAB-2\tau}{\tau(\tau+\alphaAB)}\,\zeta(\alphaB,\alphaAB,\tau)\,,\\[1.0ex]
  \nBbar & = \frac{\zeta(\alphaB,\alphaAB,\tau)}{\tau}\,,
\end{align}
while the auxiliary function $\zeta(\alphaB,\alphaAB,\tau)$ now reads as
\begin{align}
  & \hskip -0.2cm \zeta(\alphaB,\alphaAB,\tau) = \, \frac{4}{3}\tau\, +\, \frac{2}{3}\alphaAB \nonumber\\[0.0ex]
  & \hskip 1.5cm -\,\frac{1}{12}\left(1+\ri\sqrt{3}\right)\sqrt[3]{\Gamma+12\,\sqrt{\Delta}} \nonumber\\[0.0ex]
  & \hskip 1.5cm - \frac{1}{3}\,\left(1 - \ri\sqrt{3}\right)\frac{(\alphaAB+2\tau)^2}{\sqrt[3]{\Gamma+12\,\sqrt{\Delta}}}\,,\\[3.0ex]
  \Gamma & = 44\tau^3 + 120\tau^2\alphaAB+60\tau\alphaAB^2-8\alphaAB^3 \nonumber\\[1.0ex]
  & +108\alphaB\tau^2+216\tau\alphaB\alphaAB+108\alphaB\alphaAB^2\,,\\[3.0ex]
  \Delta & = 3\left(-4\alphaAB^3+27\alphaB\alphaAB^2+3\tau\alphaAB^2+54\tau\alphaB\alphaAB\right.\nonumber\\[1.0ex]
  & \left. +6\tau^2\alphaAB+27\tau^2\alphaB-5\tau^3\right)(\tau\!+\!\alphaB)(\tau\!+\!\alphaAB)^2\,.
\end{align}
As in the special case $\alphaB=\alphaAB$, a phase transition occurs for $\Delta=0$, yet this equation describes in the most general case a critical surface on the three-dimensional phase diagram $(\alphaB,\alphaAB,\nu)$. Since $\alphaB$ and $\alphaAB$ are homogeneous quantities in that they measure the effectiveness of the impulsive perturbation, we choose to represent the critical surface by solving the equation $\Delta = 0$ with respect to $\tau$. Accordingly, we find 
\begin{align}
  \hskip -0.2cm\nu_c^{-1}(\alphaB,\alphaAB) & =  \frac{9}{5}\,\alphaB+\frac{2}{5}\,\alphaAB + \frac{3}{5}\,\sqrt [3]{\Theta+10\,\sqrt {H}}\nonumber\\[0.0ex]
  & +\frac{3}{5}\,\frac{ {\alphaAB}^{2}+14\,\alphaB\alphaAB+9\,{\alphaB}^{2}}{\sqrt [3]{\Theta+10\,\sqrt{H}}}\,,
  \label{eq:critsurf}
\end{align}
\begin{align}
& \Theta = 29\,\alphaB\alphaAB^2-\alphaAB^3+63\,\alphaB^2\alphaAB+27\,\alphaB^3\,,\\[2.0ex]
&  H = -\alphaAB^5\alphaB+\alphaB^2\alphaAB^4+\alphaAB^3\alphaB^3\,.
\end{align}
Fig.~\ref{fig:figfour} shows a sequence of density plots of $\cN_c$ analogous to Fig.~\ref{fig:figthree}(a), corresponding, respectively, to $\alphaAB=2\,\alphaB,\,4\,\alphaB$ (a), (b) and $\alphaB=2\,\alphaAB\,,4\,\alphaAB$ (c), (d). It also shows the critical surface described by Eq.~(\ref{eq:critsurf}) (e).\footnote{The reader will notice that letting $\alphaB=\alphaAB=\alpha$ in Eq.(\ref{eq:critsurf}) yields, $\alpha = \frac{5}{9\cdot 2^{2/3} + 12\cdot 2^{1/3} + 11}\frac{1}{\nu_c(\alpha)}$. There is no contradiction with Eq.~(\ref{eq:hyperbole}), since it holds $\frac{1}{5\cdot 23}(9\cdot 2^{2/3} + 6\cdot 2^{1/3} -19)\cdot({9\cdot 2^{2/3} + 12\cdot 2^{1/3} + 11})=1$.} All in all, the plots suggest that the phase structure of the model is much more sensitive to $\alphaB$ than to $\alphaAB$. This is intuitively understood by recalling that each message sent by  media transfers a fraction of agents from state B to state AB and from state AB to state A. If we even switch off $\alphaAB$, part of the agents brought to state AB by media moves subsequently to state A owing to the internal dynamics of the binary NG. The continual action of media eventually leads the system to global consensus on opinion A. The only effect of choosing $\alphaAB<\alphaB$ for given $\tau$ is represented by a mild increase of $\cN_c$. By contrast, if we switch off $\alphaB$, the system stays forever in the initial state with no dynamics whatsoever. In particular, in Fig.~\ref{fig:figfour} we have
\begin{align}
  & \cN_c(\alphaB,\alphaAB,\tau)|_{\alphaB=\alphaAB}\nonumber\\[1.0ex]
  & \hskip 2.0cm\lesssim\, \cN_c(\alphaB,\alphaAB,\tau)|_{\alphaB=2\alphaAB}\nonumber\\[1.0ex]
  & \hskip 2.0cm \lesssim\, \cN_c(\alphaB,\alphaAB,\tau)|_{\alphaB=4\alphaAB}\nonumber\\[1.0ex]
  & \hskip 2.0cm \ll\, \cN_c(\alphaB,\alphaAB,\tau)|_{\alphaAB=2\alphaB}\nonumber\\[1.0ex]
  & \hskip 2.0cm \ll\, \cN_c(\alphaB,\alphaAB,\tau)|_{\alphaAB=4\alphaB}\,.
\end{align}

\subsection{Dependence upon the initial conditions}\label{sect:2.4}

\begin{figure}[t!]
  \centering
  \includegraphics[width=0.46\textwidth]{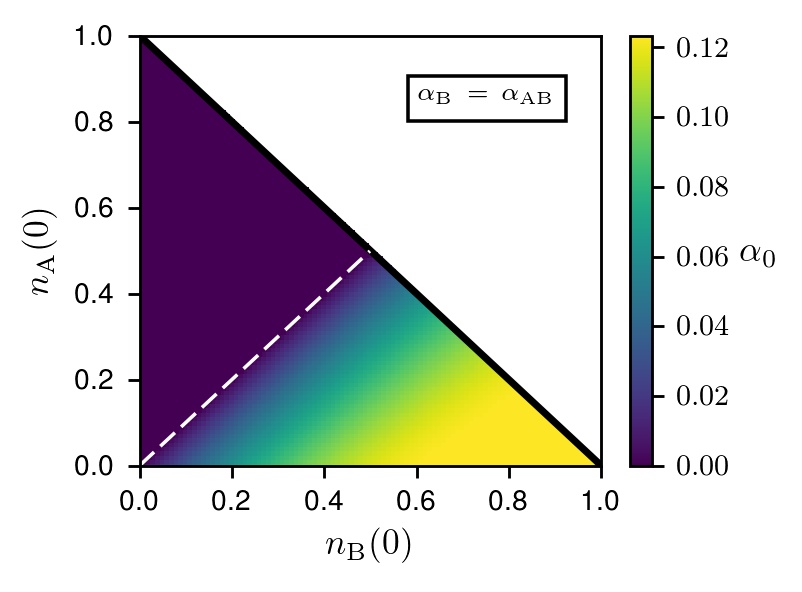}
  \vskip -0.2cm
  \caption{\footnotesize {\color{red} [color online]} density plot of $\alpha_0$ as a function of the initial densities.\label{fig:figfive}}
  \vskip -0.4cm
\end{figure}

The phase diagram for $\alphaB=\alphaAB=\alpha$ is characterized by a minimum value $\alpha_0$ of $\alpha$ below which the media campaign never succeeds, independently of $\tau\ge 1$. We should regard $\alpha_0$ as an \emph{absolute effectiveness threshold} for the action of media. Formally, $\alpha_0$ is defined by
\begin{align}
  \alpha_0 & = \min\left\{\alpha\in\dR_+:\right.\nonumber\\[0.0ex]
  & \hskip 0.5cm \left. t_*(\alphaB,\alphaAB,\tau)\bigr|_{\alphaB=\alphaAB=\alpha}<+\infty\ \text{ for }\ \tau\ge 1\right\} \nonumber\\[1.0ex]
  & = \min\left\{\alpha\in\dR_+:\right.\nonumber\\[0.0ex]
  & \hskip 0.5cm \left. t_*(\alphaB,\alphaAB,1)\bigr|_{\alphaB=\alphaAB=\alpha}<+\infty\right\} \,,
\end{align}
where the rightmost expression follows from the monotonic behavior of the critical line separating the phases of successful and unsuccessful campaigns. In Sec.~\ref{sect:2.2} we found that $\alpha_0 = 0.1237\ldots$ with good approximation. Although having $\alpha_0>0$ is phenomenologically encouraging (after all, not all media campaigns succeed in real life\ldots), its actual value looks somewhat unrealistic, since hardly could one imagine that a single real TV or radio spot is able to systematically convince about 12\% of the population. However, we argue that this value is not independent of the initial state of the system. So far, we assumed $\nB(0)=1$, which is an extreme condition. If we instead assume $\nB(0)<1$, the effort required to persuade the whole population to share opinion A decreases. Hence, the effectiveness threshold, that we now regard as a function $\alpha_0 = \alpha_0(\nA(0),\nB(0))$, decreases too. In Fig.~\ref{fig:figfive} we show a density plot of $\alpha_0$. Recall that $\nA(0) +\nB(0) = 1 - \nAB(0)\le 1$, therefore $\alpha_0$ is defined only in the lower triangular half of the plot. This region splits up into two symmetric triangles.
In the upper one we have $\nA(0)>\nB(0)$, hence the system converges certainly to consensus on opinion A with no need for a media campaign. Accordingly, we have $\alpha_0=0$ in this region. By contrast, in the lower triangle we have $\nA(0)<\nB(0)$, therefore media must do a real job to make opinion A prevail on opinion B. Indeed, here we have $\alpha_0>0$. The distribution of colors in the plot suggests that $\alpha_0$ is a smooth function of the initial densities. It also suggests that $\alpha_0$ is actually a function of $\nB(0)-\nA(0)$ with good approximation. This is reasonable, because $\nB(0)-\nA(0)$ represents the fraction of agents that must be persuaded to change opinion before the media campaign can be safely stopped. Finally, the map shows that $\alpha_0$ has its absolute maximum at $\nB(0)=1$.

Although Fig.~\ref{fig:figfive} looks perfectly intuitive and somewhat unexciting from a theoretical standpoint, its practical usefulness should not be undervalued. Consider indeed whatever real situation in which the model we are discussing might have a chance to be usefully applied. Before starting any media campaign, it would look wise to make a poll and get informed about what people actually think. Based on the outcome of the poll, Fig.~\ref{fig:figfive} provides an estimate of the lowest effectiveness the media campaign should have in order to succeed.

\subsection{Agents on complex networks}\label{sect:2.5}

\begin{figure*}[t!]
  \centering
  \mbox{
    \subfigure{\includegraphics[width=0.36\textwidth]{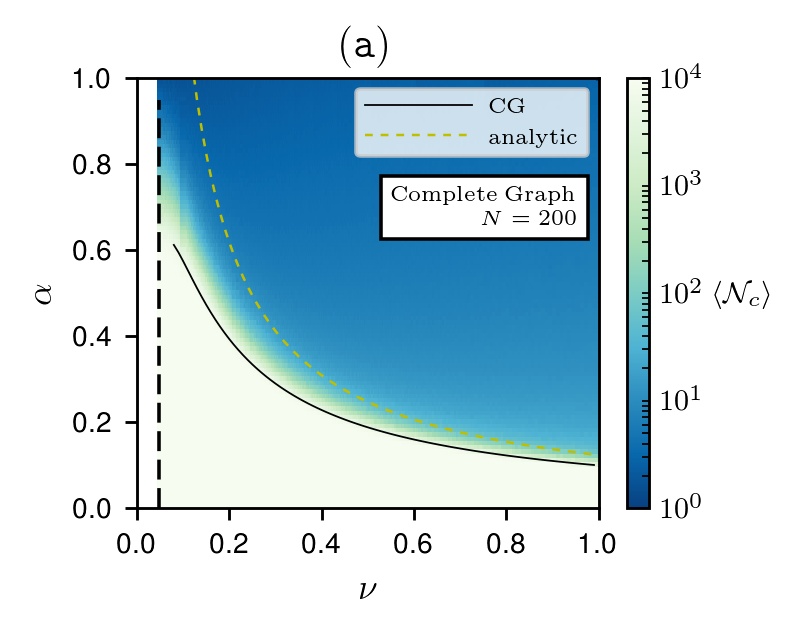}}
    \subfigure{\includegraphics[width=0.36\textwidth]{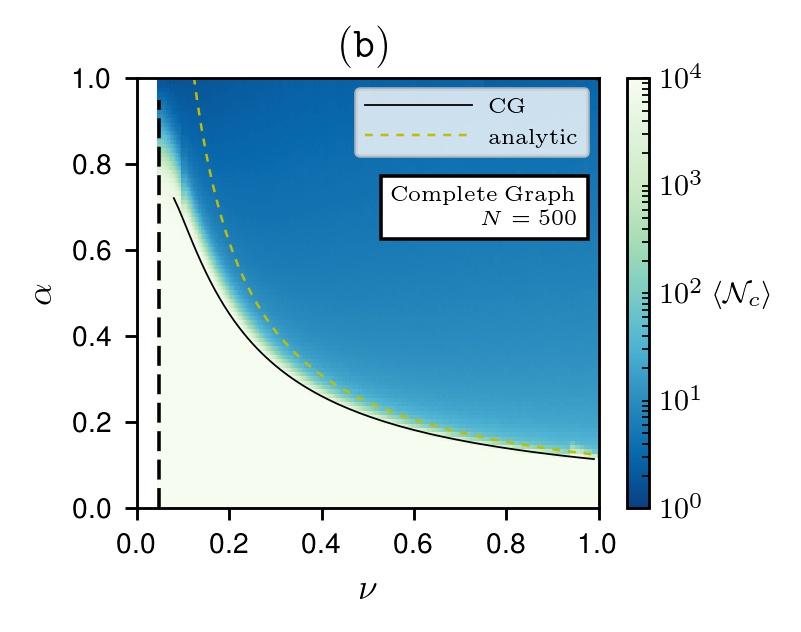}}
  }
  \vskip -0.3cm
  \mbox{
    \subfigure{\includegraphics[width=0.36\textwidth]{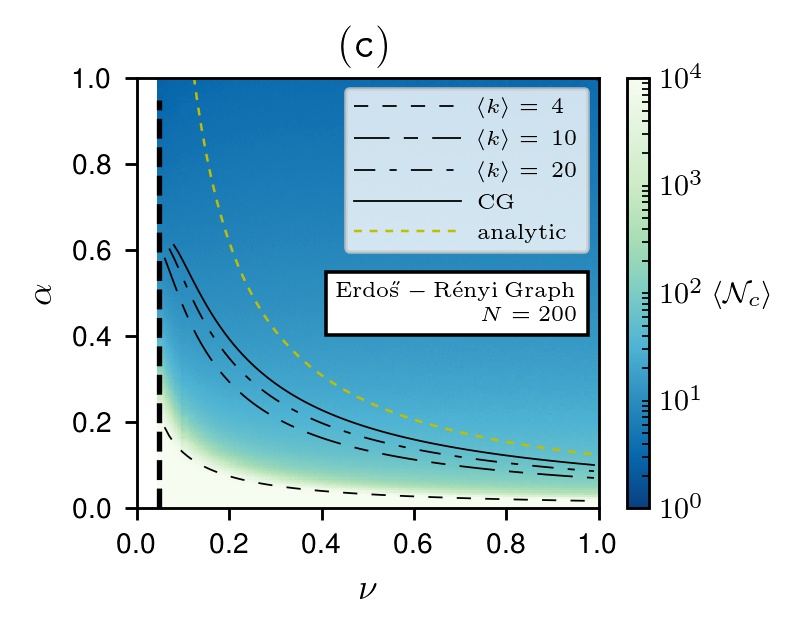}}
    \subfigure{\includegraphics[width=0.36\textwidth]{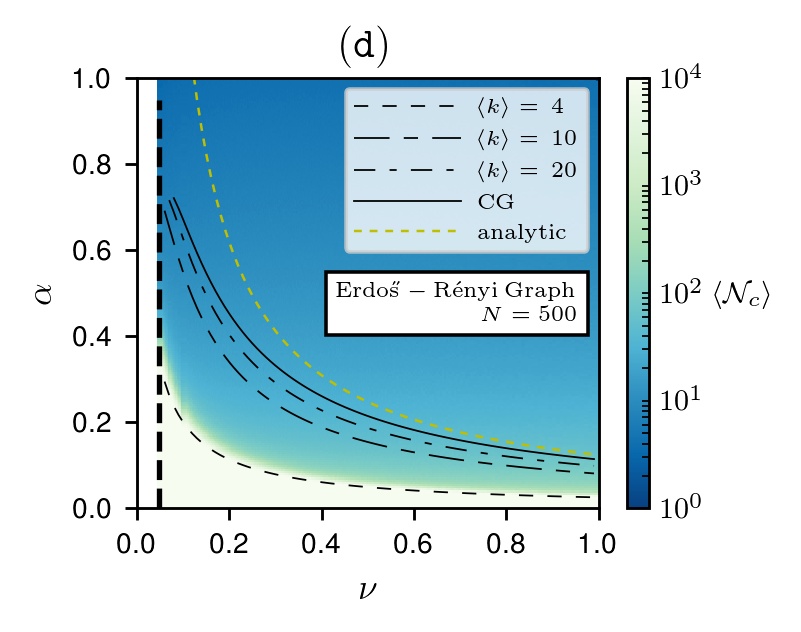}}
  }
  \vskip -0.3cm
  \mbox{
    \subfigure{\includegraphics[width=0.36\textwidth]{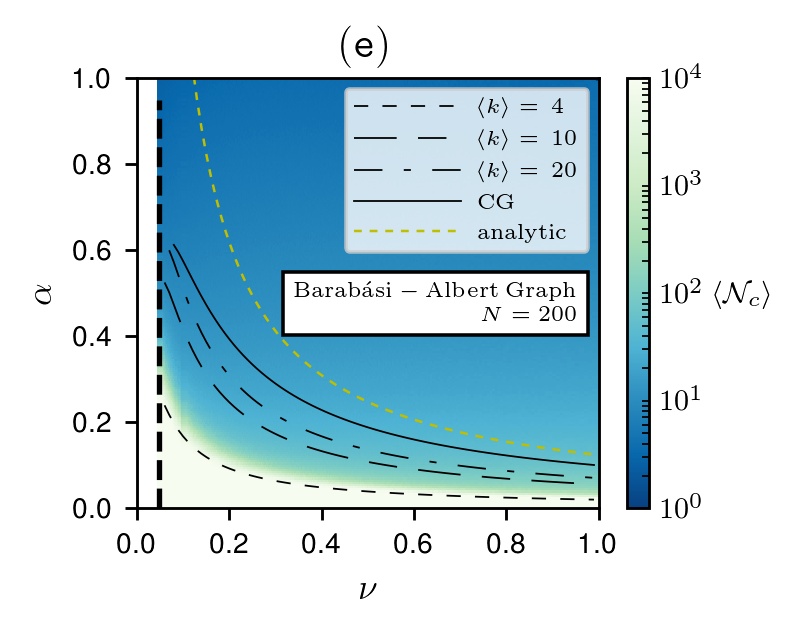}}
    \subfigure{\includegraphics[width=0.36\textwidth]{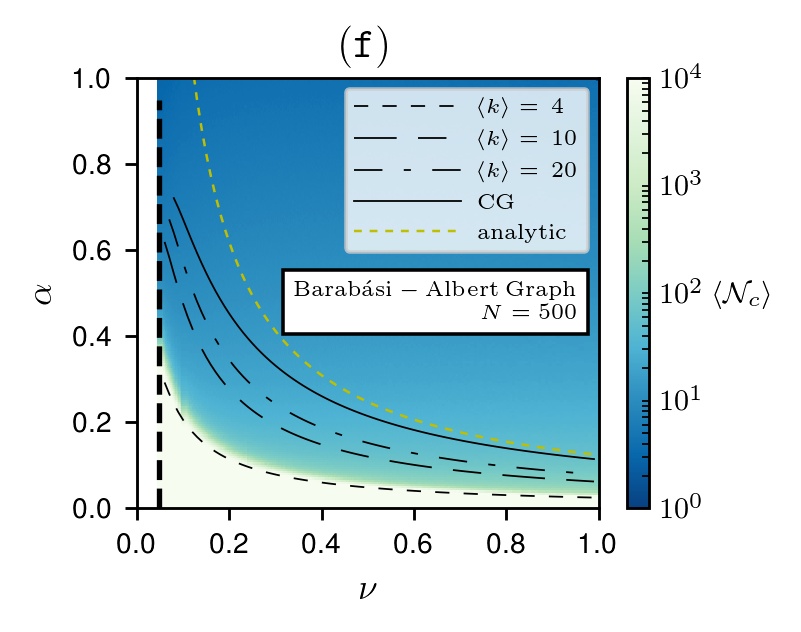}}
  }
  \vskip -0.3cm
  \mbox{
    \subfigure{\includegraphics[width=0.36\textwidth]{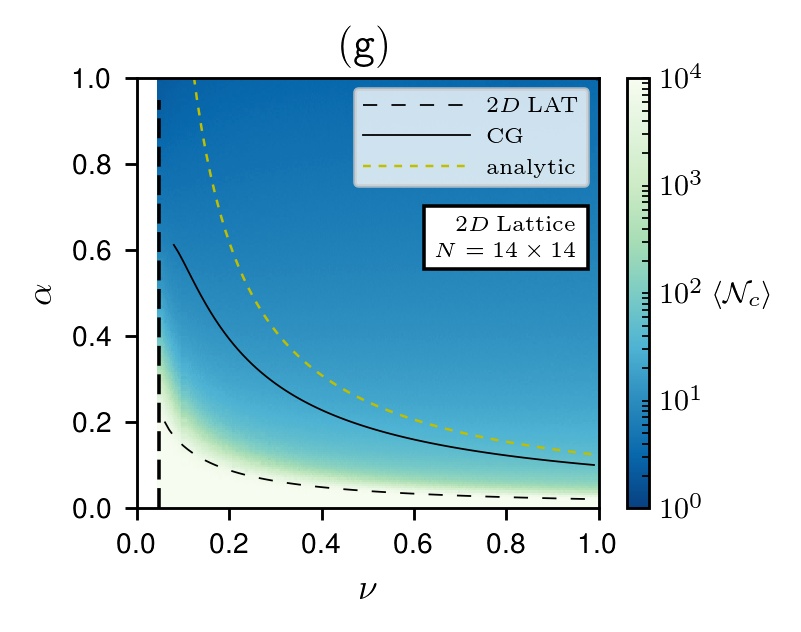}}
    \subfigure{\includegraphics[width=0.36\textwidth]{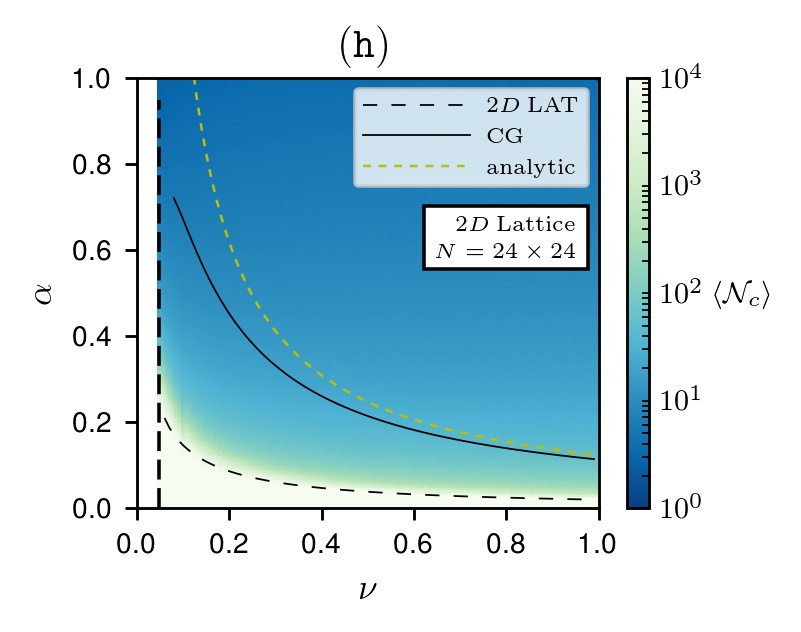}}
  }
  \vskip -0.2cm
  \caption{\footnotesize {\color{red} [color online]} density plot of the average value $\langle \cN_c\rangle$ for $\alphaB=\alphaAB=\alpha$ and for various network topologies. Plots on the left column correspond to $N=200$, whereas plots on the right column correspond to $N=500$. \label{fig:figsix}}
    \vskip -0.3cm
\end{figure*}

We finally compare the mean field approximation of the model with numerical simulations of it. In Figs.~\ref{fig:figsix}(a) and \ref{fig:figsix}(b) we report density plots of the mean value $\langle {\cal N}_c\rangle$ as obtained from simulations on a CG with $N=200$ (a) and $N=500$ (b). Each data point in the plots was produced by averaging 100 independent measures of ${\cal N}_c$. White points correspond to $\langle {\cal N}_c\rangle>10\,000$, a threshold above which simulations were forcedly arrested. The dashed vertical line in correspondence of $\nu=0.05$ represents a frequency cutoff below which simulations were too expensive given our computational budget. The yellow dotted line, reported for the reader's convenience, reproduces Eq.~(\ref{eq:hyperbole}). The light stripe separating the dark-colored (blue) domain from the white one is a narrow transition region across which $\cN_c$ surges. Its width shrinks mildly as $N$ increases, thereby signaling that the crossover to large $\cN_c$ becomes gradually steeper as the thermodynamic limit is approached. For a reason that will become clear in a while, we emphasize the lower boundary of the crossover by a continuous black line. This appears to be closer to the yellow dotted line for $N=500$ than for $N=200$. A perfect coincidence is expected for $\nu\gtrsim 0.5$ as $N\to\infty$, as Fig.~\ref{fig:figthree}(a) suggests. 

We recall that the CG is a totally unphysical network in that its nodes become infinitely connected in the thermodynamic limit. To get an idea of the phase structure in (more) realistic setups, we consider two network models for which the average number of nearest neighbors $\langle k\rangle$ keeps finite as $N\to\infty$, namely, the \Erdos-R\'enyi ({\bf ER}) and the Barab\'asi-Albert ({\bf BA}) graphs. On complex networks the imbalance between the absolute pervasiveness of the external field and the finite average connectivity of the agents results in a kind of trivialization of the phase diagram as $\langle k\rangle$ decreases: the region corresponding to large $\langle \cN_c\rangle$ [analogous to the white region in Figs.~\ref{fig:figsix}(a) and \ref{fig:figsix}(b)] keeps stable in the thermodynamic limit at fixed $\langle k\rangle$ but narrows increasingly as $\langle k\rangle$ decreases. Results are displayed in Figs.~\ref{fig:figsix}(c)-\ref{fig:figsix}(f). Plots were obtained according to the same recipe adopted for numerical simulations on the CG. For both ER and BA networks we chose $\langle k\rangle = 4,\,10,\,20$. In particular, the background colors in the plots represent $\langle\cN_c\rangle$ for $\langle k\rangle = 4$ while the black lines reproduce (from bottom to top) the lower boundary of the crossover regions corresponding to $\langle k \rangle = 4,\,10,\,20$ and the CG. For the sake of completeness, we also report the yellow dotted line representing Eq.~(\ref{eq:hyperbole}). By comparing the plots with $N=200$ (left) to those with $N=500$ (right), we just conclude that the phase transition persists in the thermodynamic limit at fixed $\langle k\rangle$. The overall convergence of the phase diagram as $N$ increases looks moderately fast, independently of whether the degree distribution is homogeneous (ER network) or heterogeneous (BA network). For both ER and BA networks the black lines shift upwards as $N$ increases. As anticipated, $\langle k\rangle$ turns out to be the most important network parameter, while the specific topology of the network connections appears to play a minor role. This is further confirmed by numerical simulations on two-dimensional square lattices ({\bf 2D LAT}) with $N$ chosen as close as possible to $N=200,\, 500$. On these networks we have $\langle k\rangle = 4$. Results are shown in Figs.~\ref{fig:figsix}(g) and \ref{fig:figsix}(h). Phases look similar to those observed on ER and BA networks with equal $\langle k\rangle$. In consideration of the exploratory character of this paper, we do not present an in-depth analysis of the residual differences of phase diagrams corresponding to networks with different topology but equal $\langle k\rangle$.

All in all, the dependence of the phase diagram upon $\langle k\rangle$ appears as an effect of the competition between the physical time scales characterizing the model, namely, the period $\tau$ of the media campaign and the time to consensus $t_\text{conv}$ of the binary NG. It is known from Ref.~\cite{Baronchelli:2} that $t_\text{conv}$ increases as $\langle k\rangle$ decreases, although the analysis of Ref.~\cite{Baronchelli:2} refers to the complete NG and not to its binary version (in the former $t_\text{conv}$ includes an initial transient during which agents invent names and a following stage during which they build up correlations, whereas in the latter these are both absent). The natural tendency to consensus of the NG dynamics is expected to play as a resilience factor hampering the media action in the beginning, when agents tend to reestablish consensus on opinion B after each interaction with the external field. The longer the time to consensus, the stronger the expected influence of media.  

\section{Periodic equilibrium with two competing external fields}\label{sect:3}

\begin{table*}[!t]
  \begin{center}
    \small
    \begin{tabular}{|c|l|c|c|l|}
      \cline{1-2}       \cline{4-5}
      before interaction   \raisebox{10pt} & after interaction & &       before interaction   \raisebox{10pt} & after interaction \\
      \cline{1-2}       \cline{4-5}
      \\[-2.8ex]
      $\MA \to {\rm L}\phantom{A}$ \raisebox{10pt} & $\MA - {\rm L}$ & & $\MB \to {\rm L}\phantom{B}$ \raisebox{10pt} & $\MB - {\rm L}$ \\
      \cline{1-2} \cline{4-5}\\[-2.8ex]
      ${\rm A}\stackbin{\rm A}{\to}{\rm A}$ \raisebox{8pt} & ${\rm A} - {\rm A}\phantom{B}$ with prob. \ 1 & & ${\rm B}\stackbin{\rm B}{\to}{\rm B}$ \raisebox{8pt} & ${\rm B} - {\rm B}\phantom{B}$ with prob. \ 1 \\
      ${\rm A}\stackbin{\rm A}{\to}{\rm B}$\, \raisebox{0pt} & ${\rm A} - {\rm AB}$ with prob. $\betaB$ & & ${\rm B}\stackbin{\rm B}{\to}{\rm A}$\, \raisebox{0pt} & ${\rm B} - {\rm AB}$ with prob. $\gammaA$ \\
      \hskip 5pt ${\rm A}\stackbin{\rm A}{\to}{\rm AB}$ \raisebox{0pt} & ${\rm A} - {\rm A}\phantom{A}$ with prob. $\betaAB$ & & \hskip 5pt ${\rm B}\stackbin{\rm B}{\to}{\rm AB}$ \raisebox{0pt} & ${\rm B} - {\rm B}\phantom{A}$ with prob. $\gammaAB$ \\
      \cline{1-2} \cline{4-5}
    \end{tabular}
    \vskip 0.0cm
    \caption{\footnotesize Table of media-listener interactions ($\MA$\,=\,media broadcasting opinion A, $\MB$\,=\,media broadcasting opinion~B, L\,=\,listener).\label{tab:competingrules}}
  \end{center}
  \vskip -0.6cm
\end{table*}

In the second part of the paper we study the system under the influence of two competing external fields. As previously, we consider $N$ agents interacting according to the dynamics of the binary NG, but now we assume that half of the agents are initially in state A and the rest are in state B. We represent media as two additional fictitious agents $\MA$ and $\MB$, both connected to the whole population and advertising opinions A and B, respectively. When interacting with the agents, media always play the role of speaker. The basic rules for the media-agent interactions are reported in Table~\ref{tab:competingrules}. We also assume that $\MA$ ($\MB$) broadcasts a message every $\TA=\tauA N$ ($\TB=\tauB N$) agent-agent interactions, with $\tauA\ge 1$ ($\tauB\ge 1$). Finally, we assume that $\MA$ broadcasts the first message at time $t=0$, while $\MB$ does it at macroscopic time $t=\tdel$ with $0\le\tdel\le\tauB$. The media action is now parametrized by the 7-tuple $(\betaB,\betaAB,\gammaA,\gammaAB,\nuA,\nuB,\phi)$ with $\nuA=1/\tauA$, $\nuB=1/\tauB$ and $\phi=\tdel/\tauB$. Owing to the high dimensionality of the parameter space, studying the model is more complicated than we did in Sec.~\ref{sect:2}. First of all, whatever choice of parameters we make the system never relaxes to a steady state. Even if consensus is reached on either opinion A or B at some point in time, the subsequent action of the other media always reintroduces disagreement among the population. In full generality, MFEs read 
\begin{equation}
  \left\{\begin{array}{ll}
  \dotnA & = \left(1-\nA-2\nB+\nB^2\right)\\[1.0ex]
  & +\betaAB\,(1-\nA-\nB)\,\delta(t\,\text{mod}\,\tauA)\\[1.0ex]
  & - \gammaA\,\nA\,\delta[(t-\tdel)\,\text{mod}\,\tauB]\,,\\[3.0ex]
  \dotnB & = \left(1-\nB-2\nA+\nA^2\right)\\[1.0ex]
  & +\gammaAB\,(1-\nA-\nB)\,\delta[(t-\tdel)\,\text{mod}\,\tauB] \\[1.0ex]
  & - \betaB\,\nB\,\delta(t\,\text{mod}\,\tauA)\,,\end{array}\right.
  \label{eq:competemfes}
\end{equation}
and the initial conditions we impose on them are 
\begin{equation}
  \nA(0) = \nB(0) = 1/2\,.
\end{equation}
Our strategy for studying the equilibrium states of the system is as follows: we first consider the plane $(\alpha,\nu)$ of highest symmetry, corresponding to $\betaB=\betaAB=\gammaA=\gammaAB=\alpha$, $\nuA=\nuB=\nu$ and $\phi = 0$; then, we depart from it by removing in turn each of these three symmetry conditions. Such an approach yields only a partial understanding of the full structure of the model, yet it is sufficient to let non-trivial phenomenological aspects of it emerge. Similarly to what we did in Sec.~\ref{sect:2}, we first study the system in the mean field approximation, then we consider agents on finite networks with CG, ER, BA and 2D LAT topologies. We anticipate that since the equilibrium densities observed on different complex networks cannot be well displayed on a single figure, such as we did in Fig.~\ref{fig:figsix} for the case of one external field, we place all plots referring to simulations with $\nuA\ne\nuB$ and $\phi\ne 0$ in Ref.~\cite{SupMat} for the sake of readability. 

\begin{figure}[t!]
  \centering
  \includegraphics[width=0.48\textwidth]{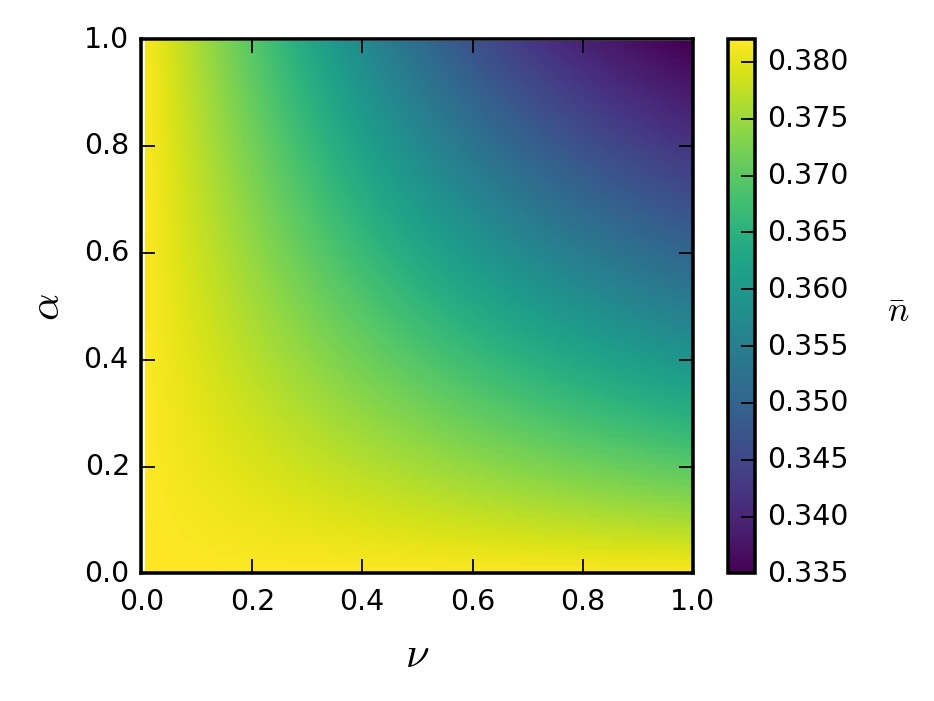}
  \vskip -0.0cm
  \caption{\footnotesize {\color{red} [color online]} Average densities $\nAbar=\nBbar\equiv \bar n$ at periodic equilibrium on the highest-symmetry plane.\label{fig:figseven}}
  \vskip -0.0cm
\end{figure}

On the highest-symmetry plane we have $\nAbar = \nBbar \equiv \bar{n}$ (by symmetry!). The analytic approximations discussed in Sec.~\ref{sect:2.2} yield a condition of periodic equilibrium, represented by the algebraic equation\break ${\bar n}^2 + (1+\alpha\nu)(1-3\bar n) = 0$. This in turn yields
\begin{equation}
  \bar n = \frac{3}{2}(1+\alpha\nu) - \frac{1}{2}\sqrt{9(1+\alpha\nu)^2 - 4(1+\alpha\nu)}\,.
  \label{eq:barn}
\end{equation}
We thus see that $\bar n$ is a function of the product $0<\alpha\nu<1$ and accordingly it belongs to the interval $ 0.35424\ldots = 3-\sqrt{7}\le\bar n\le {(3-\sqrt{5})}/{2}= 0.381964$\,\ldots. The range of $\bar n$ is extremely limited in size for all $\alpha$ and $\nu$. In Fig.~\ref{fig:figseven} we report a density plot of $\bar n$ obtained by numerically integrating Eqs.~(\ref{eq:competemfes}). The agreement with the analytic estimate is rather good but not perfect, especially for $\alpha\simeq 1$, where the ambiguity of the impulsive terms in MFEs becomes more relevant. Apart from this, the plane of highest symmetry is not particularly interesting.

\subsection{Competing external fields with different frequencies}\label{sect:3.1}

\begin{figure}[t!]
  \centering
  \includegraphics[width=0.44\textwidth]{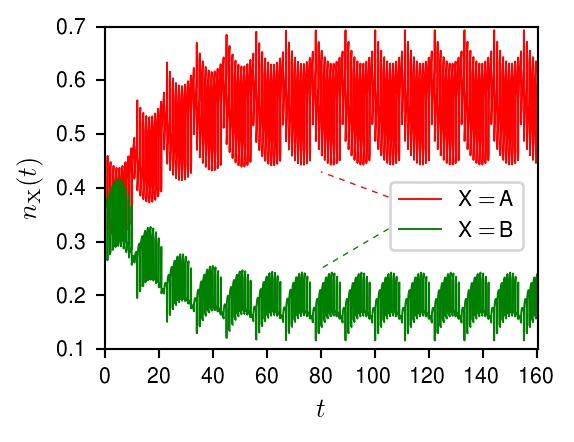}
  \vskip -0.0cm
  \caption{\footnotesize {\color{red} [color online]} Time evolution of the opinion densities for $\betaB=\betaAB=\gammaA=\gammaAB=0.3$, $\tdel=0$, $\tauA=1.0$, and $\tauB=1.1$ as obtained from numerical integration of Eqs.~(\ref{eq:competemfes}).\label{fig:figeight}}
  \vskip -0.0cm
\end{figure}

We now consider the case where $\betaB=\betaAB=\gammaA=\gammaAB=\alpha$ and $\tdel=0$ but $\tauA\ne\tauB$. For each $\alpha = \alpha^*$, the system  lives on a plane $(\nuA,\nuB)$ parallel to the plane $(\hat{\nu}_\text{\tiny A},\hat{\nu}_\text{\tiny B})$ generated by the coordinate axes $\hat{\nu}_\text{\tiny A}$ and $\hat{\nu}_\text{\tiny B}$. This plane intersects the highest-symmetry plane along the line $(\alpha^*,\nu)$. In full generality, for $\tauA\ne\tauB$ the system does not feature a periodic behavior. However, the existence of a periodic equilibrium is guaranteed whenever it is possible to find two integers $\kA$ and $\kB$ such that $\kA\tauA = \kB\tauB$, \ie, whenever $\tauA/\tauB\in\dQ$. Assuming that $\kA$ and $\kB$ are the lowest integers fulfilling this property, we conclude that the system relaxes to periodic equilibrium with period $\tau = \kA\tauA = \kB\tauB$. If $\tauA\in\dQ$ and $\tauB\in\dQ$, then $\tauA/\tauB\in\dQ$ too. In the end, since $\dQ$ is dense in $\dR$, we can safely restrict ourselves to $\tauA,\tauB\in\dQ$. Notice that $\tau$ can be much larger than both $\tauA$ and $\tauB$: for instance, if $\tauA=1.0$ and $\tauB=1.1$, then $\tau=11.0$. To make the reader confident that this is correct, in Fig.~\ref{fig:figeight} we show the behavior of the opinion densities corresponding to this choice of periods for $\alpha=0.3$ as obtained from numerical integration of Eqs.~(\ref{eq:competemfes}). For $\tauA\ne\tauB$, the exchange symmetry $\nAbar\leftrightarrow\nBbar$ is broken, but the system is still symmetric under the joint exchange $\nAbar\leftrightarrow\nBbar$, $\tauA\leftrightarrow\tauB$. Obviously, the average densities at periodic equilibrium are still defined by Eq.~(\ref{eq:periodicdens}) with $\tau$ being correctly interpreted.

For $\tauA>\tauB$ it is natural to expect that $\nAbar<\nBbar$ and the other way around. In other words, if $\MA$ broadcasts with lower frequency than $\MB$, it should also be less effective. Moreover, the larger the ratio $\tauA/\tauB$, the larger the expected ratio $\nBbar/\nAbar$. This is what we meant by \emph{naive expectation} in the Abstract of the paper. It turns out, however, that things are not always as simple as naively expected. In Fig.~SM2 of Ref.~\cite{SupMat} we report density plots of $\nAbar$ and $\nBbar$ corresponding to $\alpha=0.25,0.30,0.35$, obtained from numerical integration of Eqs.~(\ref{eq:competemfes}). Due to limitations in our computational budget, we could only perform simulations for $0.5\le \nuA,\nuB\le 1.0$. We notice the presence of two stripes, parallel and close to the symmetry diagonal, where the average opinion densities are unexpectedly swapped. The starting point of the stripes shifts progressively to lower frequencies along the main diagonal as $\alpha$ increases (in fact, the choice of $\alpha$ in Fig.~SM2 of Ref.~\cite{SupMat} was made to highlight the shifting effect). Over the rest of the plots, $\nAbar$ and $\nBbar$ are qualitatively as common sense suggests. As a check, in Fig.~SM3 of Ref.~\cite{SupMat} we plot the same quantities measured in numerical simulations on a CG with $N=16\,000$.\footnote{Whenever $\MA$ and $\MB$ broadcast their messages at the same time along the simulation, we let $\MA$ act before $\MB$ with probability $1/2$ for each agent and the other way around. This recipe guarantees coincidence at a macroscopic level and thus ensures that no advantage is given to one external field over the other. The problem does not occur in the mean field approximation.} Each data point here was produced by averaging the opinion densities along 10 periods after an appropriate transient, so as to be sure that the system was at periodic equilibrium and the stochastic fluctuations, continuously perturbing its behavior, were averaged out. We observe a smooth contour surrounding the stripes, that is absent in the mean field approximation. The contour is a finite size effect: it narrows progressively as $N$ increases, while the profile of the stripes becomes gradually sharper. The choice of $N$ in Fig.~SM2 of \cite{SupMat} follows from the need to check the predictions of the mean field approximation on the largest graph affordable with our computational resources, so as to be as close as possible to the thermodynamic limit. 

\begin{figure*}[t!]
  \centering
  \mbox{
    \subfigure{\includegraphics[width=0.435\textwidth]{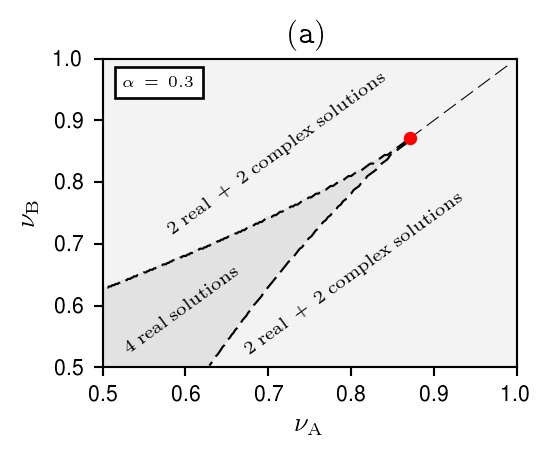}}
    \subfigure{\includegraphics[width=0.415\textwidth]{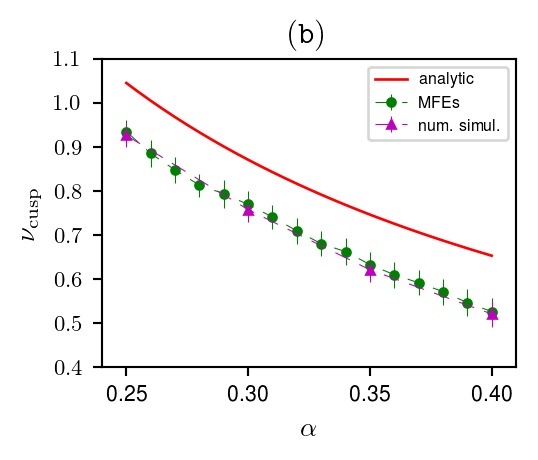}}
  }
  \vskip -0.1cm
  \caption{\footnotesize {\color{red} [color online]} (a) Structure of the solutions to Eq.~(\ref{eq:opposingquartic}) for $\alpha=0.3$; (b) $\nu_\text{cusp}$ as obtained by analytic arguments, MFEs, and numerical simulations.\label{fig:fignine}}
  \vskip -0.1cm
\end{figure*}

To obtain an estimate of the frequencies corresponding to the starting point of the stripes, we approximate Eqs.~(\ref{eq:competemfes}) at periodic equilibrium as explained in Sec.~\ref{sect:2.2}. This yields
\begin{align}
  \left\{\begin{array}{ll}
  0 & = \tau\left(1-\nAbar-2\nBbar+\nBbar^2\right)\\[1.0ex]
  & +\kA\alpha\,(1-\nAbar-\nBbar) -\kB\alpha\,\nAbar\,,\\[2.0ex]
  0 & = \tau\left(1-\nBbar-2\nAbar+\nAbar^2\right)\\[1.0ex]
  & +\kB\alpha\,(1-\nAbar-\nBbar) - \kA\alpha\,\nBbar\,.
  \label{eq:differenttau}
  \end{array}\right.
\end{align}
As already discussed, $\tau$ is an irregular function of $\tauA$ and $\tauB$, hence, the reader should not expect Eqs.~(\ref{eq:differenttau}) to yield an overall description of the system consistent with full MFEs or numerical simulations, except along the main diagonal, where $\tau\ (=\tauA=\tauB)$ changes smoothly. It is anyway interesting to see what Eqs.~(\ref{eq:differenttau}) predict off the main diagonal. By solving the first of them with respect to $\nAbar$ and by then inserting the resulting expression into the second of them, we get a quartic equation for $\nBbar$, \viz,
\vskip -0.8cm
\begin{equation}
  a_4\,\nBbar^4 + a_3\,\nBbar^3 + a_2\,\nBbar^2 + a_1\,\nBbar + a_0 = 0\,,
  \label{eq:opposingquartic}
\end{equation}
with coefficients
\begin{align}
  \left\{\begin{array}{ll}
  a_4 & = \tau^3\,,\\[2.0ex]
  a_3 & = -4\tau^3 -2\kA\tau^2\alpha\,,\\[1.5ex]
  a_2 & = 4\tau^3 + (4\kA-3\kB)\tau^2\alpha\\[0.5ex]
      & + (\kA^2-\kB^2-\kA\kB)\tau\alpha^2 \,,\\[1.5ex]
  a_1 & = -\tau^3 + 3(\kB-\kA)\tau^2\alpha\\[0.5ex]
      & - (3\kA^2+\kB^2+\kA\kB)\tau\alpha^2\\[0.5ex]
      & -(\kA^3+\kB^3+2\kA\kB^2+2\kB\kA^2)\alpha^3\,,\\[1.5ex]
  a_0 & = 2\kB^2\tau\alpha^2+\kB^2(\kA+\kB)\alpha^3\,.
  \label{eq:oppquartcoefs}
  \end{array}\right.
\end{align}
Depending on $\tauA$ and $\tauB$ this equation has four real solutions or two real and two complex solutions, as shown in Fig.~\ref{fig:fignine} (a) for $\alpha=0.3$. The region with four real solutions splits across the symmetry diagonal into two halves, roughly corresponding to the stripes of Fig.~SM2 of Ref.~\cite{SupMat}. Except for one solution that is everywhere larger than one and is thus unphysical, we have no theoretical argument to choose the right solution in this region out of the remaining three. The cusp, highlighted by a red bullet on the plot, is located at $\nuA=\nuB=\nu_\text{cusp}$. This can be calculated exactly. We find
\begin{equation}
  \hskip -0.2cm\nu_\text{cusp} = \frac{1}{1+2\sqrt{2}}\frac{1}{\alpha}\,.
  \label{eq:nucuspone}
\end{equation}
The reader is referred to Appendix~\ref{sect:appB} for a detailed discussion of Eq.~(\ref{eq:opposingquartic}) and a derivation of Eq.~(\ref{eq:nucuspone}). In Fig.~\ref{fig:fignine}~(b) we compare our analytic estimate of $\nu_\text{cusp}$ with the one obtained from numerical integration of Eqs.~(\ref{eq:competemfes}) and from numerical simulations. Results from MFEs are in excellent agreement with numerical data, while the analytic estimate is 15--20\% higher.

In Figs.~SM4--SM10 of Ref.~\cite{SupMat} we plot the average opinion densities on ER and BA networks with $N=16\,000$ and $\langle k\rangle = 4,10,20$ and also on a two-dimensional lattice with $N=100\times 100$. Similar to what we found in Sec.~\ref{sect:2}, periodic equilibria turn out to depend on the network topology mainly via the average degree of the agents, while the specific shape of the degree distribution appears to be secondary. As can be seen by looking at the figures sequentially, the starting point of the stripes shifts to lower frequencies as $\langle k\rangle$ decreases, while the overall behavior of the densities becomes smoother. In particular, for $\langle k\rangle = 4$ the stripes go off the plot region. We refer the reader to Sec.~\ref{sect:3.4} for an interpretation of the physical origin of the stripes. 

\begin{figure*}[t!]
  \vbox{
    \subfigure{\includegraphics[width=0.89\textwidth]{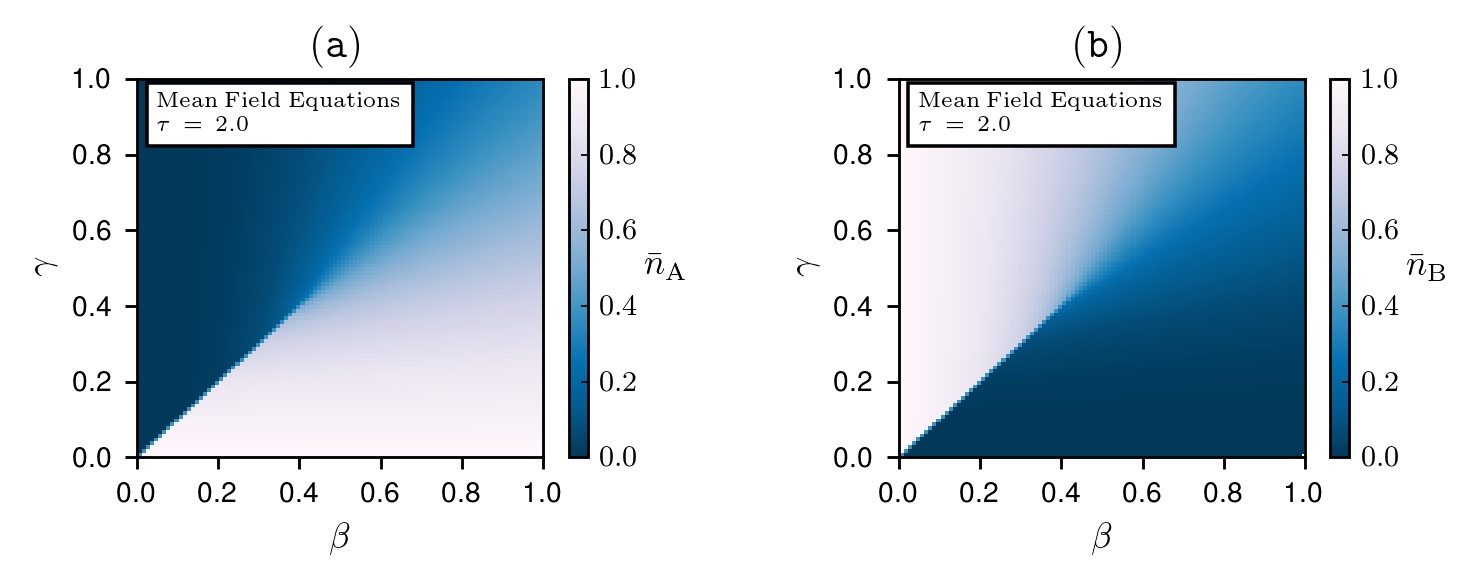}}\\[-3.0ex]
    \subfigure{\includegraphics[width=0.89\textwidth]{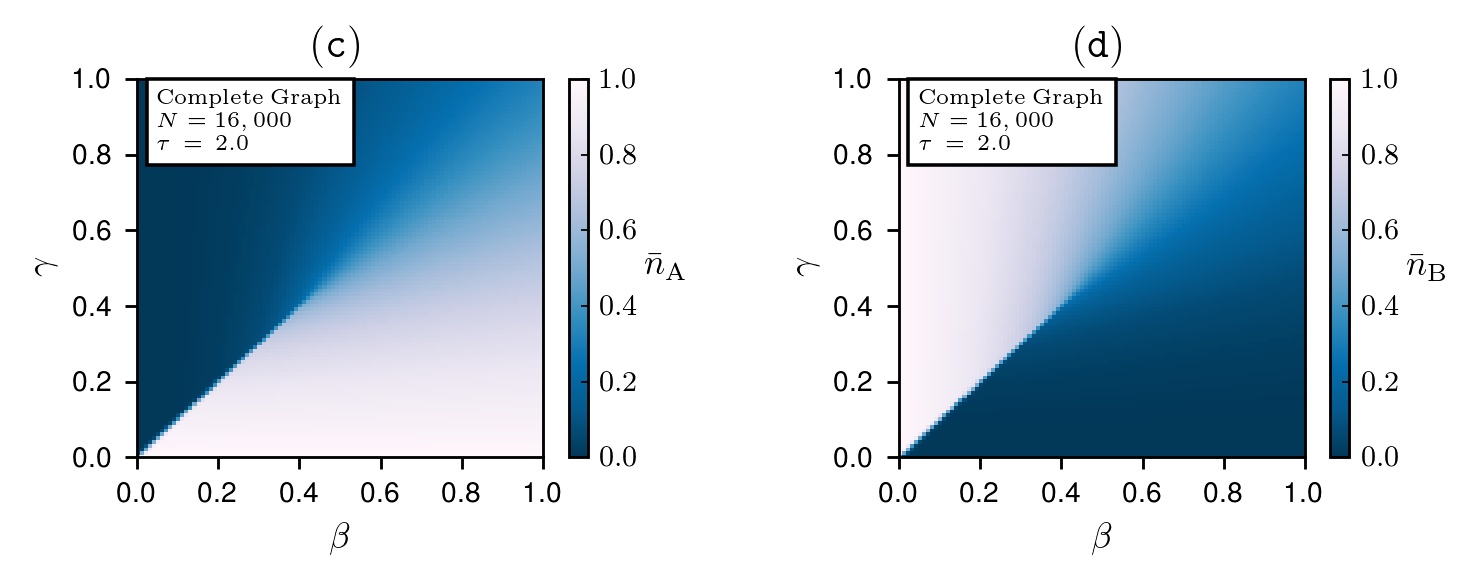}}
  }
  \vskip -0.1cm
  \caption{\footnotesize {\color{red} [color online]} (a), (b) Average densities $\nAbar$ and $\nBbar$ at periodic equilibrium for $\betaB=\betaAB=\beta$ and $\gammaA=\gammaAB=\gamma$ from numerical integration of MFEs for $\tau=2.0$; (c), (d) same quantities from numerical simulations on a CG with $N=16\,000$.\label{fig:figten}}
  \vskip 0.1cm
\end{figure*}

\subsection{Competing external fields with different effectiveness}\label{sect:3.2}

We then consider the case where $\tauA=\tauB=\tau=1/\nu$ and $\tdel=0$ but $\betaB=\betaAB = \beta$ and $\gammaA=\gammaAB=\gamma$ with $\beta\ne\gamma$. For each $\nu=\nu^*$, the system lives on a plane $(\beta,\gamma)$ which is neither parallel to the plane $(\hat {\beta}_\text{\tiny B},\hat {\gamma}_\text{\tiny A})$ generated by the coordinate axes $\hat{\beta}_\text{\tiny B}$ and $\hat{\gamma}_\text{\tiny A}$ (this would correspond to having $\betaAB=\text{const.}$ and $\gammaAB=\text{const.}$) nor parallel to the plane $(\hat {\beta}_\text{\tiny AB},\hat {\gamma}_\text{\tiny AB})$ generated by the coordinate axes $\hat{\beta}_\text{\tiny AB}$ and $\hat{\gamma}_\text{\tiny AB}$ (this would correspond to having $\betaB=\text{const.}$ and $\gammaA=\text{const.}$). Anyway, this plane intersects the highest-symmetry plane along the line $(\alpha,\nu^*)$ for $\beta=\gamma=\alpha$. In this setup the action of the external fields is perfectly synchronous.

In Fig.~\ref{fig:figten} we report density plots of $\nAbar$ and $\nBbar$ at periodic equilibrium, from numerical integration of Eqs.~(\ref{eq:competemfes}) [Figs.~\ref{fig:figten}(a) and \ref{fig:figten}(b)] and from numerical simulations on a CG with $N=16\,000$ [Figs.~\ref{fig:figten}(c) and \ref{fig:figten}(d)], both corresponding to $\tau=2.0$. In this case, the behavior of the system is qualitatively as one would naively expect: in the area above the symmetry diagonal, where $\MB$ is more effective than $\MA$, we have $\nBbar>\nAbar$ and vice versa. No region with swapped densities is observed. Interestingly, we notice a discontinuity of the average densities across the symmetry diagonal, which is very strong for small values of $\beta=\gamma$. The discontinuity fades out for larger values of the effectiveness parameters. The edge of it, i.e. the point of largest effectiveness along the main diagonal for which it can be observed, depends on~$\tau$: both MFEs and numerical simulations show that it shifts upwards as $\tau$ increases. It is possible to estimate the coordinates $\beta=\gamma=\beta_\text{disc}$ of this point in the same way as we did for $\nu_\text{cusp}$ in Sec.~\ref{sect:3.1}. The analytic approximations discussed in Sec.~\ref{sect:2.2} now yield the system
\begin{align}
  \left\{\begin{array}{ll}
  0 & = \tau\left(1-\nAbar-2\nBbar+\nBbar^2\right)\\[1.0ex]
    & +\beta\,(1-\nAbar-\nBbar) - \gamma\,\nAbar\,,\\[3.0ex]
  0 & = \tau\left(1-\nBbar-2\nAbar+\nAbar^2\right)\\[1.0ex]
    & +\gamma\,(1-\nAbar-\nBbar) - \beta\,\nBbar\,.
  \label{eq:differenteff}
  \end{array}\right.
\end{align}
By solving the first of Eqs.~(\ref{eq:differenteff}) with respect to $\nAbar$ and by then inserting the resulting expression into the second of them, we obtain another quartic equation for $\nBbar$, namely,
\begin{equation}
  \tilde a_4\nBbar^4 + \tilde a_3\nBbar^3 + \tilde a_2\nBbar^2 + \tilde a_1 \nBbar + \tilde a_0 = 0\,,
\end{equation}
with coefficients reading as
\begin{align}
  \left\{\begin{array}{ll}
  \tilde a_4 & = \tau^3\,,\\[2.0ex]
  \tilde a_3 & = -4\tau^3-2\tau^2\beta\,,\\[2.0ex]
  \tilde a_2 & = 4\tau^3 + \tau^2(4\beta-3\gamma)+\tau(\beta^2-\gamma^2-\beta\gamma)\,,\\[2.0ex]
  \tilde a_1 & = -\tau^3+3\tau^2(\gamma-\beta)-\tau(3\beta^2+\gamma^2+\gamma\beta)\\[1.0ex]
             & -(\beta^3+\gamma^3+2\beta\gamma^2+2\gamma\beta^2)\,,\\[2.0ex]
  \tilde a_0 & = 2\tau\gamma^2+\gamma^2(\beta+\gamma)\,.
  \end{array}\right.
\end{align}
We can solve this equation in the same way as explained in Appendix~\ref{sect:appB}. In doing it, we meet the same kind of problems we discussed in relation to Eq.~(\ref{eq:opposingquartic}). In particular, the geometric structure of the solutions is qualitatively similar to Fig.~\ref{fig:appbone} with $\nuA$ and $\nuB$ replaced by $\beta$ and $\gamma$, respectively. We notice that letting $\gamma=\beta$ in Eqs.~(\ref{eq:differenteff}) yields the same system as obtained by letting $\tauA=\tauB$ in Eqs.~(\ref{eq:differenttau}). Therefore, the point $\beta_\text{disc}$ of largest effectiveness for which the average densities are discontinuous across the symmetry diagonal is immediately seen to be given by
\begin{equation}
  \hskip -0.2cm\beta_\text{disc} = \frac{\tau}{1+2\sqrt{2}}\,.
  \label{eq:betadisc}
\end{equation}
For instance, for $\tau=2.0$ we have $\beta_\text{disc}=0.522\ldots$, which is about 20\% higher than the value observed in Fig.~\ref{fig:figten}, in perfect analogy with the results of Sec.~\ref{sect:3.1}. Owing to the absence of unexpected results in MFEs and numerical simulations on the CG, in this case we did not extend the analysis to complex networks.

\subsection{Competing external fields with relative time shift}\label{sect:3.3}

We finally consider the case where $\tauA=\tauB=\tau=1/\nu$ and $\betaB=\betaAB=\gammaA=\gammaAB=\alpha$ but $\phi = \tdel/\tau\ne 0$. For each $\nu=\nu^*$, the system lives on a plane $(\alpha,\phi)$ that is not parallel to any coordinate plane, just like we had in Sec.~\ref{sect:3.2}. This plane intersects the highest-symmetry plane along the line $(\alpha,\nu^*$). 

\begin{figure}[t!]
  \centering
  \includegraphics[width=0.46\textwidth]{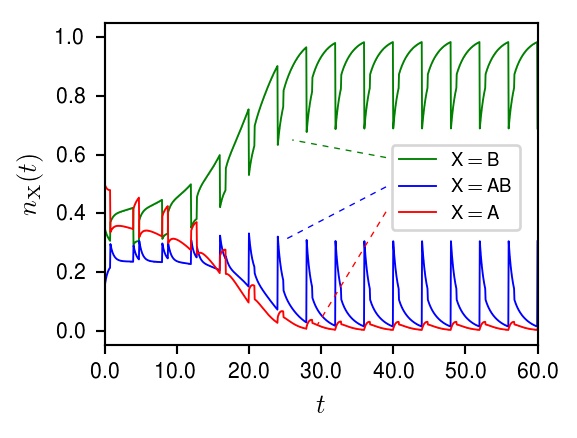}
  \vskip -0.3cm
  \caption{\footnotesize {\color{red} [color online]} time evolution of the opinion densities for $\betaB=\betaAB=\gammaA=\gammaAB=0.3$, $\tauA=\tauB=4$ and $\tdel=0.8$ as obtained by numerical integration of Eqs.~(\ref{eq:competemfes}).\label{fig:figeleven}}
  \vskip -0.5cm
\end{figure}

For $0<\phi<1$, $\MA$ broadcasts first while $\MB$ broadcasts second within each period $\cI_\ell = [t_\ell,t_{\ell+1})$. As a consequence, $\MA$ enjoys a competitive advantage over $\MB$. At the beginning of the game the advantage is quantified by the difference $\nA(0^+)-\nB(0^+) = \alpha/2$. Since the action of the external fields is periodic, it is natural to expect that the initial imbalance of the population will persist in time and will ultimately result in a predominance of opinion A over opinion B as $t\to\infty$. It turns out, actually, that also this naive expectation is wrong. As an example, in Fig.~\ref{fig:figeleven} we show the time evolution of the opinion densities for $\alpha = 0.3$, $\tau = 4.0$, and $\tdel=0.8$ from numerical integration of Eqs.~(\ref{eq:competemfes}). We notice that $\nA(0^+)=0.5$ while $\nB(0^+)=0.35$, in agreement with the formula given above. Within each period we observe two discontinuities affecting the opinion densities, one pointing upwards and one downwards. Of course, they are produced by the alternate action of the external fields. Following a transient the system relaxes to periodic equilibrium. For this choice of parameters it takes exactly three periods for $\MB$ to definitely fill the starting gap and make $\nB(t)>\nA(t)$ for the rest of the time. 

To clarify when the initial advantage of $\MA$ is unprofitable, in Fig.~SM11 of Ref.~\cite{SupMat} we report density plots of $\nAbar$ and $\nBbar$ corresponding to $\tau=1.0,\,2.0,\,4.0$, obtained by numerically integrating MFEs. All plots split qualitatively into two halves: on the left we have $\nBbar>\nAbar$, on the right we have the opposite sign. A dashed black curve separates these two regions. The splitting is exactly at $\phi = 1/2$ for sufficiently large~$\alpha$. The lowest point for which the dashed curve is vertical shifts upwards as $\tau$ increases. Below this point the average opinion densities have a discontinuity, whereas above it they feature a smooth behavior. Finally, the banded regions in the plots represent domains of absolute majority, i.e. here we have $\nXbar>0.5$, X = A, B. As a check, in Fig.~SM12 of Ref.~\cite{SupMat} we plot the same quantities measured in numerical simulations on a CG with $N=16\,000$. Each data point was produced according to the same recipe described in Sec.~\ref{sect:3.1}. The discontinuity affecting the average opinion densities for small $\alpha$ is now replaced by a smooth transition as a result of the finite size of the graph. The boundaries of the banded areas are more irregular than in Fig.~SM11 of Ref.~\cite{SupMat}. This is also likely to be a finite size effect and a result of stochastic fluctuations. 

Unfortunately, we cannot use MFEs to derive analytic predictions for this setup of parameters: the~approximations described in Sec.~\ref{sect:2.2} are insensitive to any relative time shift of the impulsive perturbations, as the reader can easily figure out. We have no alternative strategy at present but to study the system numerically.

In Figs.~SM13--SM19 of Ref.~\cite{SupMat} we plot the average opinion densities on ER and BA networks with $N=16\,000$ and $\langle k\rangle = 4,10,20$ and also on a two-dimensional lattice with $N=100\times 100$. Once again, on complex networks~$\langle k\rangle$ turns out to be the most important parameter affecting the equilibria of the system. In particular, the vertical part of the dashed curve extends downwards as $\langle k\rangle$ decreases, while the observed discontinuity~shrinks. Interestingly, on the 2D LAT topology the discontinuity is absent for all values of $\tau$, \ie, for small $\alpha$ we have $\nAbar \simeq \nBbar$ independently of $\phi$. In Table~\ref{tab:tabthree} we report the size of the areas of relative and absolute majority on the various network topologies. We notice that $\text{meas}(\nBbar>\nAbar)$ approaches $0.5$ as $\langle k\rangle$ decreases, while $\text{meas}(\nBbar>0.5)$ reduces progressively in the same limit. All in all, the system appears to be more balanced when the agents have low connectivity. Again, we refer the reader to Sec.~\ref{sect:3.4} for an interpretation of the results.

\begin{table*}[t!]
\hskip 2.0cm
  \begin{tabular}{c|c|c|c|c|cc}
    \hhline{======}    
    $\tau$ & \text{topology} & $\langle k \rangle$ & $\text{meas}(\nBbar>\nAbar)$ & $\text{meas}(\nAbar>0.5)$ & $\text{meas}(\nBbar>0.5)\phantom{\int^{\int^a}}$ \\[0.3ex]
    \cline{1-6}\cline{1-6}
\multirow{10}{*}{1.0}  & MFEs & --- & 0.449 & 0.376 & 0.283 & $\phantom{\int^{a\int^a}}$ \\
                       & CG   & --- & 0.457 & 0.372 & 0.275 & \\
    \cline{2-6}
           & ER & 20 & 0.461 & 0.355 & 0.264 & $\phantom{\int^{\int^a}}$ \\
           & ER & 10 & 0.457 & 0.343 & 0.245 &  \\
           & ER &  4 & 0.481 & 0.265 & 0.213 & \\
    \cline{2-6}
           & BA & 20 & 0.468 & 0.331 & 0.257 &  $\phantom{\int^{\int^a}}$ \\
           & BA & 10 & 0.474 & 0.311 & 0.245 & \\
           & BA &  4 & 0.488 & 0.236 & 0.198 & \\
    \cline{2-6}
           & 2D LAT & --- & 0.489 & 0.229 & 0.198 &  $\phantom{\int^{\int^a}}$ \\
    \hhline{======}
\multirow{10}{*}{2.0}  & MFEs & --- & 0.418  & 0.507 & 0.349 & $\phantom{\int^{\int^a}}$  \\
                       & CG   & --- & 0.422  & 0.504 & 0.343 & \\
    \cline{2-6}
           & ER & 20 & 0.428 & 0.483 & 0.339 &  $\phantom{\int^{\int^a}}$ \\
           & ER & 10 & 0.425 & 0.476 & 0.322 & \\
           & ER &  4 & 0.464 & 0.387 & 0.308 & \\
    \cline{2-6}
           & BA & 20 & 0.439 & 0.463 & 0.338 &  $\phantom{\int^{\int^a}}$ \\
           & BA & 10 & 0.449 & 0.440 & 0.334 & \\
           & BA &  4 & 0.478 & 0.358 & 0.307 & \\
    \cline{2-6}
           & 2D LAT & --- & 0.478 & 0.333 & 0.290 & $\phantom{\int^{\int^a}}$ \\
    \hhline{======}
\multirow{10}{*}{4.0} & MFEs & --- & 0.353 & 0.629 & 0.340 & $\phantom{\int^{\int^a}}$ \\
                      & CG   & --- & 0.356 & 0.625 & 0.338 & \\
    \cline{2-6}
           & ER & 20 & 0.367 & 0.608 & 0.341 & $\phantom{\int^{\int^a}}$ \\
           & ER & 10 & 0.363 & 0.601 & 0.327 & \\
           & ER &  4 & 0.427 & 0.488 & 0.340 & \\
    \cline{2-6}
           & BA & 20 & 0.382 & 0.583 & 0.346 & $\phantom{\int^{\int^a}}$ \\
           & BA & 10 & 0.400 & 0.552 & 0.350 & \\
           & BA &  4 & 0.452 & 0.447 & 0.348 & \\
    \cline{2-6}
           & 2D LAT & --- & 0.457 & 0.420 & 0.334 & $\phantom{\int^{\int^a}}$ \\
    \hhline{======}
  \end{tabular}
  \vskip 0.2cm
  \caption{\footnotesize Size of the regions of relative and absolute majority for $\tdel\ne 0$ on various networks.}
  \label{tab:tabthree}
  \vskip -0.3cm
\end{table*}

\subsection{First vs. second mover. Who wins?}\label{sect:3.4}

Marketing strategists know that firms can (and often do) enjoy a competitive advantage over rivals when they pioneer a new technology or enter a market with innovative products. They call this \emph{the first-mover advantage}. Sometimes, first movers lose the competition for various reasons, including free-riding of competitors, inertia in reacting to environmental changes, etc. When this happens, they refer to the overtaking of rivals as \emph{the second-mover advantage}. The idea was first formalized in a renowned paper~\cite{Lieberman} and nowadays a huge literature (both scientific and popular) on the subject exists. In our theoretical setup, we have a competition among two media and a gauge to measure which of them is more convincing. In Secs.~\ref{sect:3.1} and~\ref{sect:3.3} we describe two different situations that can be traced back to a first-vs-second-mover dynamics. On varying the model parameters we observe periodic equilibrium states corresponding to a predominance of the first mover or the second one. Where do these come from? Since a full analytic solution of MFEs appears to be unattainable, we propose a heuristic interpretation based on the limited information ensuing from our numerical experiments.

Let us start with the setup of Sec.~\ref{sect:3.3}, where $\MA$ ($\MB$) plays as first (second) mover. We split each time period $\cI_\ell$ into two disjoint intervals $\cI_{\ell\rm\scriptscriptstyle A}=(t_\ell,t_\ell+\tdel)$ and $\cI_{\ell\rm\scriptscriptstyle B}=(t_\ell+\tdel,t_{\ell+1})$. No media-agent interactions occur during each of them, hence the evolution of the opinions is \emph{free}, \ie, it is exclusively governed by the internal rules of the NG. Since $\MA$ broadcasts its $\ell$th message at $t=t_\ell$, we regard $\cI_{\ell\rm\scriptscriptstyle A}$ as a time during which the population reacts to the persuasive power of $\MA$. The same holds for $\cI_{\ell\rm\scriptscriptstyle B}$ and $\MB$. Fig.~SM11 of Ref.~\cite{SupMat} tells us that, for sufficiently large~$\alpha$, the second (first) mover prevails for $\phi<1/2$ ($\phi>1/2$). We could call this dichotomous behavior the ``$\phi=1/2$'' rule. Since $|\cI_{\ell\rm\scriptscriptstyle A}|=\tdel$ and $|\cI_{\ell\rm\scriptscriptstyle B}|=\tau-\tdel$, we conclude that the first mover wins provided $|\cI_{\ell\rm\scriptscriptstyle A}|>|\cI_{\ell\rm\scriptscriptstyle B}|$ and the other way around. In other words, the winner is the external field that has more time to influence the agents along their \emph{free} evolution. Alternatively, we could say that the second mover wins whenever he behaves like a \emph{fast follower}, to borrow another common expression from the jargon of business strategy.

The above interpretation seems to be contradicted by the behavior of the system for small $\alpha$. In this limit, the first mover predominates for $\phi>\phi^*$, with $\phi^*=\phi^*(\alpha,\tau)<1/2$. The violation of the $\phi=1/2$ rule follows from the natural tendency of the internal dynamics of the NG to unbalance the opinions densities through the spread of consensus, while the competition among the external fields tends to exert an overall balancing effect on them\footnote{Recall that media contributions to MFEs are linear in the opinion densities, whereas agent-agent interactions contribute by quadratic terms.}. When the influence of the external fields lessens (\eg, for $\alpha\ll 1$) the internal dynamics becomes predominant and non-linear effects emerge: the crossover between the regions of first- and second-mover advantage turns into a phase transition and the system relaxes more easily to a state of first-mover advantage [nonetheless $\phi^*(\alpha,\tau)>0$ for all $\alpha>0$ and $\tau\ge1$]. We have a hint that this picture is correct upon looking at the behavior of the system for large~$\tau$: as $\tau$ increases, the influence of media similarly weakens, because opinions have more time to freely evolve after each message is broadcast. Fig.~SM11 of Ref.~\cite{SupMat} shows that for larger $\tau$ the region where the $\phi=1/2$ rule is violated expands, while Table~\ref{tab:tabthree} confirms that the area of first-mover advantage increases. The edge of the discontinuity of the opinion densities raises proportionally to $\tau$. An invariance mechanism is at work here. Indeed, to restore the loss of influence media undergo for larger $\tau$, we need to increase $\alpha$ proportionally, hence the edge of the discontinuity shifts correspondingly. Another sign that we are on the right track comes from the behavior of the system on complex networks with small~$\langle k\rangle$: as~$\langle k\rangle$ decreases, the influence of media strengthens because social cohesion lessens (indeed the time to consensus in the NG with no media increases). Figs.~SM13--SM18 of Ref.~\cite{SupMat} show that for smaller $\langle k\rangle$ the region where the~$\phi = 1/2$ rule is violated shrinks, while Table~\ref{tab:tabthree} confirms that the area of first-mover advantage decreases [recall that $\text{meas}(\nAbar>\nBbar) = 1 - \text{meas}(\nBbar>\nAbar)$].  

To summarize, we started from a naive expectation, according to which the first mover always wins because he enjoys a competitive advantage over the second one. A partial analysis of our simulations suggested a different picture, in which it is not really important who broadcasts first: what is important is that after each broadcasting the population has sufficient time to positively react and propagate consensus ($\phi=1/2$ rule). A complete analysis of all available data finally showed that the $\phi=1/2$ rule holds when both media exert a strong influence on the system, while non-linear effects emerge as a result of  the internal dynamics of the NG as the influence of media becomes too weak. 

Similar arguments allow to explain the origin of the stripes of swapped densities observed in Figs.~SM2--SM10 of Ref.~\cite{SupMat}. In the setup of Sec.~\ref{sect:3.2}, the agent dynamics is symmetric under the exchange $\nA\leftrightarrow\nB$ for $0\le t<\min\{\tauA,\tauB\}$, while the symmetry gets definitely broken for $t\ge\min\{\tauA,\tauB\}$. For $\tauA<\tauB$ we can regard $\MB$ as broadcasting with the same frequency as $\MA$ but with positive time delay $t_{{\rm\scriptscriptstyle del},\ell} = \ell(\tauB-\tauA)$, increasing at each subsequent period, \ie, for $\ell=1,2$\,\ldots. Therefore, for $t\ge\tauA$, $\MA$ ($\MB$) plays as first (second) mover. If $\tauB-\tauA\ll \tauA/2$, the increase rate of $t_{{\rm\scriptscriptstyle del},\ell}$ is very slow. In this case $\MB$ behaves like a fast follower over many periods. For the reasons explained in the previous paragraphs the system relaxes to a periodic equilibrium in which $\MB$ prevails over $\MA$ (second-mover advantage), while the imbalance of the opinion densities becomes eventually too large to be reversed after a number $\ell$ of periods such that $t_{{\rm\scriptscriptstyle del},\ell}\simeq\tauA/2$. By contrast, if $\tauB-\tauA\lesssim \tauA/2$, the increase rate of $t_{{\rm\scriptscriptstyle del},\ell}$ is faster. In this case, a small number $\ell$ of periods is sufficient to have $t_{{\rm\scriptscriptstyle del},\ell}\simeq\tauA/2$. If this occurs before the system has relaxed to its ultimate equilibrium, then $\MA$ has sufficient time to let the internal dynamics of the agents work in its favor and $\MB$ has no chance to overtake $\MA$ (first-mover advantage). Fig.~SM11 of Ref.~\cite{SupMat} tells us that the transition from the former to the latter behavior takes place sharply for $\tauB-\tauA = \tau^*$, with $\tau^*\ll \tauA/2$. The above reasoning can be repeated for $\tauB<\tauA$ by simply exchanging $\MA\leftrightarrow\MB$.

What can we say about the dependence of the stripes upon $\alpha$ and $\langle k\rangle$? By increasing $\alpha$, the influence of the external fields strengthens. According to our previous considerations, the system should be more balanced. Indeed, Fig.~SM2 of Ref.~\cite{SupMat} shows that the starting point of the stripes shifts to lower frequencies, while the behavior of the average opinion densities across the main diagonal becomes progressively smoother. The invariance mechanism explained above works here as well. We have already noticed that the system is governed by two drivers: on the one hand we have the media action, exerting a linearizing effect on the opinions; on the other we have the internal dynamics of the NG, working to unbalance the system through the spread of consensus. By equally rescaling $\alpha$, $\tauA$, and $\tauB$, the overall influence of media is left nearly invariant and the physics does not change. Correspondingly, the starting point of the stripes shifts downwards along the main diagonal of Fig.~SM2 of Ref.~\cite{SupMat}. Similarly, by decreasing $\langle k\rangle$ the influence of media strengthens. Again, the starting point of the stripes shifts to lower frequencies by the same invariance mechanism.

To summarize, the stripes of swapped densities can be regarded as domains of second-mover advantage. They are non-linear effects, emerging when the influence of media is weak. The above considerations allow to explain how the starting point of the stripes shifts on varying~$\alpha$. Yet, they cannot explain why they do have a starting point. A full understanding of the dynamical mechanism generating the stripes requires $\tau^* = \tau^*(\alpha,\tauA,\tauB)$ to be analytically known. Unfortunately, exact solutions to MFEs are unattainable, whereas numerical ones are of little help. 

To conclude, we stress that the specific shape of the domains observed in Figs.~SM2--SM19 of Ref.~\cite{SupMat} depends expressly on the microscopic dynamics of the NG. It is reasonable that other discrete-opinion models will feature differently shaped regions of first- and second-mover advantage under the periodic action of competing external fields. Our study describes a complex system in which a non-trivial first-vs-second-mover dynamics emerges naturally as a result of the collective behavior of interacting agents. 

\section{Conclusions}\label{sect:4}

In this paper we have investigated the influence of periodic external fields, representing mass media, on the dynamics of agents playing the binary naming game (NG). Our major motivation for studying this was understanding to what extent consensus can be exogenously directed in the framework of an empirically grounded opinion model. In this regard, we recall that theoretical predictions derived from the NG have been shown to reproduce correctly experimental results in Web-based live games with controlled design~\cite{Centola}. In our study, we have found that (i) a single external field is able to convert an entire population of agents within a finite time provided it acts with sufficiently high frequency and effectiveness and (ii) two competing external fields, contending for supremacy, lead the population to a complex periodic equilibrium. We have worked out the phase structure of the model in the mean field approximation, thereby providing evidence that the large-scale dynamics is characterized by non-linear effects and discontinuities in the parameter space. Upon writing mean field equations, we have soon realized that the impulsive terms representing the action of the external fields make exact analytic solutions unattainable. We have found, yet, that integrating mean field equations over a time period allows to asymptotically simplify their algebraic structure, thus paving the way to approximate predictions, that we have systematically derived. We have finally studied the phase structure of the model beyond the mean field approximation via numerical simulations on a complete graph (to evaluate finite size effects) and on various complex networks (to examine how changing the network topology impacts on results). 

Although nontrivial, the dynamics in the presence of one external field is rather intuitive. The system is driven by two opposing forces: on the one hand the external field compels the agents to abandon their original opinion and to adopt the advertised one, on the other the internal dynamics of the NG tends to restore the initial configuration of the system. The clash between these two forces results in a sharp transition from a phase where opposite opinions stay forever in a state of periodic equilibrium to one where consensus on the advertised opinion is reached within a finite time. The existence of an absolute threshold for the effectiveness parameter, below which the action of the external field turns out to be unsuccessful, seems to be the most important result of our analysis, especially in relation to prospective applications of the model. Social cohesion, measured by the average degree of the agents, turns out to be the main factor of resilience to the pervasiveness of the external field. It sets the absolute effectiveness threshold within small corrections related to the specific topology of the network links (\eg, the shape of the degree distribution).

More complex appears to be the dynamics in the presence of two competing external fields. The system~is now driven by three distinct forces: the field actions, compelling the population to adopt one advertised opinion or the other, and the internal dynamics of the NG, which dynamically amplifies consensus on the opinion prevailing at any given time. Studying the system in full generality looks more difficult. Therefore, we~have examined three ideal situations: (i)~synchronous external fields with different effectiveness; (ii)~external fields with equal effectiveness and frequency but relative time shift; (iii)~external fields with equal effectiveness but different frequency and no relative time shift. Case (i) is plain vanilla: the most effective field always prevails over its competitor. Cases (ii) and (iii) are more interesting. The condition of equal effectiveness corresponds phenomenologically to a situation of optimal competition: press agents on each side are the best and work to prepare the best ads. Timing is now crucial. We have found two different dynamical regimes. When media exert a strong influence (\eg, they feature high effectiveness, high frequency, and/or social cohesion is weak), the internal dynamics of the agents becomes secondary: the system lives in a kind of \emph{linear} regime. In case (ii) the opinion prevailing at equilibrium is the one that has systematically more time to spread across the network before the opposite opinion is newly advertised. For a reason explained in Sec.~\ref{sect:3.4}, we have called this the ``$\phi=1/2$'' rule. In case (iii) the field with higher frequency prevails over its competitor.   By contrast, when media exert a weak influence (\eg, they feature low effectiveness, low frequency, and/or social cohesion is strong) the internal dynamics of the agents becomes predominant: the system lives in a \emph{nonlinear} regime.  Phase transitions (discontinuities in the parameter space) and non-linear effects emerge.

In industrial competitions for which timing is a relevant discriminating factor, it sounds pretty natural to adopt a terminology, commonly used by marketing strategists, according to which two competitors are regarded as \emph{first} and \emph{second mover}, depending on the sequential order of their actions. The main message emerging from our analysis is that in order to evaluate first-vs-second-mover dynamics, considerations based on well-established mechanisms (such as technological leadership, preemption of scarce assets, switching costs, and buyer choice under uncertainty~\cite{Lieberman}) could be insufficient in specific conditions, because they do not take into account non-linear effects related to the spontaneous emergence of social consensus as a result of population dynamics. In this regard, the model discussed in this paper represents hopefully a useful abstraction.

\section*{Acknowledgments}

F.~P. and S.~F. acknowledge E. Mollona for stimulating discussions about the relation between regional policies in the EU and the perception of the EU institutions by Europeans (see discussion in Sec.~\ref{sect:1} and Ref.~\cite{perceive}). The computing resources used for our numerical simulations and the related technical support have been provided by the CRESCO/ENEA\-GRID High Performance Computing infrastructure and its staff~\cite{Ponti}. CRESCO ({\color{red}C}omputational  {\color{red} RES}earch centre on {\color{red} CO}mplex systems) is funded by ENEA and by Italian and European research programmes.

\begin{appendices}

\section{Reality of $\zeta(\alpha,\tau)$ for $\Delta<0$}\label{sect:appA}

We want to show that $\zeta(\alpha,\tau)\in\dR$ for $(\alpha,\tau)$ belonging to phase II. We assume $\Delta = -|\Delta|<0$. Accordingly, we have
\begin{align}
  & \sqrt[3]{\Gamma + 12\sqrt{\Delta}}  = \left(\Gamma + 12\,\ri\sqrt{|\Delta|}\right)^{1/3} \nonumber\\[0ex]
  & = \sqrt[6]{\Gamma^2+144|\Delta|}\cdot\left[\cos\left(\frac{1}{3}\arctan\frac{12\sqrt{|\Delta|}}{\Gamma}\right)\right.\nonumber\\[0ex]
    &\hskip 2.62cm \left. + \ri\sin\left(\frac{1}{3}\arctan\frac{12\sqrt{|\Delta|}}{\Gamma}\right)\right] \nonumber\\[0ex]
  & = \sqrt[6]{\Gamma^2+144|\Delta|}\cdot\left(C + \ri\,S\right)\,,
\end{align}
where $C$ and $S$ denote, respectively, the cos and sin functions. It is not difficult to show by explicit calculation that
\begin{equation}
  \Gamma^2+144|\Delta| = \left[4(\alpha+2\tau)^2\right]^3\,.
\end{equation}
Hence, it follows
\begin{equation}
  \sqrt[3]{\Gamma + 12\sqrt{\Delta}} = 2(\alpha+2\tau)\cdot(C+\ri S)\,,
\end{equation}
Inserting this expression into $\zeta(\alpha,\tau)$ yields
\begin{align}
  \zeta(\alpha,\tau) & = \frac{2}{3}(\alpha + 2\tau)\\[1.0ex]
  & -\frac{1}{6}(\alpha+2\tau)\left(1+\ri\sqrt{3}\right)(C+\ri S)\nonumber\\[1.0ex]
  & -\frac{1}{6}(\alpha+2\tau)\left(1-\ri\sqrt{3}\right)(C-\ri S)\nonumber\\[2.0ex]
  & = \frac{1}{3}(\alpha + 2\tau)\left(2 -C + \sqrt{3}S\right).
  \label{eq:zetapc}
\end{align}
We thus see that the imaginary part of $\zeta(\alpha,\tau)$ vanishes provided $\Delta<0$. Eq.~(\ref{eq:zetapc}) can be used for numerical computations.

\begin{figure*}[t!]
  \includegraphics[width=0.775\textwidth]{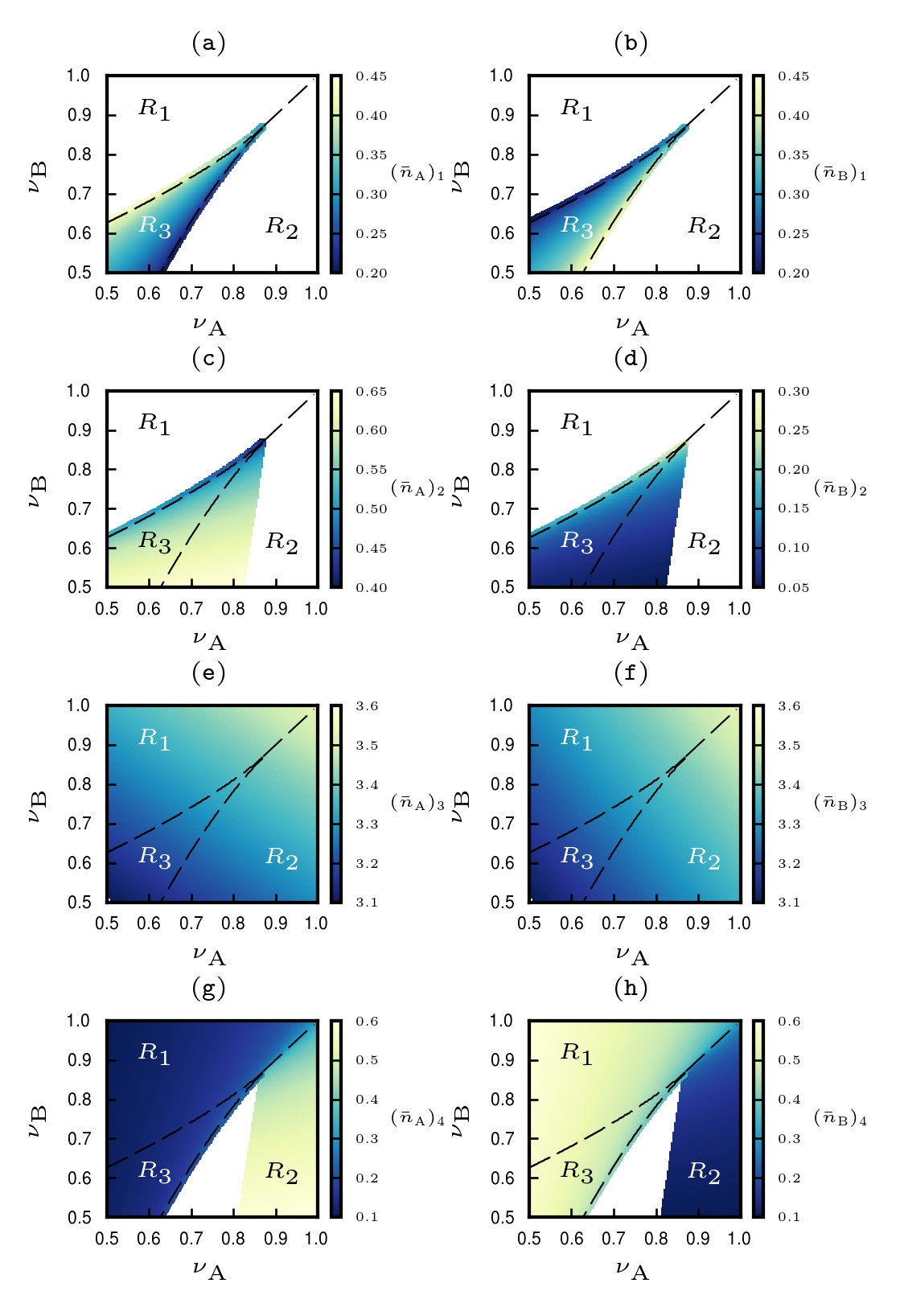}
  \vskip-0.4cm
  \caption{\footnotesize {\color{red} [color online]} Solutions to Eqs.~(\ref{eq:differenttau}) for $\alpha=0.3$ in the regions where they are real.\label{fig:appbone}}
\end{figure*}

\section{Structure of the solutions to Eqs.~(\ref{eq:differenttau})}\label{sect:appB}

We recall that general formulas for the roots of an algebraic equation of fourth degree such as Eq.~(\ref{eq:opposingquartic}) (for which $a_4\ne 0$) are
\begin{align}
  (\nBbar)_{1,2} = -\frac{a_3}{4a_4} - \Lambda \pm \frac{1}{2}\sqrt{\Sigma_+}\,,\\[2.0ex]
  (\nBbar)_{3,4} = -\frac{a_3}{4a_4} + \Lambda \pm \frac{1}{2}\sqrt{\Sigma_-}\,,
\end{align}
with the coefficients $\Sigma_\pm$ and $\Lambda$ being given by
\begin{align}
  \Sigma_\pm & = -4\Lambda^2-2\mu \pm \frac{\nu}{\Lambda}\,,\nonumber\\[1.0ex]
  \Lambda & = \frac{1}{2}\sqrt{-\frac{2}{3}\mu + \frac{1}{3a_4}\left(\Omega + \frac{\Delta_0}{\Omega}\right)}
\end{align}
and with
\begin{align}
  \mu & = \frac{8a_4a_2-3a_3^2}{8a_4^2}\,,\\[2.0ex]
  \nu & = \frac{a_3^3-4a_4a_3a_2+8a_4^2a_1}{8a_4^3}\,,
\end{align}
\begin{align}
  \Omega & = \left(\frac{\Delta_1 + \sqrt{\Delta_1^2-4\Delta_0^3}}{2}\right)^{1/3}\,,\\[3.0ex]
  \Delta_0 & = a_2^2-3a_1a_3+12a_4a_0\,,\\[1.0ex]
  \Delta_1 & = 2a_2^3-9a_3a_2a_1+27a_3^2a_0\nonumber\\[0.5ex]
  & +27a_4a_1^2-72a_4a_2a_0\,.
\end{align}

By inserting the coefficients reported in Eq.~(\ref{eq:oppquartcoefs}) into the above formulas, we compute the four solutions $(\nBbar)_{1,\ldots,4}$ to Eq.~(\ref{eq:opposingquartic}) numerically and then obtain the corresponding densities $(\nAbar)_{1,\ldots,4}$ by substitution in the first of Eqs.~(\ref{eq:differenttau}). The resulting numbers are real or complex depending on $\tauA$, $\tauB$ and $\alpha$. In Fig.~\ref{fig:appbone} we report density plots of the solutions for $\alpha = 0.3$ in the regions where they are real. For convenience, we have split the parameter plane $(\nuA,\nuB)$ into three regions $R_1$, $R_2$, and $R_3$. We observe that $(\nXbar)_1$ (X\,$=$\,A, B) is real within $R_3$, $(\nXbar)_2$ is real within $R_3$ and in part of $R_2$, $(\nXbar)_3$ is always real and $(\nXbar)_4$ is real within $R_1$, $R_3$ and in part of $R_2$. Since $(\nXbar)_2$ is real in $R_2$ where $(\nXbar)_4$ is complex and the other way around, we conclude that we have two real and two complex solutions in $R_1$ and $R_2$, while  we have four real solutions in $R_3$. We also notice that $(\nXbar)_3>1$ for all $\nuA$ and $\nuB$, hence, we can leave it out. Accordingly, we are left with just one real solution in $R_1$ and $R_2$, representing indeed the physical average densities at periodic equilibrium.

In principle, $(\nXbar)_1$, $(\nXbar)_2$ and $(\nXbar)_4$ are all acceptable in $R_3$ (they are real, positive, and less than one). Moreover, $R_3$~contains part of the symmetry diagonal, where we know that $\nAbar=\nBbar=\bar n$, with $\bar n$ given by Eq.~(\ref{eq:barn}). It is clear from Fig.~\ref{fig:appbone} that $(\nAbar)_2\ne(\nBbar)_2$ and $(\nAbar)_4\ne(\nBbar)_4$ for $\nuA=\nuB$, while it can be checked with numerical precision that $(\nAbar)_1=(\nBbar)_1=\bar n$ for all $\nuA=\nuB$ within $R_3$. Therefore we conclude that the correct solution along the symmetry diagonal in $R_3$ is represented by $(\nXbar)_1$. Out of the main diagonal we cannot choose the right solution based on Fig.~\ref{fig:appbone} alone, \ie, without \emph{a priori} information. For instance, we could split $R_3$ into two halves along the symmetry diagonal, then choose $(\nXbar)_4$ in the upper half and $(\nXbar)_2$ in the lower one. This would yield no stripes of swapped densities but a discontinuity along the symmetry diagonal. Otherwise, we could choose $(\nXbar)_2$ in the upper half and $(\nXbar)_4$ in the lower one. This would yield two regions of strongly reversed densities, resembling those observed in numerical simulations. Finally, we could choose $(\nXbar)_1$ in all $R_3$. This would yield two stripes of weakly reversed densities, likewise resembling the physical stripes. All these choices are equally acceptable. We conclude that the analytic approximation is ambiguous in $R_3$. 

In order to locate the cusp of $R_3$, we consider $(\nXbar)_1$ in the special case where $\tauA=\tauB=\tau$, for which $\kA=\kB=1$. It takes a few pages of scratch paper to work out $\Lambda$ and $\Sigma_+$ along the symmetry diagonal. We find
\begin{align}
  \Lambda & =  \frac{1}{2 \sqrt{6\tau^3}}\sqrt{\pi_0 + \frac{\pi_1+\left[2\left(\pi_2 + 3 \sqrt{\pi_3}\right)\right]^{2/3}}{ \left(\pi_2 + 3 \sqrt{\pi_3}\right)^{1/3}}}\,,
\end{align}
\begin{align}
  \Sigma_+ & =  \frac{1}{6 \tau ^3}\biggl[2\pi_0 -\frac{\pi_1+\left[ 2\left(\pi_2+3 \sqrt{\pi_3}\right)\right]^{2/3}}{\left(\pi_2+3 \sqrt{\pi_3}\right)^{1/3}}\nonumber\\
    & \hskip 0.2cm  -\frac{12 \sqrt{6}\,\tau^{3/2} (2 \alpha +\tau )^3 }{\displaystyle{\sqrt{\pi_0+ \frac{\pi_1 +\left[2\left(\pi_2+3 \sqrt{\pi_3}\right)\right]^{2/3}}{(\pi_2+3 \sqrt{\pi_3})^{1/3}}}}}\biggr]\,,
\end{align}
with 
\begin{align}
 \pi_0 & = 2 \tau  \left(5 \alpha ^2+10 \alpha  \tau +4 \tau ^2\right)\\[2.7ex]
 \pi_1 & = -2^{4/3} \tau ^2 \left(7 \alpha ^2+2 \alpha  \tau -\tau ^2\right)\nonumber\\[1.0ex]
 & \hskip 1.5cm \cdot \left(5 \alpha ^2+10 \alpha  \tau +4 \tau ^2\right)\,,\\[2.3ex]
 \pi_2 & = \tau ^3 \left(7 \alpha ^2+2 \alpha  \tau -\tau ^2\right)^2\nonumber\\[1.0ex]
 & \hskip 0.6cm \cdot \left(22 \alpha ^2+32 \alpha  \tau +11 \tau ^2\right)\,,\\[2.3ex]
 \pi_3 & = 3\tau ^6 (2 \alpha +\tau )^4 \left(7 \alpha^2+2 \alpha  \tau -\tau ^2\right)^3\nonumber\\[1.0ex]
 & \hskip 2.25cm \cdot \left(9 \alpha ^2+14 \alpha  \tau +5 \tau ^2\right) 
\end{align}
We checked these expressions against numerical results from standard math libraries for solving quartic equations. In particular, $\Sigma_+$ vanishes for $7\alpha^2+2\alpha\tau-\tau^2=0$, \ie\ for $\tau=1/\nu_\text{cusp}$ with $\nu_\text{cusp}$ given by Eq.~(\ref{eq:nucuspone}), as a consequence of
\begin{align}
  \pi_1|_{\tau=1/\nu_\text{cusp}} & = \pi_2|_{\tau=1/\nu_\text{cusp}} \nonumber\\[1.0ex]
  & = \pi_3|_{\tau=1/\nu_\text{cusp}} = 0\,,\\[2.0ex]
  & \hskip -1.8cm 2\pi_0^{3/2}|_{\tau=1/\nu_\text{cusp}} = 12\sqrt{6}\tau^{3/2}(2\alpha+\tau)^3|_{\tau=1/\nu_\text{cusp}}\,.
\end{align}


\end{appendices}

\newpage
 



\section*{Supplementary material}

# Supplemental Material: "Influence of periodic external fields in multi-agent models with language dynamics"

Filippo Palombi,[1, *] Stefano Ferriani,[2] and Simona Toti[3]

[1]ENEA—Italian National Agency for New Technologies, Energy and Sustainable Economic Development
Via Enrico Fermi 45, 00044 Frascati – Italy

[2]ENEA—Italian National Agency for New Technologies, Energy and Sustainable Economic Development
Via Martiri di Monte Sole 4, 40129 Bologna – Italy

[3]ISTAT—Italian National Institute of Statistics
Via Cesare Balbo 16, 00184 Rome – Italy

* filippo.palombi@enea.it

We collect here all density plots for the cases of two competing external fields with different frequencies and relative time shift. In Fig. SM1 the reader may find a synoptic table of networks and simulation parameters with corresponding figure numbers.

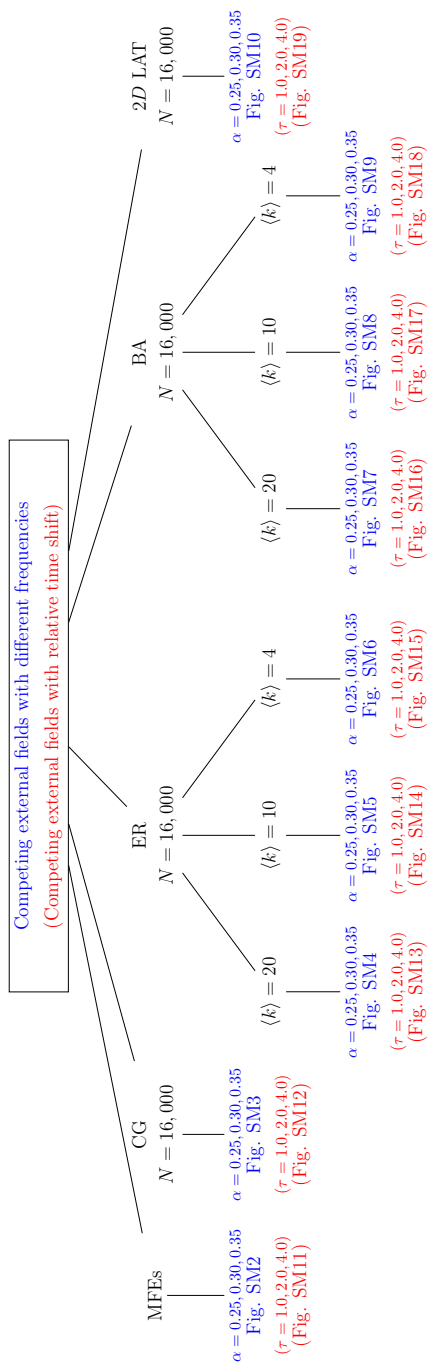

**Fig. SM1** − [color online] Synoptic table of networks and simulation parameters with figure numbers.

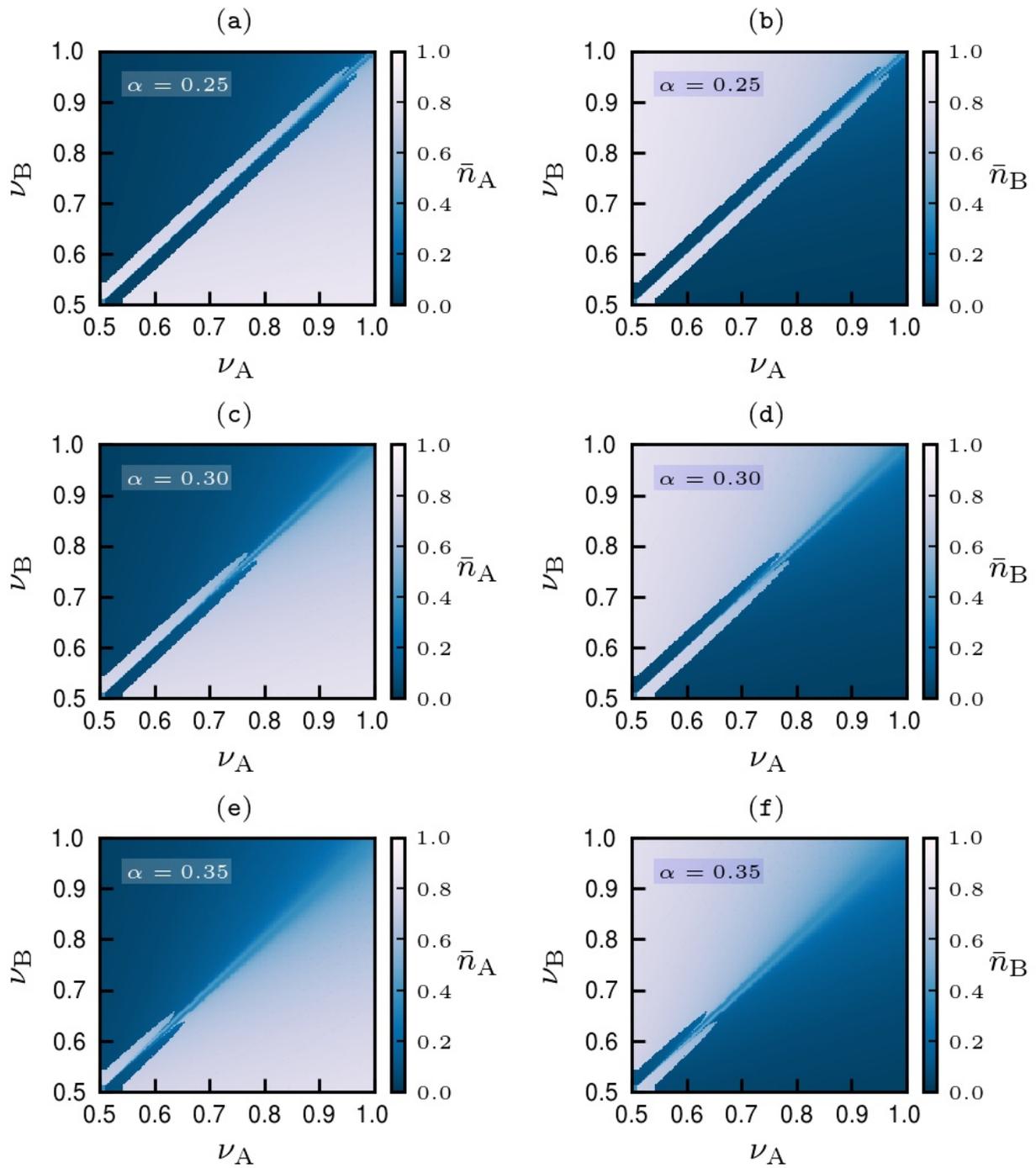

**Fig. SM2** – [color online] Average densities $\bar{n}_A$ and $\bar{n}_B$ at periodic equilibrium for $\tau_A \neq \tau_B$ from numerical integration of mean field equations in the explicit scheme.

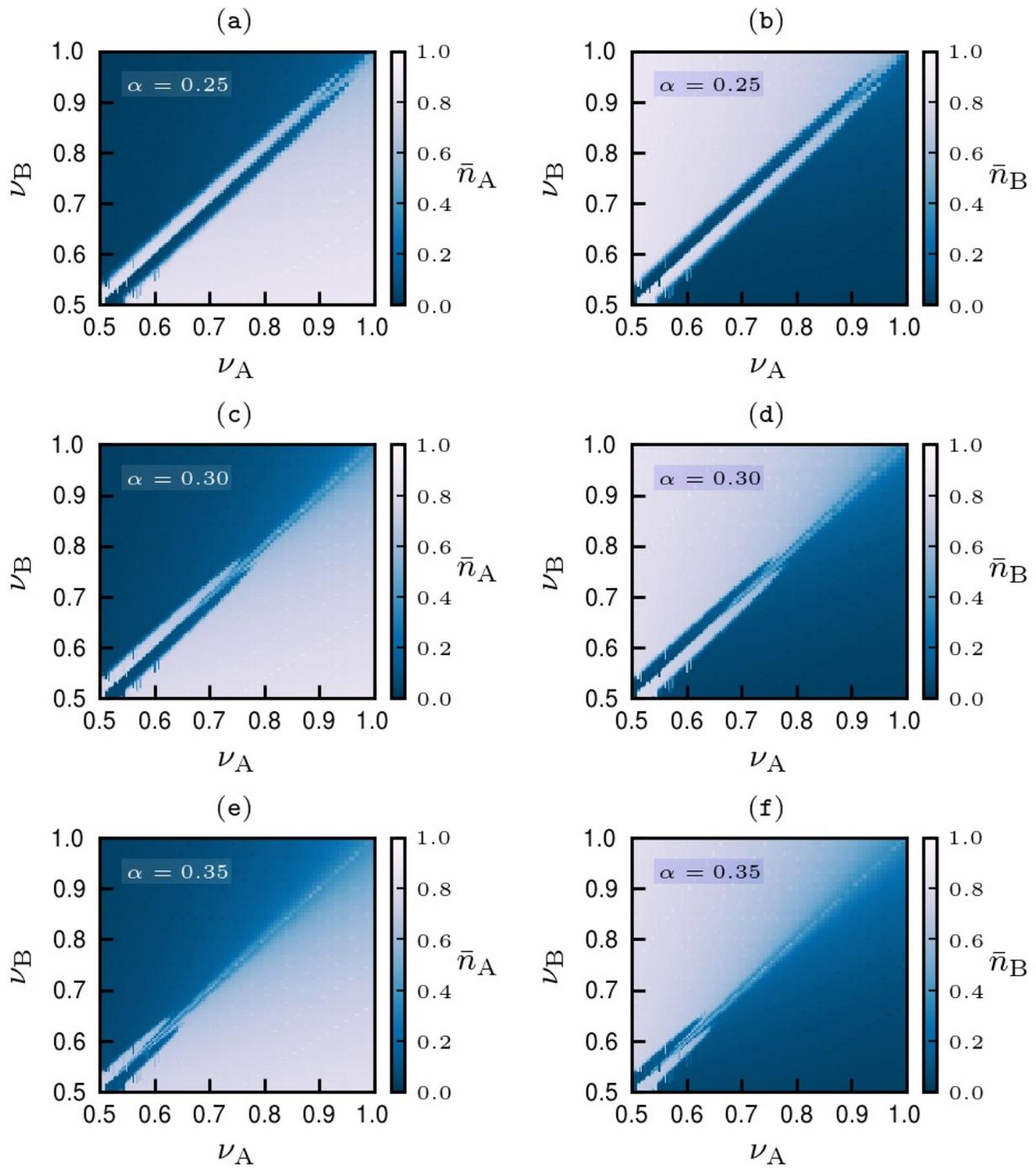

**Fig. SM3** – [color online] Average densities $\bar{n}_\mathrm{A}$ and $\bar{n}_\mathrm{B}$ at periodic equilibrium for $\tau_\mathrm{A} \neq \tau_\mathrm{B}$ from numerical simulations on a complete graph with $N = 16\,000$.

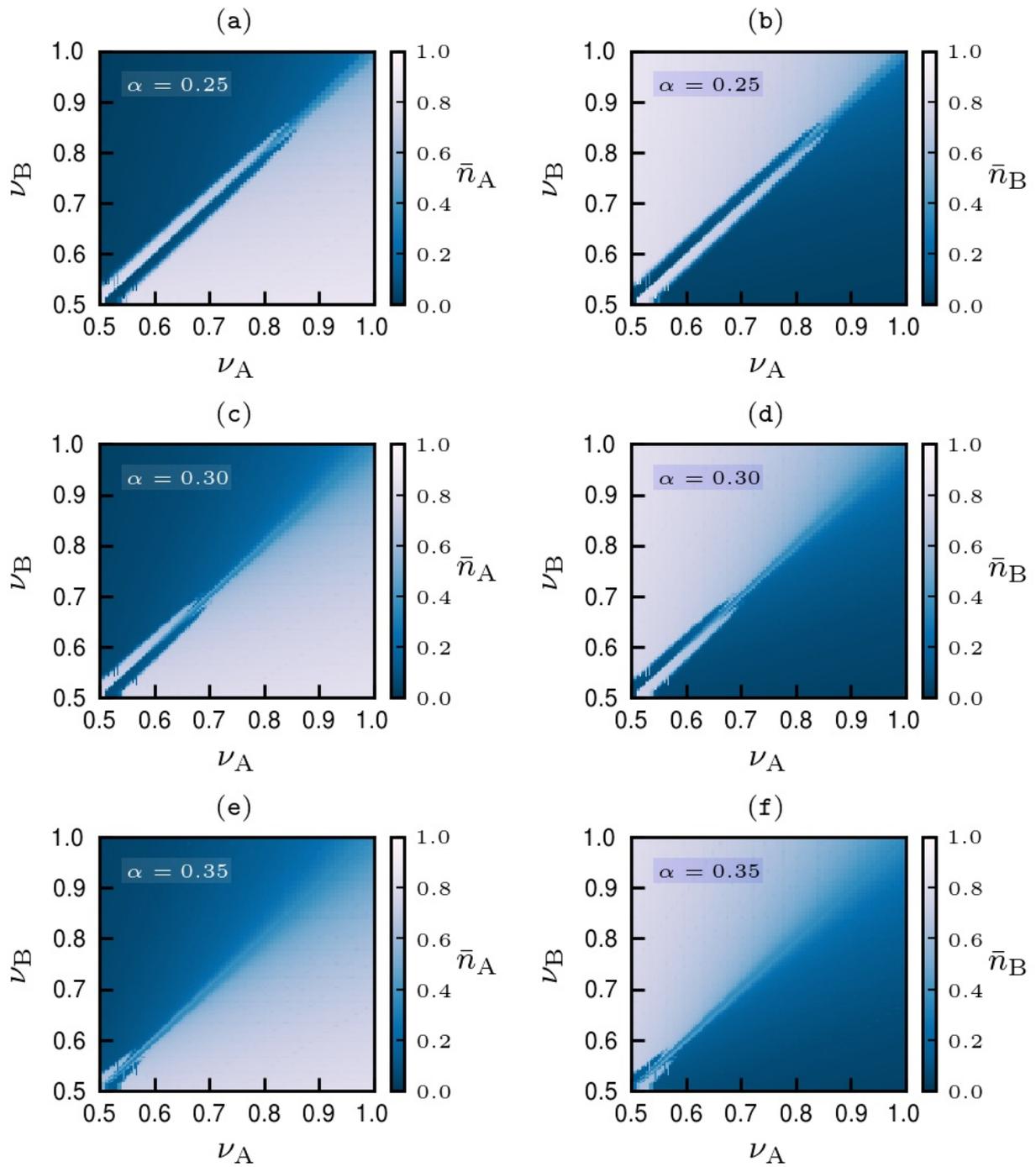

**Fig. SM4** – [color online] Average densities $\bar{n}_A$ and $\bar{n}_B$ at periodic equilibrium for $\tau_A \neq \tau_B$ from numerical simulations on Erdős-Rényi networks with $N = 16\,000$ and $\langle k \rangle = 20$.

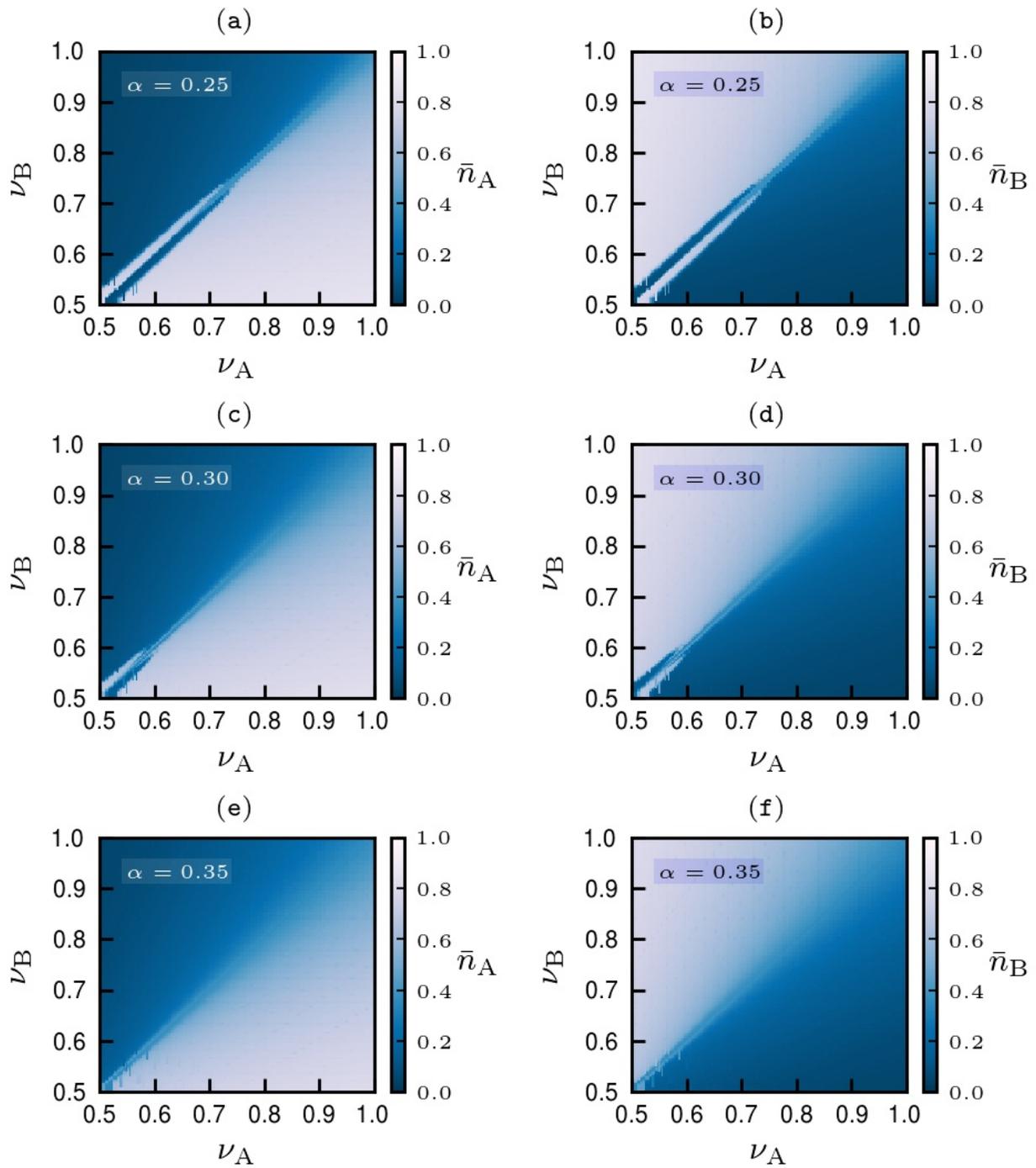

**Fig. SM5** – [color online] Average densities $\bar{n}_A$ and $\bar{n}_B$ at periodic equilibrium for $\tau_A \neq \tau_B$ from numerical simulations on Erdős-Rényi networks with $N = 16\,000$ and $\langle k \rangle = 10$.

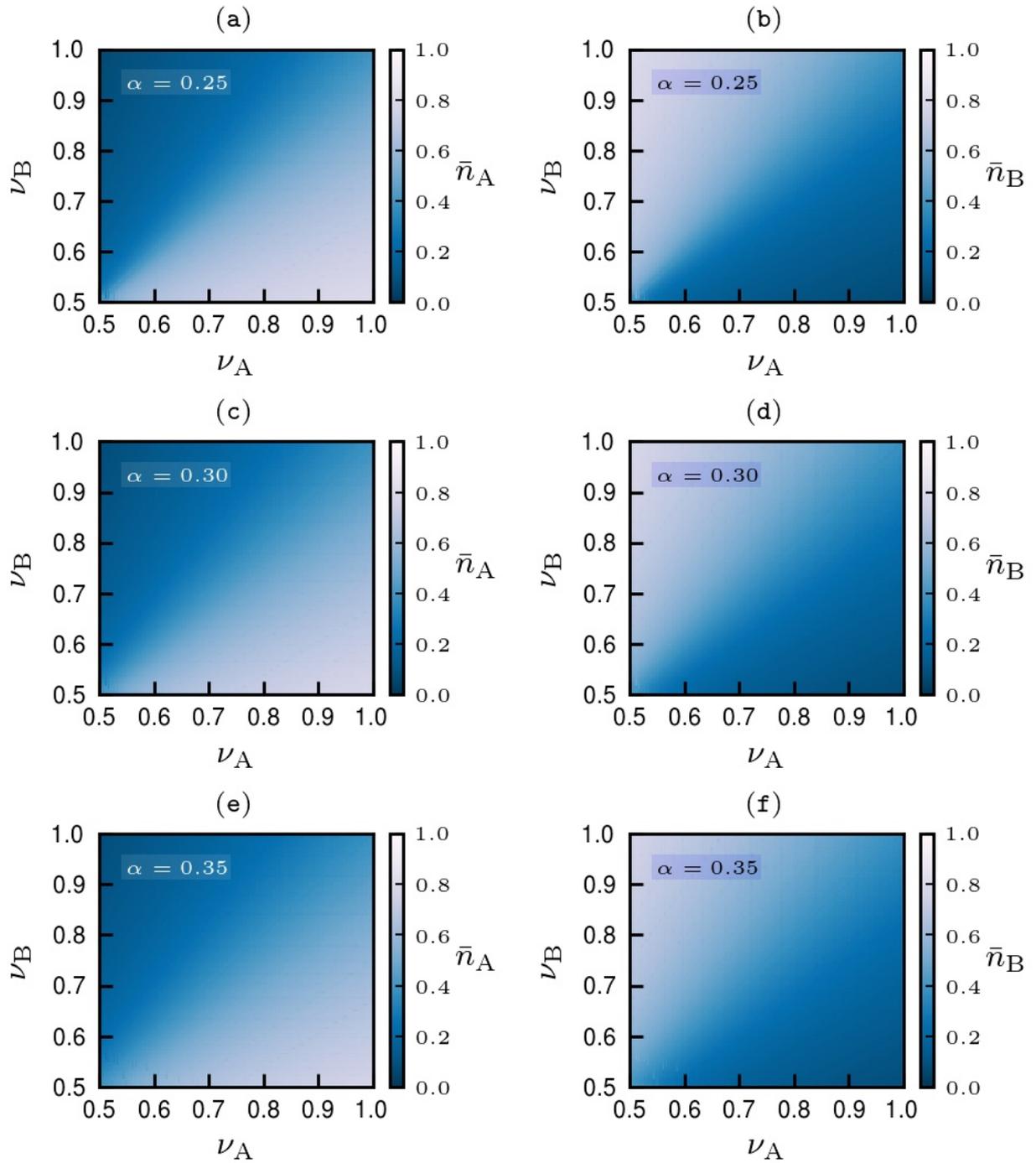

**Fig. SM6** – [color online] Average densities $\bar{n}_A$ and $\bar{n}_B$ at periodic equilibrium for $\tau_A \neq \tau_B$ from numerical simulations on Erdős-Rényi networks with $N = 16\,000$ and $\langle k \rangle = 4$.

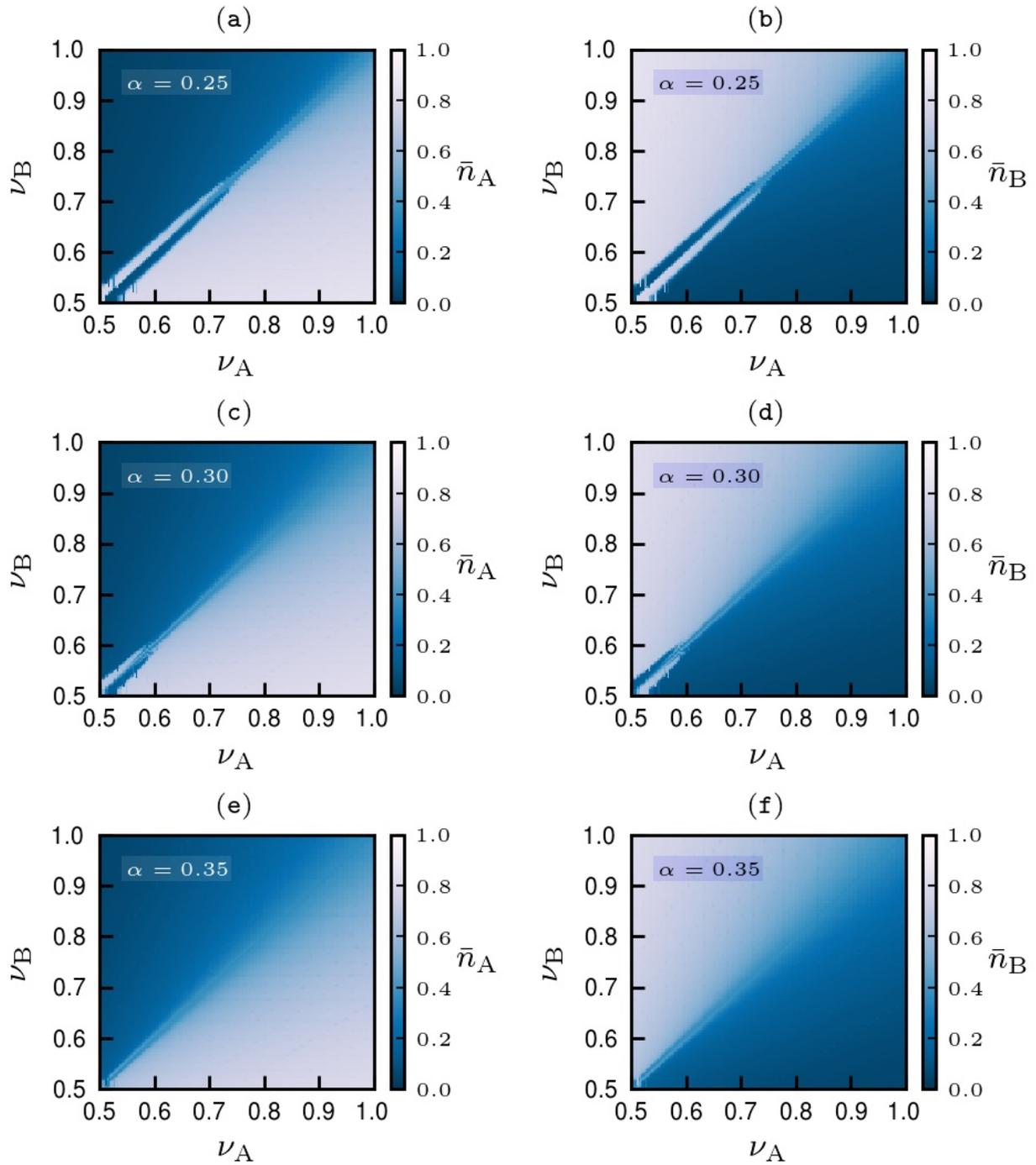

**Fig. SM7** – [color online] Average densities $\bar{n}_\mathrm{A}$ and $\bar{n}_\mathrm{B}$ at periodic equilibrium for $\tau_\mathrm{A} \neq \tau_\mathrm{B}$ from numerical simulations on Barabási–Albert networks with $N = 16\,000$ and $\langle k \rangle = 20$.

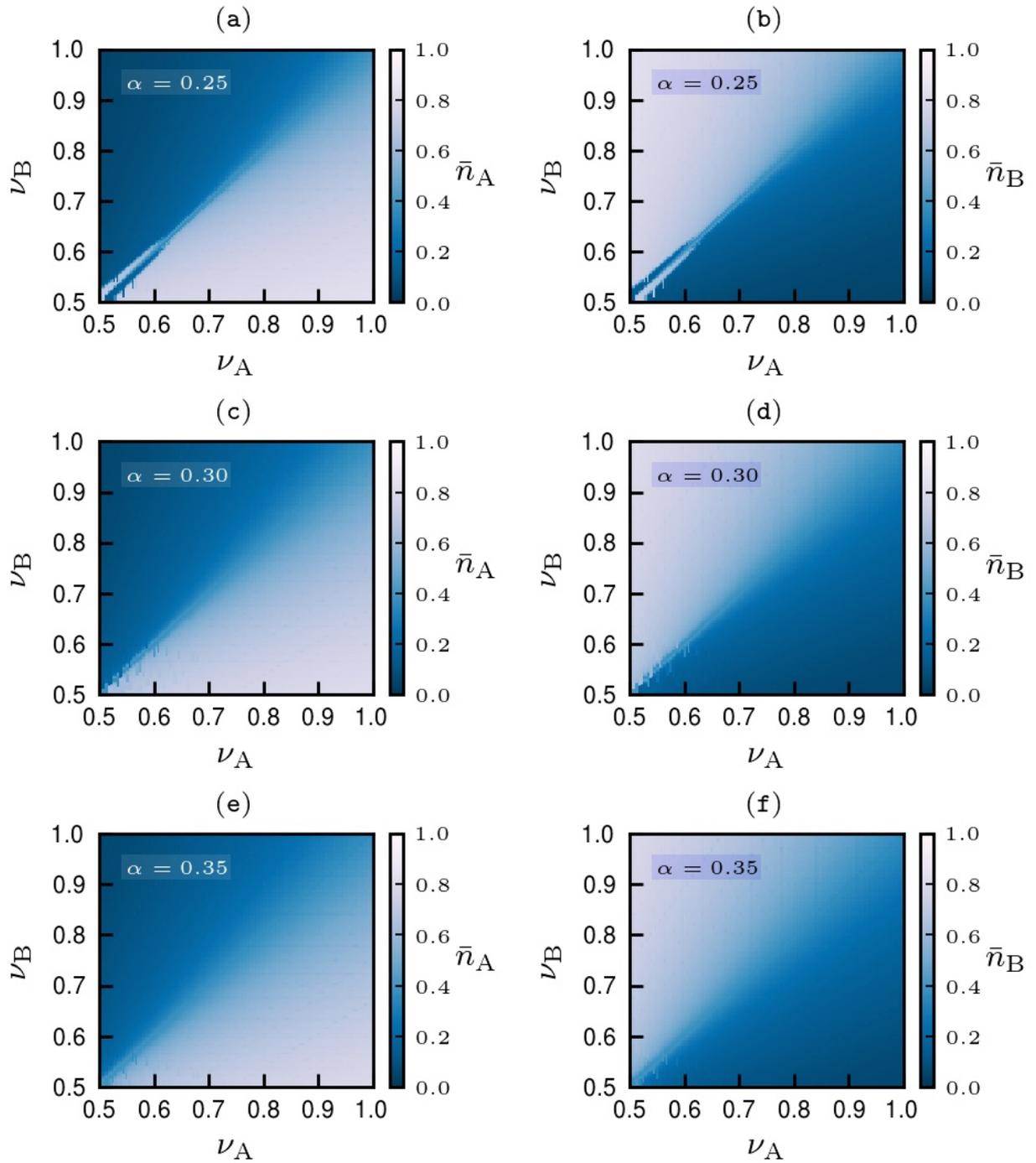

**Fig. SM8** – [color online] Average densities $\bar{n}_A$ and $\bar{n}_B$ at periodic equilibrium for $\tau_A \neq \tau_B$ from numerical simulations on Barabási–Albert networks with $N = 16\,000$ and $\langle k \rangle = 10$.

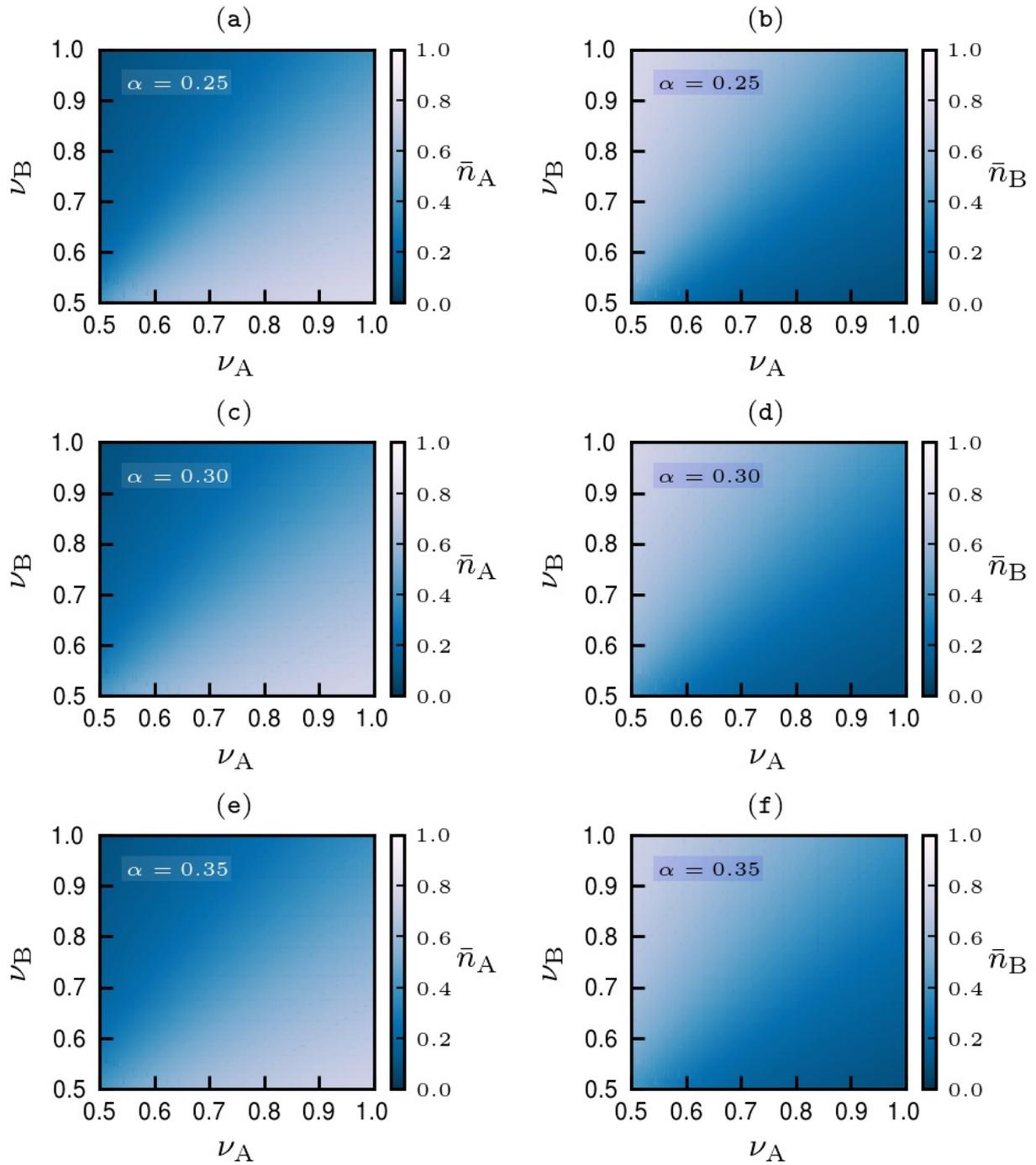

**Fig. SM9** – [color online] Average densities $\bar{n}_A$ and $\bar{n}_B$ at periodic equilibrium for $\tau_A \neq \tau_B$ from numerical simulations on Barabási–Albert networks with $N = 16\,000$ and $\langle k \rangle = 4$.

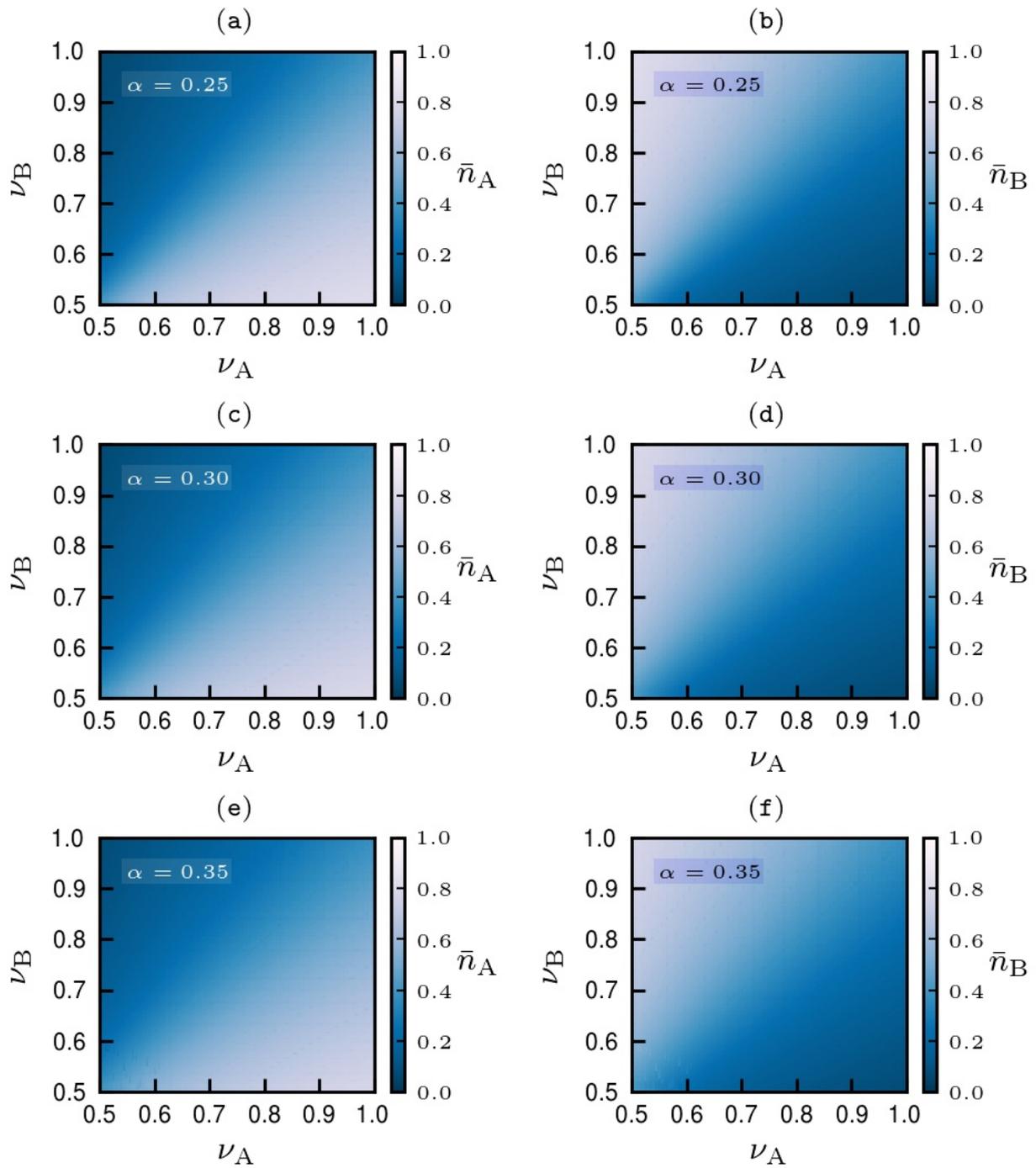

**Fig. SM10** – [color online] Average densities $\bar{n}_\mathrm{A}$ and $\bar{n}_\mathrm{B}$ at periodic equilibrium for $\tau_\mathrm{A} \neq \tau_\mathrm{B}$ from numerical simulations on a two-dimensional lattice with $N = 100 \times 100$.

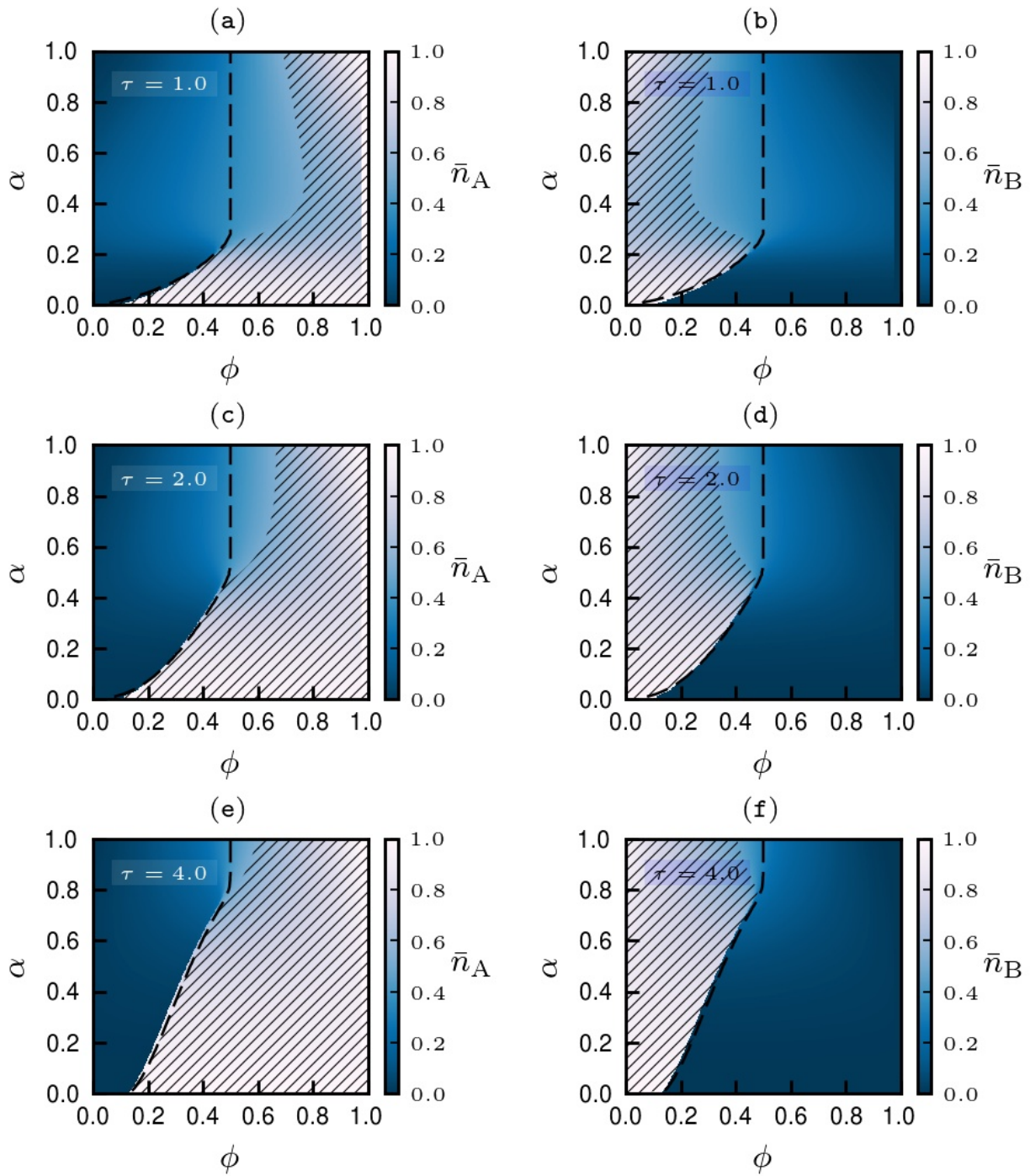

**Fig. SM11** – [color online] Average densities $\bar{n}_\mathrm{A}$ and $\bar{n}_\mathrm{B}$ at periodic equilibrium for $t_\mathrm{del} \neq 0$ from numerical integration of MFEs in the explicit scheme. The black dashed curve separates regions of opposite relative majority. The banded regions represent domains of absolute majority.

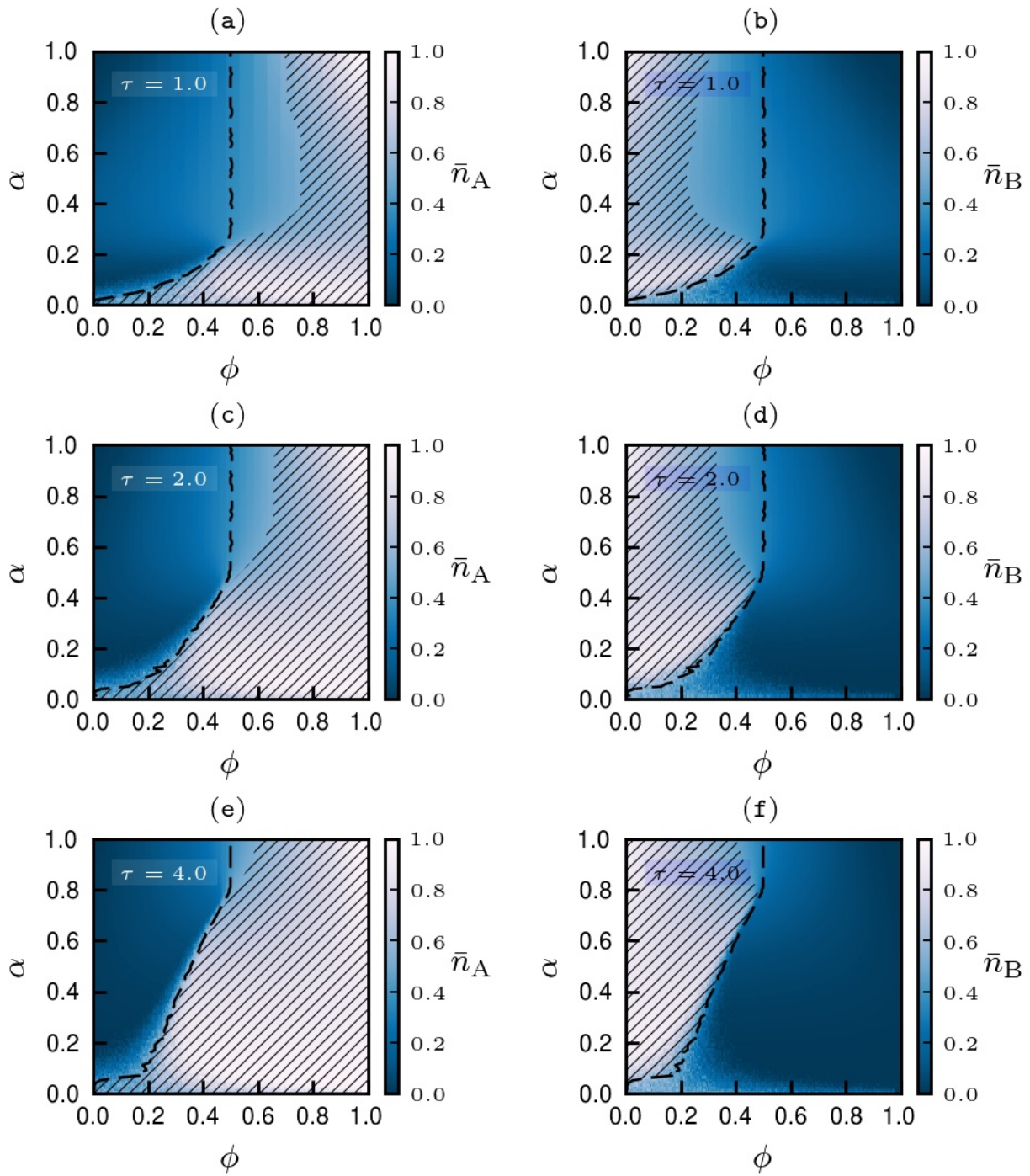

**Fig. SM12** – [color online] Average densities $\bar{n}_A$ and $\bar{n}_B$ at periodic equilibrium for $t_{del} \neq 0$ from numerical simulations on a complete graph with $N = 16\,000$. The black dashed curve separates regions of opposite relative majority. The banded regions represent domains of absolute majority.

14

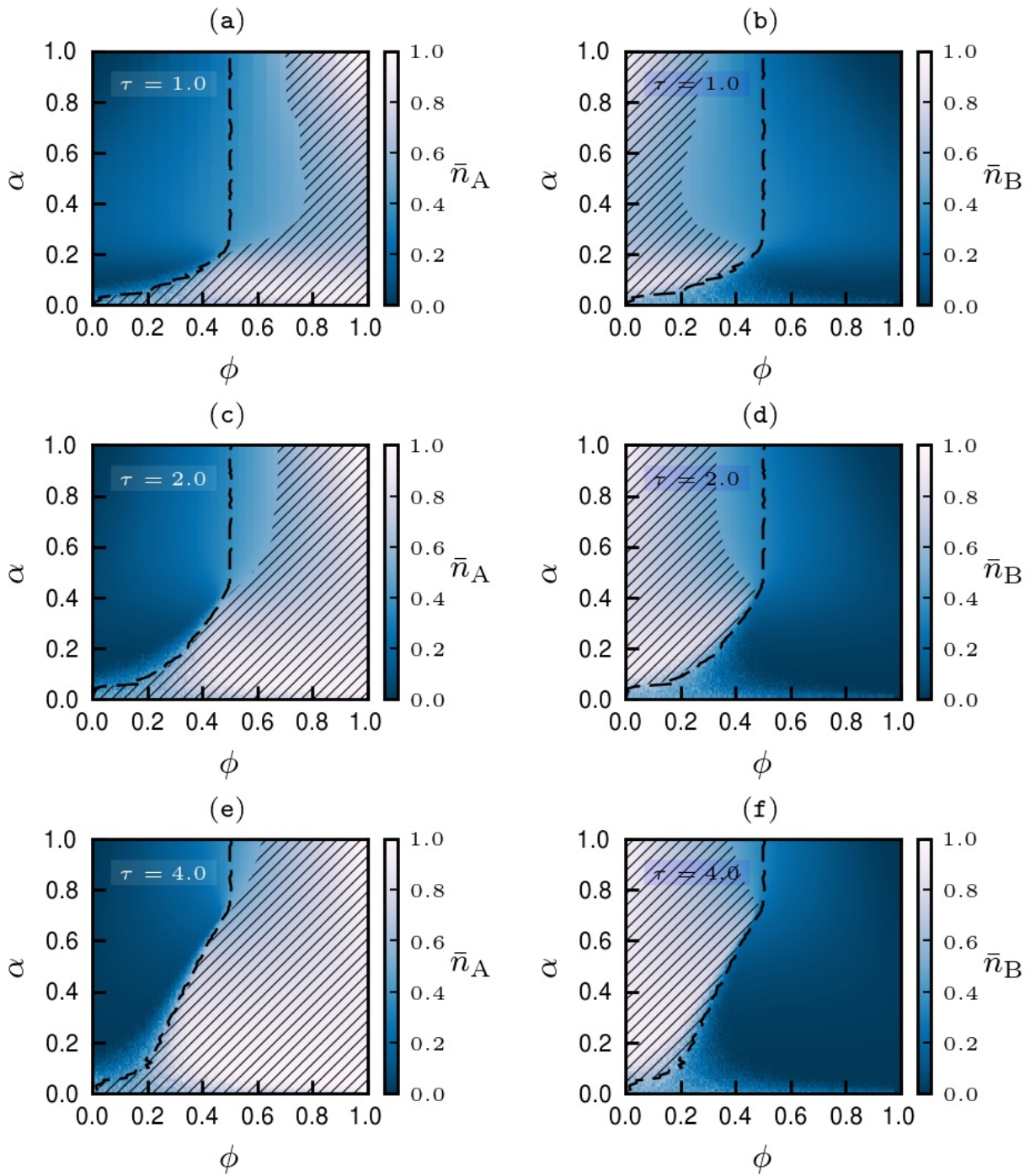

**Fig. SM13** – [color online] Average densities $\bar{n}_A$ and $\bar{n}_B$ at periodic equilibrium for $t_{\mathrm{del}} \neq 0$ from numerical simulations on Erdős-Rényi networks with $N = 16\,000$ and $\langle k \rangle = 20$. The black dashed curve separates regions of opposite relative majority. The banded regions represent domains of absolute majority.

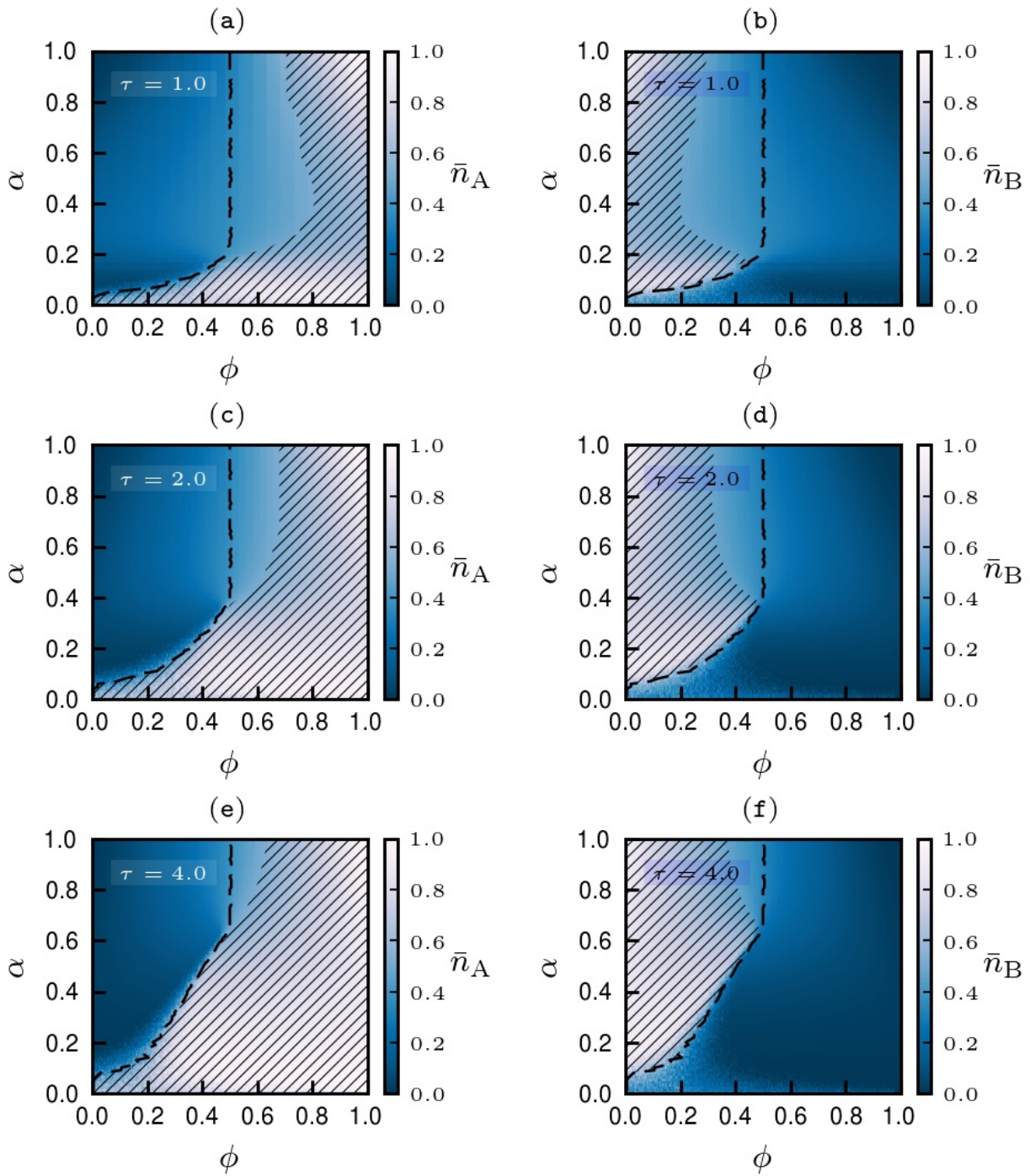

**Fig. SM14** – [color online] Average densities $\bar{n}_A$ and $\bar{n}_B$ at periodic equilibrium for $t_{del} \neq 0$ from numerical simulations on Erdős-Rényi networks with $N = 16\,000$ and $\langle k \rangle = 10$. The black dashed curve separates regions of opposite relative majority. The banded regions represent domains of absolute majority.

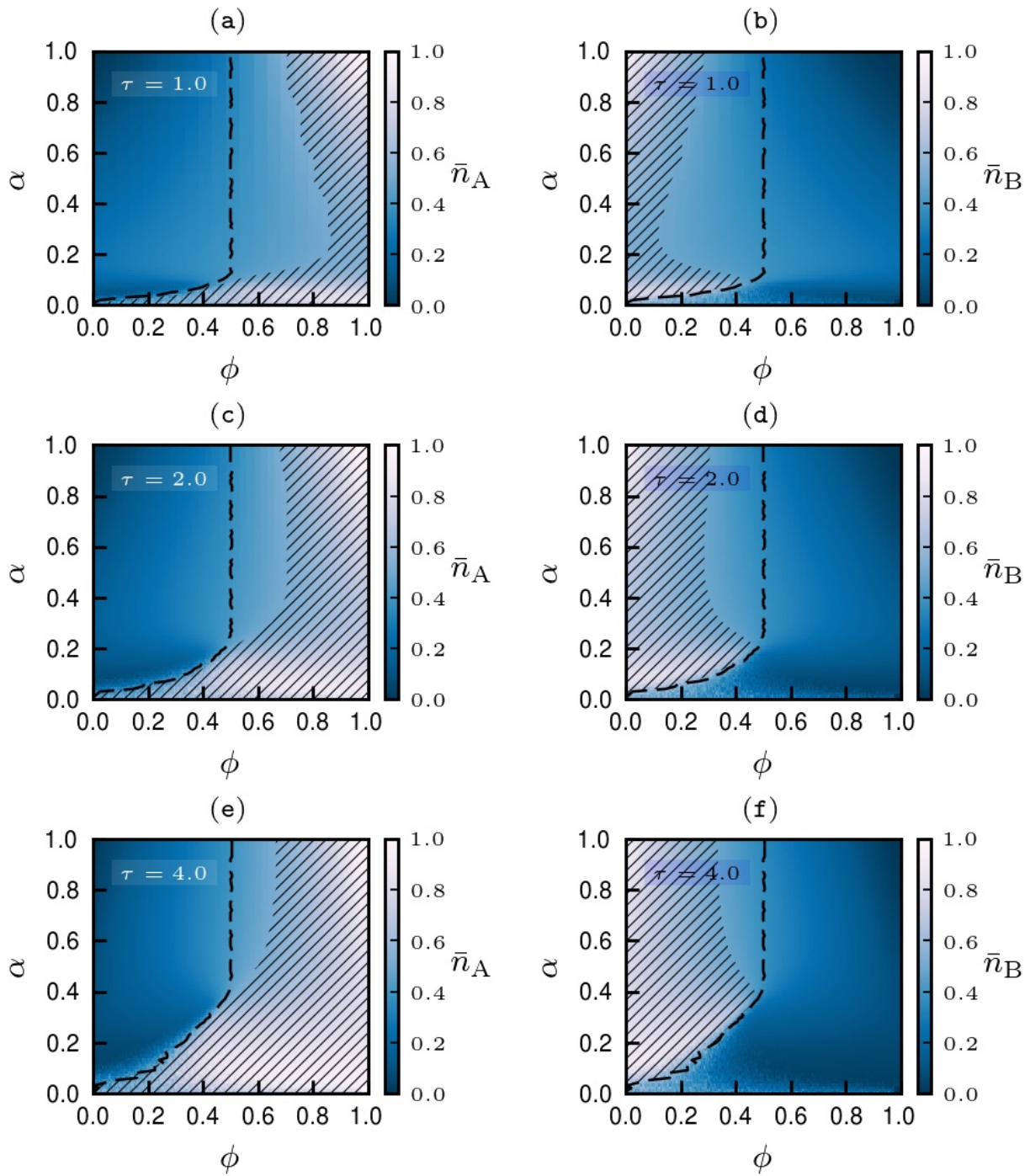

**Fig. SM15** – [color online] Average densities $\bar{n}_A$ and $\bar{n}_B$ at periodic equilibrium for $t_{del} \neq 0$ from numerical simulations on Erdős-Rényi networks with $N = 16\,000$ and $\langle k \rangle = 4$. The black dashed curve separates regions of opposite relative majority. The banded regions represent domains of absolute majority.

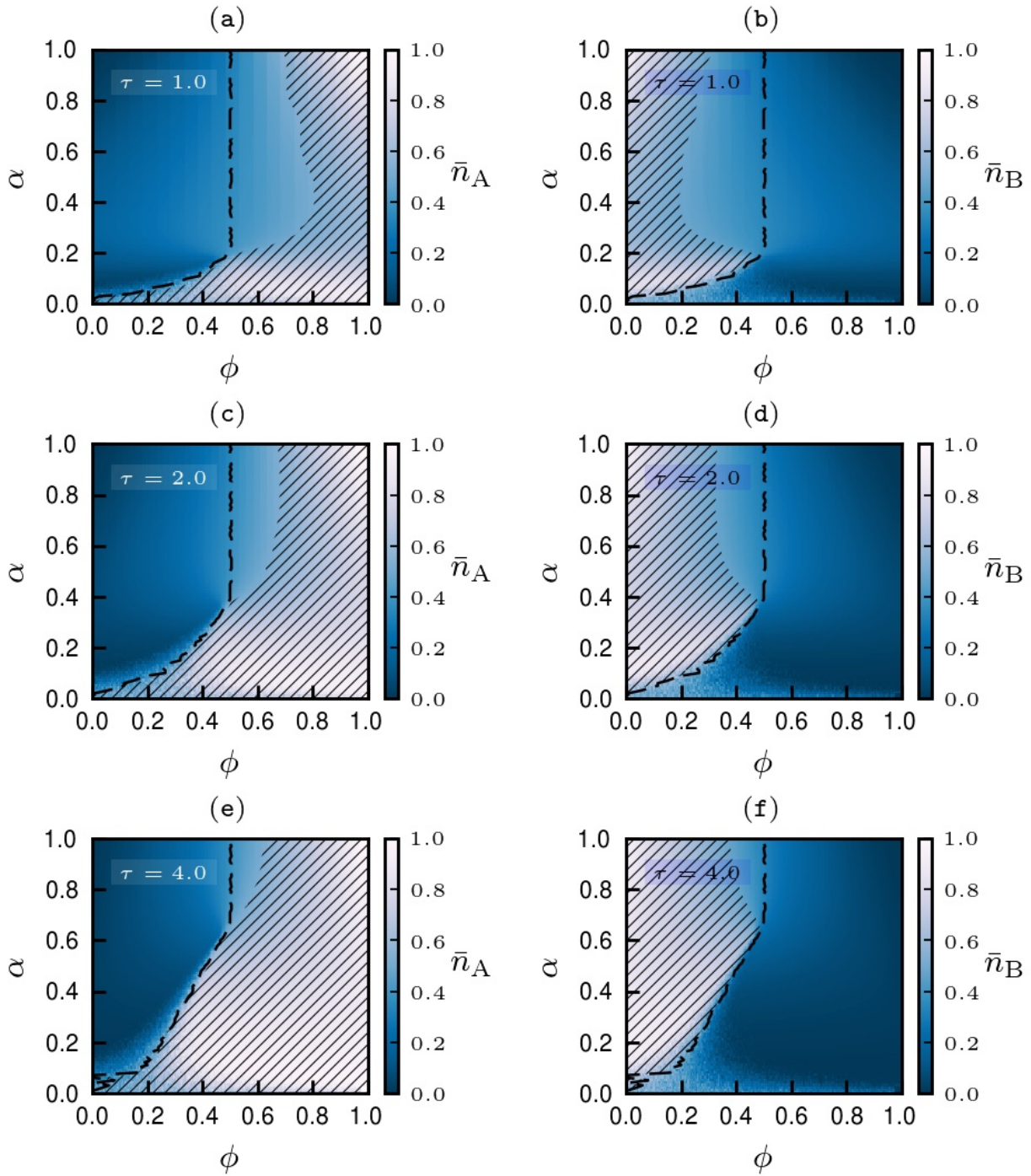

**Fig. SM16** – [color online] Average densities $\bar{n}_{\mathrm{A}}$ and $\bar{n}_{\mathrm{B}}$ at periodic equilibrium for $t_{\mathrm{del}} \neq 0$ from numerical simulations on Barabási–Albert networks with $N = 16\,000$ and $\langle k \rangle = 20$. The black dashed curve separates regions of opposite relative majority. The banded regions represent domains of absolute majority.

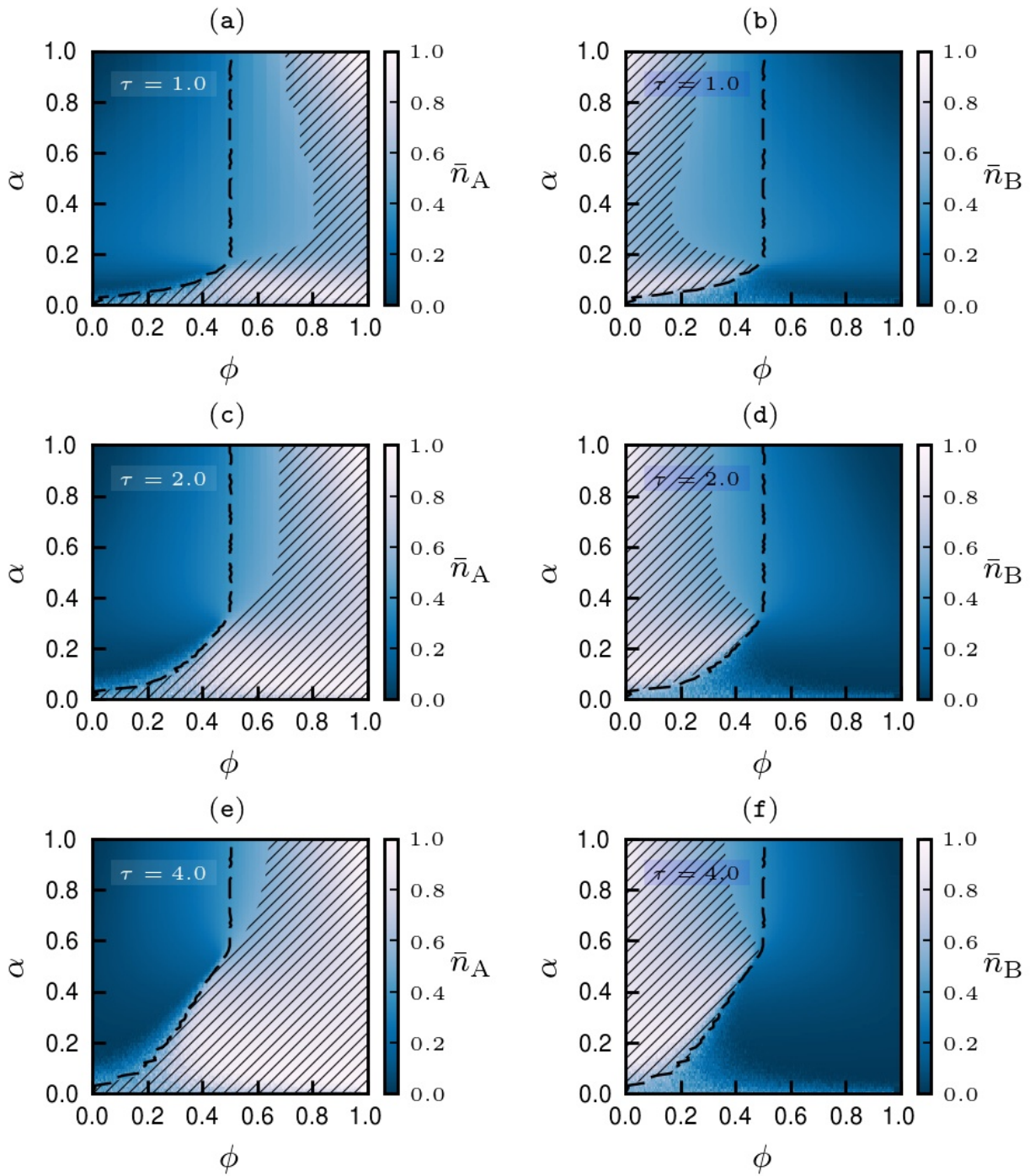

**Fig. SM17** – [color online] Average densities $\bar{n}_A$ and $\bar{n}_B$ at periodic equilibrium for $t_{\text{del}} \neq 0$ from numerical simulations on Barabási–Albert networks with $N = 16\,000$ and $\langle k \rangle = 10$. The black dashed curve separates regions of opposite relative majority. The banded regions represent domains of absolute majority.

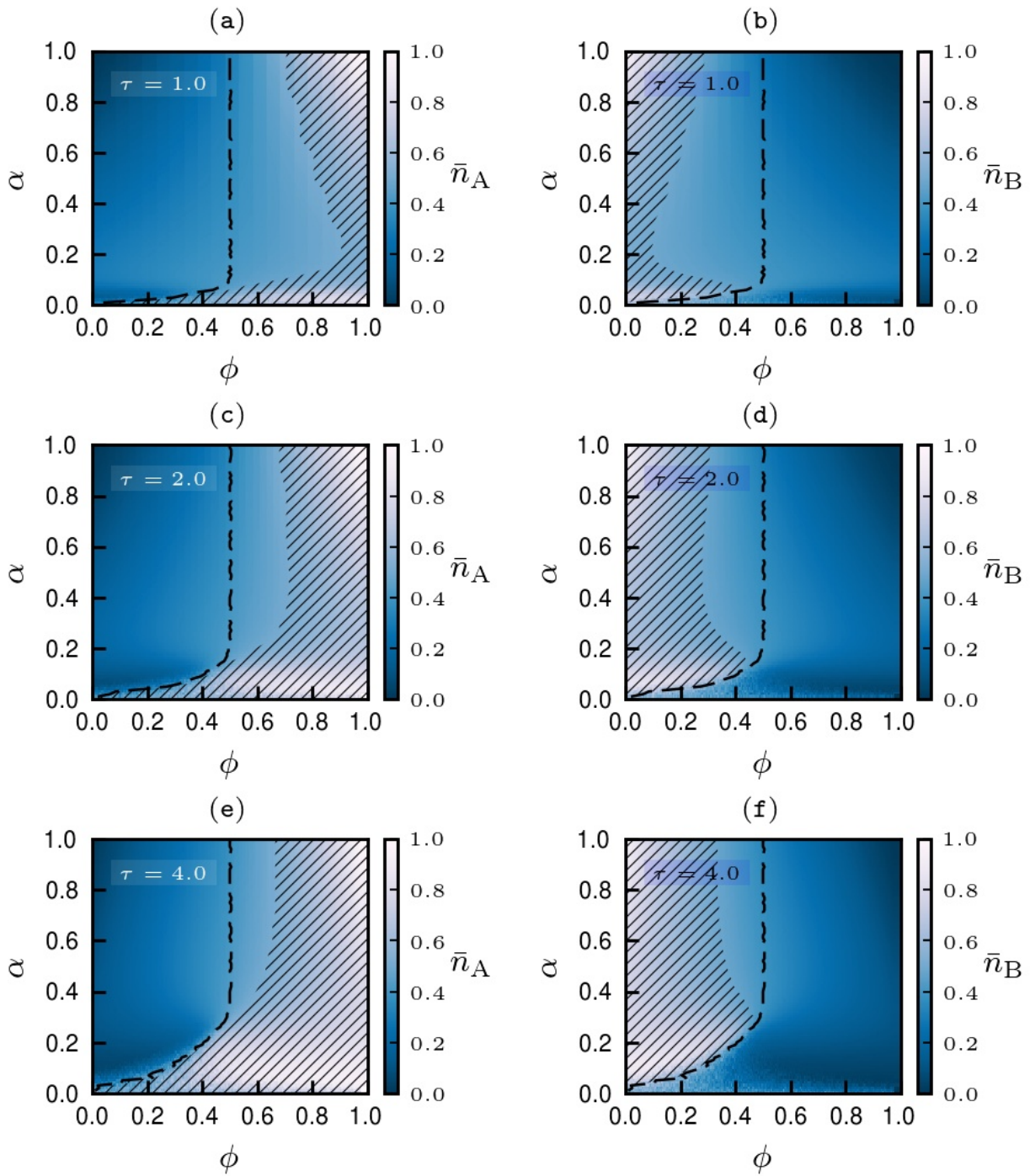

**Fig. SM18** – [color online] Average densities $\bar{n}_{\mathrm{A}}$ and $\bar{n}_{\mathrm{B}}$ at periodic equilibrium for $t_{\mathrm{del}} \neq 0$ from numerical simulations on Barabási–Albert networks with $N = 16\,000$ and $\langle k \rangle = 4$. The black dashed curve separates regions of opposite relative majority. The banded regions represent domains of absolute majority.

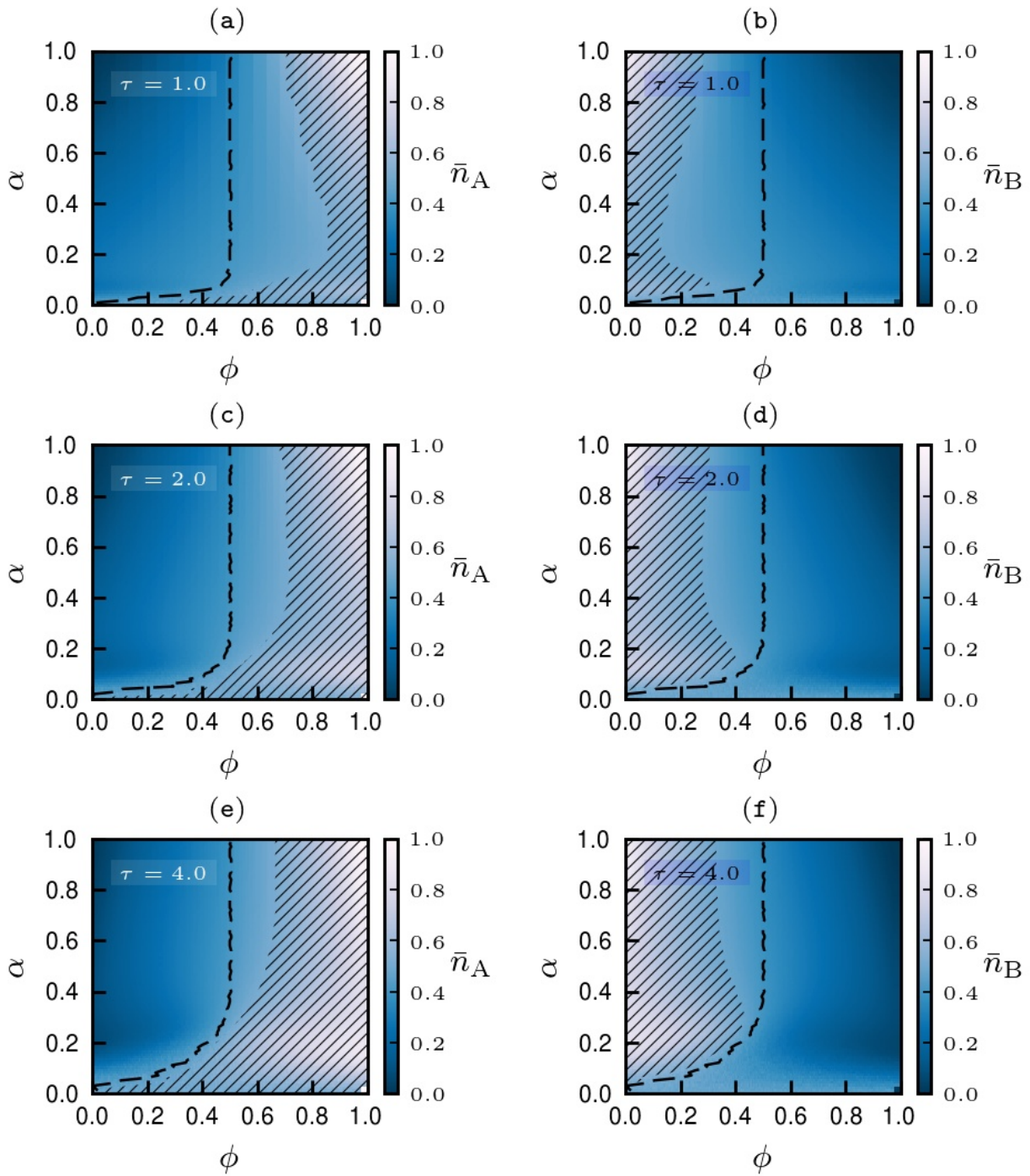

**Fig. SM19** − [color online] Average densities $\bar{n}_A$ and $\bar{n}_B$ at periodic equilibrium for $t_{\text{del}} \neq 0$ from numerical simulations on a two-dimensional lattice with $N = 100 \times 100$. The black dashed curve separates regions of opposite relative majority. The banded regions represent domains of absolute majority.



\end{document}